\pdfoutput=1
\documentclass{aastex61_noPage}
\usepackage[T1]{fontenc}
\usepackage{ae,aecompl}
\usepackage{graphicx}
\usepackage[caption=false]{subfig}
\usepackage{newtxtext,newtxmath}
\usepackage[T1]{fontenc}
\usepackage{ae,aecompl}

\usepackage{graphicx}
\usepackage{epsf}
\usepackage{longtable}
\usepackage{rotating}
\usepackage{bm}
\usepackage{color}
\usepackage{amsmath}

\usepackage{booktabs}
\usepackage{braket}
\usepackage{multirow}
\usepackage{natbib}

\begin{document}

\title{Non-Gaussianity of Secondary Anisotropies  from ACTPol and Planck}
\correspondingauthor{William R. Coulton}
\email{wcoulton@princeton.edu}
\author{William R. Coulton}
\affiliation{Joseph Henry Laboratories, Princeton University, Princeton, NJ 08544, USA}
\author{Simone Aiola}
\affiliation{Joseph Henry Laboratories, Princeton University, Princeton, NJ 08544, USA}
\author{Nicholas~Battaglia}
\affiliation{Department of Astrophysical Sciences, Princeton University,Peyton Hall, Princeton, NJ 08544, USA}
\affiliation{Center for Computational Astrophysics, Flatiron Institute,162 5th Avenue, 10010, New York, NY, USA}
\author{Erminia~Calabrese}
\affiliation{School of Physics and Astronomy, Cardiff University, The Parade, Cardiff, CF24 3AA, UK}
\author{ Steve K. Choi}
\affiliation{Joseph Henry Laboratories, Princeton University, Princeton, NJ 08544, USA}
\author{Mark J. Devlin}
\affiliation{Department of Physics and Astronomy, University of Pennsylvania, 209 South 33rd Street, Philadelphia, PA, USA 19104}
\author{Patricio A. Gallardo}
\affiliation{Department of Physics, Cornell University, Ithaca, NY 14853, USA}
\author{ J.~Colin~Hill}
\affiliation{Center for Computational Astrophysics, Flatiron Institute,162 5th Avenue, 10010, New York, NY, USA}
\affiliation{School of Natural Sciences, Institute for Advanced Study, Olden Lane, Princeton, NJ 08540}
\author{ Adam~D.~Hincks}
\affiliation{Department of Physics, University of Rome ``La Sapienza'', Piazzale Aldo Moro 5, I-00185 Rome, Italy}
\affiliation{Department of Physics and Astronomy, University of British Columbia, Vancouver, BC, Canada V6T 1Z4}
\author{Johannes Hubmayr}
\affiliation{NIST Quantum Devices Group, 325 Broadway, Mailcode 817.03, Boulder, CO 80305, USA}
\author{John~P.~Hughes}
\affiliation{Department of Physics and Astronomy, Rutgers,  The State University of New Jersey, Piscataway, NJ USA 08854-8019}
\author{Arthur~Kosowsky}
\affiliation{Department of Physics and Astronomy, University of Pittsburgh, Pittsburgh, PA, USA 15260}
\affiliation{Pittsburgh Particle Physics, Astrophysics, and Cosmology Center, University of Pittsburgh, Pittsburgh PA 15260}
\author{Thibaut~Louis}
\affiliation{Laboratoire de l'Acc\'el\'erateur Lin\'eaire, Univ. Paris-Sud,CNRS/IN2P3, Universit\'e Paris-Saclay, Orsay, France}
\author{Mathew S. Madhavacheril}
\affiliation{Department of Astrophysical Sciences, Princeton University,Peyton Hall, Princeton, NJ 08544, USA}
\author{Lo{\"i}c Maurin}
\affiliation{Instituto de Astrof\'isica and Centro de Astro-Ingenier\'ia, Facultad de F\'isica, Pontificia Universidad Cat\'olica de Chile, Av. Vicu\~na Mackenna 4860, 7820436 Macul, Santiago, Chile}
\author{Sigurd~Naess}
\affiliation{Center for Computational Astrophysics, Flatiron Institute,162 5th Avenue, 10010, New York, NY, USA}
\author{Federico Nati}
\affiliation{Department of Physics and Astronomy, University of Pennsylvania, 209 South 33rd Street, Philadelphia, PA, USA 19104}
\author{Michael D. Niemack}
\affiliation{Department of Physics, Cornell University, Ithaca, NY 14853, USA}
\author{Lyman~A.~Page}
\affiliation{Joseph Henry Laboratories, Princeton University, Princeton, NJ 08544, USA}
\author{Bruce~Partridge}
\affiliation{Department of Physics and Astronomy, Haverford College, Haverford, PA, USA 19041}
\author{Blake D. Sherwin}
\affiliation{Department of Applied Mathematics and Theoretical Physics, University of Cambridge, Wilberforce Road, Cambridge CB3 0WA}
\affiliation{Berkeley Center for Cosmological Physics, University of California, Berkeley, CA 94720}
\author{David N. Spergel}
\affiliation{Department of Astrophysical Sciences, Princeton University,Peyton Hall, Princeton, NJ 08544, USA}
\affiliation{Center for Computational Astrophysics, Flatiron Institute,162 5th Avenue, 10010, New York, NY, USA}
\author{Suzanne T. Staggs}
\affiliation{Joseph Henry Laboratories, Princeton University, Princeton, NJ 08544, USA}
\author{Alexander~Van~Engelen}
\affiliation{Canadian Institute for Theoretical Astrophysics, University of Toronto, Toronto, ON, Canada M5S 3H8}
\author{Edward~J.~Wollack}
\affiliation{NASA/Goddard Space Flight Center, Greenbelt, MD, USA 20771}

\begin{abstract}
Most secondary sources of cosmic microwave background anisotropy (radio sources, dusty galaxies, thermal Sunyaev Zel'dovich distortions from hot gas, and gravitational lensing) are highly non-Gaussian. Statistics beyond the power spectrum are therefore potentially important sources of information about the physics of these processes. We combine data from the Atacama Cosmology Telescope and with data from the \textit{Planck} satellite (only using \textit{Planck} data in the overlapping region) to constrain the amplitudes of a set of theoretical bispectrum templates from the thermal Sunyaev-Zeldovich (tSZ) effect, dusty star-forming galaxies (DSFGs), gravitational lensing, and radio galaxies. We make a strong detection of radio galaxies (>5$\sigma$) and have hints of non-Gaussianity arising from the tSZ effect, DSFGs, from cross-correlations between the tSZ effect and DSFGs and from cross-correlations among the tSZ effect, DSFGs and radio galaxies.  These results suggest that the same halos host radio sources, DSFGs, and have tSZ signal. We present a new method to calculate the non-Gaussian contributions to the template covariances. Using this method we find significant non-Gaussian contributions to the variance and covariance of our templates, with templates involving the tSZ effect most effected.  Strong degeneracies exist between the various sources at the current noise levels. In light of these degeneracies, combined with theoretical uncertainty in the templates, these results are a demonstration of this technique. With these caveats, we demonstrate the utility of future bispectrum measurements by using the tSZ bispectrum measurement to constrain a combination of the amplitude of matter fluctuations and the matter density to be $\sigma_8 \Omega_m^{0.17}=0.65^{+0.05}_{-0.06}$. Improvements in signal to noise from upcoming  Advanced ACT, SPT-3G, Simons Observatory, and CMB-S4 observations will enable the separation of bispectrum components and robust constraints on cosmological parameters.

\end{abstract}

\section{Introduction}
Secondary sources of cosmic microwave background (CMB) anisotropy occur as photons propagate from the surface of last scattering to us. Lensing distortions arise as the paths of CMB photons are deflected by intervening matter \citep{Blanchard1987}. The thermal Sunyaev Zel'dovich (tSZ) effect is induced by the inverse Compton scattering of CMB photons as they travel through ionized gas, distorting the frequency spectrum of the photons \citep{Sunyaev1970}. The additional kinematic Sunyaev-Zel'dovich (kSZ) effect produces a near-blackbody temperature shift if the ionized gas possesses a bulk velocity along the line of sight \citep{Sunyaev1972}. Radio galaxies and dusty star-forming galaxies (DSFGs) are the two dominant populations of extragalactic emissive sources at millimeter wavelengths. Radio galaxies have large levels of synchrotron emission, primarily from active galactic nuclei, and in dusty star-forming galaxies the radiation arises from thermal dust emission. 

Many of these secondary sources have been intensely studied. The power spectrum of the tSZ effect was measured by \cite{planck2014-a28} and its amplitude on small angular scales has been measured \citep[e.g.][]{dunkley2013,george2015,planck2014-a13}. The thermal SZ effect in galaxy clusters has been studied through cluster number counts \citep[e.g.][]{Staniszewski2009,HasselfieldClust2013,deHaan2016,planck2014-a30} and through cross-correlations with numerous other tracers  \citep[e.g.][]{planck2014-a29,hill2014,Hojjati2017,Waerbeke2014}. The gas properties in these clusters have been studied at low redshifts through X-ray measurements \citep{arnaud2010}. The dusty star-forming galaxies have been measured through their number counts with \textit{Herschel} and \textit{Spitzer} \citep[e.g.][]{Glenn2010,berta2011,bethermin2010} and power spectrum \citep{planck2013-pip56} as well as through cross-correlations with lensing and other tracers \citep[e.g.][]{planck2013-p13,planck2014-a29,vanEngelen2015}. Similarly the radio galaxies' number counts have been measured \citep{deZotti2010} and hints of their correlation with the tSZ effect have been seen \citep{Gralla2013,Gupta2017}. Finally the kSZ effect has been studied through a wide variety of different methods and different data sets \citep[e.g.][]{hand2012,DeBernardis2017,planck2015-XXXVII,Soergel2016, planck2017-LIII,schaan2016,Hill2016}. All of these measurements have provided information about the astrophysics of these sources, but the non-Gaussian nature of these sources means that we can gain complementary information by analyzing the higher order correlation functions \cite[e.g.][]{wilson2012,Hill2013,Crawford2014,hill2014b,Cooray2000,Timmons2017}.

Here we focus on the bispectrum from secondary sources. The bispectrum is the harmonic-space transform of the spatial three-point correlation function, which is zero for a purely Gaussian random field. Numerous observed and hypothesized sources give a nonzero bispectrum contribution in the microwave sky. These fall into four categories: primordial sources, secondary sources, second-order gravitational terms and galactic foregrounds. Primordial non-Gaussianity offers another handle on constraining the physics of the early universe and could provide the ability to differentiate between many classes of inflationary theories, though so far it has been found to be consistent with zero \citep{komatsu2011,planck2013-p09a,planck2014-a19}; see \cite{liguori2010Review,Yadav2010review,chen2010review} for reviews of primordial non-Gaussianity. Second-order gravitational terms arise from non-linearities in the transfer function from the initial perturbations to the late time temperature fluctuations in CMB \citep{Pettinari2013,Bartolo2006,bartolo2007,Pitrou2010}. We do not discuss these here as they are below the sensitivity of our experiments but they will be important for future, more sensitive measurements \citep{Pettinari2014}. Galactic foreground sources include thermal dust emission and synchrotron emission; these are typically most important on the largest scales (degree scales). As we restrict our analysis to small angular scales ($\ell>200$), these sources of non-Gaussianity are not considered in this work \citep{gold2011}. We focus on investigating the bispectrum from CMB secondary sources.

The large scale DSFG bispectrum has been measured in \cite{planck2013-pip56} as has the large scale bispectrum from the tSZ effect \citep{planck2014-a28} and the bispectrum from correlations between the Integrated Sachs-Wolfe and gravitational lensing \citep{planck2014-a19,planck2013-p05b}. The skewness of the tSZ effect was examined in \cite{wilson2012} and \cite{Hill2013} and the one point one point probability density function (PDF) of the tSZ effect has been studied in \cite{hill2014b}. The bispectrum from the kSZ effect and galaxies has been probed by pairwise estimators \citep{hand2012,DeBernardis2017,Soergel2016, planck2015-XXXVII}, by combining stacked clusters with their reconstructed velocities \citep{schaan2016}, and through projected field estimators \citep{Hill2016}. \cite{Crawford2014} explored the bispectrum from secondary sources constraining the small-scale tSZ effect, DSFG and Poisson contributions to the bispectrum. Our work builds on aspects of these previous works; in particular we extend these analyses to include the one-halo contributions to the cross-correlations among the tSZ effect, DSFGs and radio galaxies, such as tSZ-tSZ-DSFG and radio-DSFG-tSZ bispectra. Our sensitivity to these terms is enhanced compared to \cite{Crawford2014} due to our inclusion of cross frequency bispectra.  By jointly constraining these sources we hope to clearly identify the contributions arising from the differing components.

Bispectrum measurements of secondary sources could provide a wealth of astrophysical information. One halo tSZ bispectrum measurements probe gas pressure profiles on small scales, which provide insight into galaxy cluster processes such as AGN feedback \citep{Battaglia2012a}. \citet{Hurier2017} have used measurements of the tSZ bispectrum, power spectrum and cluster counts to constrain  the hydrostatic mass bias, the matter density and the amplitude of fluctuations. Cross-correlations among the tSZ effect, radio galaxies and DSFGs can disentangle their contributions to the temperature power spectrum measurements, which could lead to a better understanding of the observed deficit of small-scale tSZ power \citep{planck2014-a28}. The cross-bispectra from the tSZ and optical weak-lensing convergence maps can help constrain hydrostatic mass bias \citep{nelson2014,rasia2006}, a limiting systematic in tSZ cluster cosmology, as well as constraining cosmological parameters \citep{bhattacharya2012,Crawford2014,wilson2012}. Beyond the tSZ effect, the small-scale DSFG bispectrum can constrain the spatial distribution of DSFGs and  the cross-bispectrum from the tSZ and dusty galaxies or radio galaxies probes the masses of haloes these galaxies occupy. The bispectrum provides a flux-weighted count of the total number of these galaxies which can be combined with other measures of galaxy counts to constrain their population properties. DSFGs are a tracer of star formation \citep{elbaz2007} and bispectrum measurements can elucidate the interplay between cluster physics and star formation. Bispectrum measurements can help constrain the power spectrum of the kSZ effect \citep{Reichardt2012}, and joint measurements of the kSZ bispectrum and power spectrum can probe cluster thermodynamics and the distribution of baryons \citep{battaglia2017,schaan2016}. Finally, bispectrum measurements can be used to characterize extragalactic biases to CMB lensing power spectra \citep{vanEngelen2014,Osborne2014}.

In this work we analyze microwave intensity data from the Atacama Cosmology Telescope Polarimeter (ACTPol) at 148 GHz over a sky area of  $\sim550$ deg$^2$ \citep{Thornton2016,Louis2017}, combined with \textit{Planck} 100 GHz and 217 GHz intensity data \citep{planck2013-p03} in the same region. The ACTPol data have an effective angular resolution of $1.4$ arcmin, compared to Planck's effective resolution of  $9.7$ arcmin (100 GHz) and $5.0$ arcmin (217 GHz). For all our computations we use cosmological parameters from \textit{Planck} Collaboration \citep{planck2014-a15}, in particular we use $\Omega_c h^2= 0.1188$, $\Omega_m h^2= 0.02230$, $n_s= 0.9667$, $H_0=67.74$ and $\sigma_8= 0.8159$.

In section \ref{section:Estimators} of this paper we review the Komatsu, Spergel and Wandelt (KSW) estimator \citep{komatsu2005} and its flat-sky limit and section \ref{section:templates} provides an overview of our secondary CMB anisotropy templates.  Sections \ref{section:DataSets} and \ref{section:Pipeline} briefly describe the data sets used here and the analysis pipeline applied to the data.. We present our results and conclusions in sections \ref{section:Results} and \ref{section:conclusions}.  Appendix  \ref{app:appendixFlatToFull} derives the correspondence between the full-sky and flat-sky bispectrum estimators, while Appendix  \ref{app:bispectrumTemplates} contains a detailed derivation of the bispectrum templates for the major non-Gaussian microwave components. The details of our pipeline validation tests are described in Appendix \ref{app:CodeValidation}. Appendix \ref{app:TemplateErrors} contains the calculation of the non-Gaussian contribution to our estimator errors and Appendix \ref{app:NorderHODmoments} describes how to calculate the DSFG N-point functions used in calculating the non-Gaussian errors. In Appendix F \ref{app:sig8} we provide the details of how we extract a constraint on the amplitude of fluctuations and the matter density from our tSZ bispectrum measurement.

\section{Bispectrum estimators in the flat-sky approximation}\label{section:Estimators} 
The ACTPol and \textit{Planck} experiments measure fluctuations in the specific intensity, $\Delta I_{\nu}(\bm{n})$, of CMB across the sky at a frequency $\nu$. These intensity fluctuations are then related to a temperature fluctuations $\Delta T(\bm{n})$ via
\begin{align}
\Delta T(\bm{n})=\frac{k_B T^2c^2(e^{h\nu /(k_B T)}-1)^2}{2h^2\nu^4}\Delta I_{\nu}(\bm{n}),
\end{align}
where $k_B$ is the Boltzmann's constant, $T$ is the CMB temperature, $h$ is the Planck's constant and $c$ is the speed of light. In this work we focus on small regions of sky, so a flat-sky approximation is valid. In this regime the temperature anisotropies are decomposed as Fourier modes 
\begin{align}
\frac{\Delta T(\bm{n})}{T}=\int \frac{\mathrm{d}\ell^2}{4\pi^2} a_{\bm{\ell}} \,e^{i \bm{n} \cdot\bm{\ell}}.
\end{align}
This is analagous to full-sky analyses where the temperature fluctuations are expanded into spherical harmonics, $Y_{\ell, m}$
\begin{equation}
\frac{\Delta T(\bm{n})}{T}=\sum\limits_\ell \sum\limits_{-\ell< m<\ell} Y_{\ell, m}(\bm{n}) a_{\ell,m}.
\end{equation}
The flat-sky approximation is accurate to better than $1\%$ for $\ell>200$ \citep{loverde2008}. We will focus our analysis on the bispectrum, which is equal to the ensemble average of three $ a^{X}_{\bm{\ell}}$ (where the superscript X denotes maps from different frequencies or telescope arrays)
\begin{equation}
B^{(X_1,X_2,X_3)}({\bm{\ell}_1},{\bm{\ell}_2},{\bm{\ell}_3})=\left\langle a^{X_1}_{\bm{\ell}_1}a^{X_2}_{\bm{\ell}_2}a^{X_3}_{\bm{\ell}_3} \right \rangle.
\end{equation}
Under the assumption of rotational invariance, the flat-sky bispectrum can be expressed as \citep{goldberg1999,wHu2000}
\begin{equation}
B^{X_1,X_2,X_3}({\bm{\ell}_1},{\bm{\ell}_2},{\bm{\ell}_3})=4\pi^2 \delta^{(2)}(\bm{\ell}_1+\bm{\ell}_2+\bm{\ell}_3) b^{X_1 X_2 X_3}_{\ell_1 \ell_2 \ell_3},
\end{equation}
where $b^{X_1 X_2 X_3}_{\ell_1 \ell_2 \ell_3}$ is the reduced bispectrum. The assumption of rotational invariance is justified as the bispectra considered in this work are from extragalactic sources. The full sky bispectrum has a corresponding form for rotational invariance; see Appendix \ref{app:appendixFlatToFull} for a more detailed discussion of the correspondance. 

In the case of weak non-Gaussianity, we can obtain an optimal (minimum-variance) estimator for the amplitude $A$ of a non-Gaussian component with a given reduced bispectrum, $b^{\rm th}_{\ell_1 \ell_2\ell_3}$. In full-sky analyses the optimal estimator can be written as \citep{babich2004, Creminelli2006,Senatore2010,Verde2013}
\begin{align}\label{eq:optimalestimatorfull}
\hat{A} = & \frac{1}{\mathcal{N}} \sum_{X_i,}\sum_{X'_i} \sum_{\ell_i,m_i} \sum_{\ell'_i, m'_i} \mathcal{G}^{\,\,\ell_1\; \ell_2\; \ell_3}_{m_1 m_2 m_3 } b^{X_1 X_2 X_3, \, \rm th}_{\ell_1 \ell_2 \ell_3} \left\{ \left[ (C^{-1}_{\ell_1 m_1, \ell_1' m_1'})^{X_1 X'_1} a^{X'_1}_{\ell_1'm_1'}\, (C^{-1}_{\ell_2 m_2, \ell_2' m_2'} )^{X_2 X'_2} a^{X'_2}_{\ell_2'm_2'} \right. \right. \nonumber \left. (C^{-1}_{\ell_3 m_3, \ell_3' m_3'})^{X_3 X'_3} a^{X'_3}_{\ell_3'm_3'} \right] \nonumber \\
 & - 3\left. \left[ \,(C^{-1}_{\ell_1 m_1, \ell_2 m_2})^{X_1 X_2} (C^{-1}_{\ell_3 m_3,\ell_3' m_3'})^{X_3 X'_3} a^{X'_3}_{\ell_3'm_3'} + \text{ cyclic} \right] \right\} \; , 
\end{align}
where $ X$ parameterizes the different sky maps, $(C^{-1})^{X_iX_j}$ is the inverse of the covariance matrix $C^{X_iX_j}= \left\langle a^{X_i}_{\ell _1m_1} a^{X_j}_{\ell_2 m_2} \right\rangle$, $\mathcal{N}$ is the normalization and $ \mathcal{G}^{\ell_1 \ell_2 \ell_3}_{m_1 m_2 m_3}$ is the Gaunt integral over the product of three spherical harmonics (see equation \ref{eq:GauntIntegral} for the Gaunt integral definition). The normalization is chosen such that a unity amplitude is returned for an input map with a reduced bispectrum equal to the theoretical reduced bispectrum. The covariance matrix includes two components
\begin{equation}
C^{\mathrm{Total}}_{\ell,m,\ell',m'}= \mathrm{w}_{\ell,m}\mathrm{w}_{\ell',m'} C^{\mathrm{Signal}}_{\ell,m,\ell',m'} +\mathrm{N}_{\ell,m,\ell',m'}.
\end{equation}
The $\mathrm{w}_{\ell,m}$ carries the information of the beam profile and the pixelization function, $C^{\mathrm{Signal}}_{\ell,m,\ell',m'} $ is the covariance of the sky signal and $N_{\ell,m,\ell',m'}$ is the noise covariance. 

The flat-sky estimator, which can be derived in a nearly identical manner to the full-sky estimator, is
\begin{align}\label{eq:estFlatSky}
\hat{A} =\frac{1}{\mathcal{N}}\bar{A} = &\frac{1}{\mathcal{N}} \sum_{X_i} \sum_{X'_i}  \int \prod_{i}\mathrm{d^2}\bm{\ell}_i \,4\pi^2 b^{X_1 X_2 X_3, \, \rm th}_{{\ell_1} {\ell_2} {\ell_3} } \delta^{(2)} 
( \bm{\ell}_1+\bm{\ell}_2+\bm{\ell}_3)\
\Bigl( (C^{-1}_{\bm{\ell}_1,\bm{\ell'_1}})^{X_1X'_1} a^{X'_1}_{\bm{\ell'_1}}\, (C^{-1}_{\bm{\ell}_2,\bm{\ell'_2}})^{X_2X'_2} a^{X'_2}_{\bm{\ell'_2}}\, 
 (C^{-1}_{\bm{\ell}_3,\bm{\ell'_3}})^{X_3X'_3} a^{X'_3}_{\bm{\ell'_3}} \nonumber \\
 & -3\bigl((C^{-1}_{\bm{-\ell_1},\bm{\ell}_2})^{X_1X_2} (C^{-1}_{\bm{\ell}_3,\bm{\ell'_3}})^{X_3X'_3} a^{X'_3}_{\bm{\ell'_3}} + ... \text{ cyclic} \bigr) \Bigr),
\end{align}
where $\mathcal{N}$ is the normalization term as before, $\bar{A}$ is the unnormalized template amplitude, $\hat{A}$ is the estimated template amplitude and, analogously to the full sky,
\begin{equation}
C_{\bm{\ell},\bm{\ell'}}= \mathrm{w}_{\bm{\ell}}\mathrm{w}_{\bm{\ell'}} C^{\mathrm{Signal}}_{\bm{\ell},\bm{\ell'}} +\mathrm{N}_{\bm{\ell},\bm{\ell'}}.
\end{equation}
 For simplicity, we use the following notation to denote inverse covariance filtered maps
\begin{align}
{C^{-1}_{\bm{\ell}}}^Xa \equiv \sum\limits_{X'}\int\mathrm{d}^2\ell'(C^{-1}_{\bm{\ell},\bm{\ell'}})^{X,X'} a^{X'}_{\bm{\ell'}}.
\end{align}
The last term in equation \ref{eq:estFlatSky}, which depends only on one $a^{X}_{\bm{\ell}}$, is known as the linear term \citep{Yadav2008,Creminelli2006}. We calculate the linear term and the estimator normalization by ensemble averages, which is described in section \ref{sec:estLinearAndNorm}. In the optimal case the variance of this estimator is related to the normalization by
\begin{equation}
\mathrm{Var}(\hat{A})=\frac{1}{\mathcal{N}}.
\end{equation}
In our analysis we apply a real space mask, $M(\bm{n})$, and this alters this formula to 
\begin{equation}
\mathrm{Var}(\hat{A}) = \frac{f_{\mathrm{sky}}^{(6)}}{{f_{\mathrm{sky}}^{(3)}}^2 \mathcal{N}}
\end{equation}
where the proportionality constant depends on $f^{(n)}_{\mathrm{sky}}= 1/(N_{\mathrm{pix}})\int\mathrm{d}^2\bm{n} M(\bm{n})^n$ and $N_{\mathrm{pix}}$ is the total number of pixels. 
We will fit multiple templates simultaneously so we will use an extended version of this estimator
\begin{equation}\label{eq:estJointFit}
\hat{A}^{\mathrm{joint}}_i=\sum\limits_{j}{\mathcal{N}^{-1}}_{i,j}\bar{A}_j
\end{equation}
where $\mathcal{N}_{i,j}$ is a generalized normalization constant and the covariance of the estimators is given by
\begin{equation}\label{eq:estGausCov}
\mathrm{Cov}(\hat{A}^i,\hat{A}^j)=\frac{f_{\mathrm{sky}}^{(6)}}{{f_{\mathrm{sky}}^{(3)}}^2}{\mathcal{N}^{-1}}_{i,j}.
\end{equation}
Whilst this estimator is derived in the limit of weak non-Gaussianity, it is unbiased, but not optimal, for large non-Gaussianity. For large non-Gaussianity there could be additional contributions to the template covariances and these are discussed in Appendix \ref{app:TemplateErrors}. Hereafter the covariance matrix in equation \ref{eq:estGausCov} is called the `Gaussian' covariance and the covariance containing the contributions described in Appendix \ref{app:TemplateErrors} is called the `non-Gaussian' covariance. Note that for the non-Gaussian case, we do not change the estimator. Thus the measured values will be the same for the `Gaussian' and `non-Gaussian' cases, only the error bars will differ.

\subsection{KSW estimator}
The Komatsu, Spergel and Wandelt (KSW) estimator \citep{komatsu2005} offers the ability to compute the amplitude of separable or nearly separable bispectra in an efficient manner. The technique uses the separability to significantly reduce the computational cost. Consider the following reduced bispectrum
\begin{equation}
b_{{\ell_1},{\ell_2},{\ell_3}}= \mathrm{w}_{\ell_1}\mathrm{w}_{\ell_2}\mathrm{w}_{\ell_3} \int dr \, \left(\alpha_{\ell_1}(r) \beta_{\ell_2} (r)\gamma_{\ell_3} (r)+\alpha_{\ell_2}(r)\beta_{\ell_3}(r)\gamma_{\ell_1}(r) +\, 4 \,\mathrm{permutations}\,\mathrm{of}\,\ell_1,\,\ell_2\, \mathrm{and}\,\ell_3 \right)
\end{equation}
where $\mathrm{w}_{\ell}$ is again the window function with pixelation window and the integral over $\alpha, \, \beta$ and $\gamma$ is the theoretical signal bispectrum. Many types of primordial bispectra have reduced bispectra with this structure. We have assumed that we have only one map so that, for simplicity, we can drop the $X_i$ superscripts. Inserting this form into our estimator and rearranging we attain the following
\begin{align}
\hat{A} \propto& \int d r \int d^2 \mathbf{n} \Bigl( \int\mathrm{d^2}\bm{\ell}_1\, \mathrm{w}_{\ell_1} \alpha_{\ell_1} (r) C_{\bm{\ell}_1}^{-1} a e^{-i \bm{n}\cdot \bm{ \ell_1}} \int\mathrm{d^2}\bm{\ell}_2\, \mathrm{w}_{\ell_2}\beta_{\ell_2}(r) C_{\bm{\ell}_2} ^{-1} a e^{-i \bm{n}\cdot\bm{ \ell_2}} \int\mathrm{d^2}\bm{\ell}_3\, \mathrm{w}_{\ell_3} \gamma_{\ell_3}(r) Ca_{\bm{\ell}_3}^{-1}  e^{-i \bm{n}\cdot\bm{ \ell_3}} + .... \nonumber \\ & + \left\langle\int\mathrm{d^2}\bm{\ell}_1\, \mathrm{w}_{\ell_1} \alpha_{\ell_1} (r) C_{\bm{-\ell_1}}^{-1} a e^{-i \bm{n}\cdot\bm{ \ell_1}} \int\mathrm{d^2}\bm{\ell}_2\, \mathrm{w}_{\ell_2}\beta_{\ell_2}(r) C_{\bm{\ell}_2} ^{-1}a e^{-i \bm{n}\cdot\bm{ \ell_2}} \right \rangle \, \int\mathrm{d^2}\bm{\ell}_3\, \mathrm{w}_{\ell_3} \gamma_{\ell_3}(r) C_{\bm{\ell}_3}^{-1} a e^{-i \bm{n}\cdot\bm{ \ell_3}} + ... \Bigr),
\end{align}
where $\ell\equiv |\bm{\ell}|$. Note that we also used the relation $C^{-1}_{\bm{\ell},\bm{\ell'}}=\left\langle C_{\bm{\ell}}^{-1} a  C_{\bm{\ell'}}^{-1} a \right\rangle$ to reformulate the linear term. The separation of the integrals allows an efficient calculation of the estimator, and this method can be extended for templates with double integrals as we will describe in the next section.

When applied to real data we discretise the integrals over $\ell$'s and the resulting sums are efficiently calculated via fast Fourier transforms, for which we use the FFTW package \citep{FFTW3}.

\subsection{Estimator Normalization and Linear Term}\label{sec:estLinearAndNorm}
We compute the estimator's normalization and linear term using the ensemble average method described in \cite{Smith2011}. We outline this method here and refer the reader to the original paper for more details. First we generate sets of Gaussian simulations with power spectrum $C^{\mathrm{Total}}_{\ell}$. To achieve this we use the CAMB \citep{LewisCAMB} code to generate the underlying CMB power spectrum and apply the beam and pixel window function, as has been presented in \cite{hasselfield2013} for ACTPol and in \cite{planck2013-p03c} for \textit{Planck}. We add secondary sources to our CAMB power spectrum with amplitudes as determined in \cite{dunkley2013}. For \textit{Planck} , the noise power is estimated from simulations \citep{planck2014-a14}. For the ACTPol maps we have four splits of the data with uncorrelated noise, created by using data taken on every fourth night (see also Section \ref{section:DataSets}). From these, we estimate the noise power, assuming it is diagonal, via the following equation
\begin{equation}
\label{eq:noisePower}
\mathrm{N}_{\bm{\ell}} \approx \frac{1}{4} \sum_i a^{i}_{\bm{\ell}} {a^{i}_{\bm{\ell}}}^* - \frac{1}{12}\sum_{i\ne j} a^{i}_{\bm{\ell}} {a^{j}_{\bm{\ell}}}^*,
\end{equation}
where $a^{i}_{\bm{\ell}}$ is the power in the $\mathrm{i^{\mathrm{th}}}$ split of the data. We assume that the noise is uncorrelated between the different arrays. We apply the same masks as used in the data to our simulations, when masking is required. The real noise is anisotropic and we model the anisotropy in our simulations by weighting the noise simulations in real space by the square root of the hits map.

For each Gaussian simulation, we then calculate the following quantity
\begin{align}\label{eq:estNormalization}
\nabla^{i}_{\bm{\ell}} \,T(a^X)=\frac{1}{2} \sum\limits_{X_1,X_2}\int\int \mathrm{d}^2\ell_1\mathrm{d}^2\ell_2 4\pi^2\delta^{(2)}(\bm{\ell}+\bm{\ell}_1+\bm{\ell}_2)b^{i,X,X_1,X_2}(\ell_1,\ell_2,\ell_3) {a^*}^{X_1}_{\bm{\ell}_1}{a^*}^{X_2}_{\bm{\ell}_2},
\end{align}
where the index $i$ labels the type of non-Gaussianity. We then use this quantity to calculate the normalization
\begin{align}
\mathcal{N}_{i,j}\propto \int\int\mathrm{d}^2\ell_1\mathrm{d}^2\ell_2 \left(\frac{1}{3}\left\langle\nabla^{i}_{\bm{\ell}_1}\,T({C^{-1}a}^{X_1})C^{X_1,X_2}_{\bm{\ell}_1,\bm{\ell}_2} (\nabla^{j}_{\bm{\ell}_2}\,T({C^{-1}a}^{X_2}))^*\right \rangle-\frac{1}{3}\left\langle\nabla^{i}_{\bm{\ell}_1}\,T({C^{-1}a}^{X_1})\right \rangle C^{X_1,X_2}_{\bm{\ell}_1,\bm{\ell}_2}\left\langle(\nabla^{j}_{\bm{\ell}_2}\,T({C^{-1}a}^{X_2}))^*\right \rangle\right)
\end{align}
and the linear term
\begin{align}
\int \prod_{i}\mathrm{d^2}\bm{\ell}_i \,4\pi^2\delta^{(2)}(\bm{\ell}_1+\bm{\ell}_2+\bm{\ell}_3)b^{i,X_1,X_2,X_3}(\ell_1,\ell_2,\ell_3){C^{-1}}^{X_1,X_2}_{-\bm{\ell}_1,\bm{\ell}_2} {C^{-1}_{\bm{\ell}_3}}^{X_3}a= \int \mathrm{d}^2\ell_3\left\langle\nabla^{i}_{\bm{\ell}_3}\,T(C^{-1}a^X_3) \right \rangle {C^{-1}_{\bm{\ell}_3}}^{X_3}a.
\end{align}
The normalization proportionality constant depends on $f_{\rm{sky}}^{(n)}$. We use this method to calculate the normalization as it was found to be more efficient than a more direct approach. 

The linear term is only important when the covariance matrix has large off diagonal terms or for highly squeezed configurations. This can be seen by considering the case when ${C^{-1}_{\bm{\ell},\bm{\ell'}}}^{X,X'}$ is diagonal ($\bm{\ell}=\bm{\ell'}$). In this case the contribution from the linear term vanishes. In our pipeline (described in section \ref{section:Pipeline} ) we take steps to reduce the impact of mode coupling and inhomogeneous noise, which are the main sources of off-diagonal covariance terms. These steps mean that the covariance matrix only has significant off-diagonal contributions when $\bm{\ell}\sim\bm{\ell'}$  and thus the main contributions to the linear term will be from squeezed configurations (configurations with two large $\ell$ and one small $\ell$). The configuration dependence of the linear term is discussed in \citet{Creminelli2006} and \citet{Bucher2016} demonstrate that the linear term is a minor effect for non-squeezed configurations. As discussed in section \ref{section:Pipeline} we mask the lowest $\ell$ modes to avoid ground contamination and in doing so we strongly suppress squeezed configurations and therefor the linear term. For the templates used here we found that the linear term contributed negligibly to our results and that when it was included we required many more simulations for the ensemble averages to converge than if we neglected the linear term. After verifying it was negligible for all the templates we proceeded to neglect its contribution.

\section{Bispectrum Templates}\label{section:templates} 
\begin{figure}
 \centering
 \includegraphics[width=1.\textwidth]{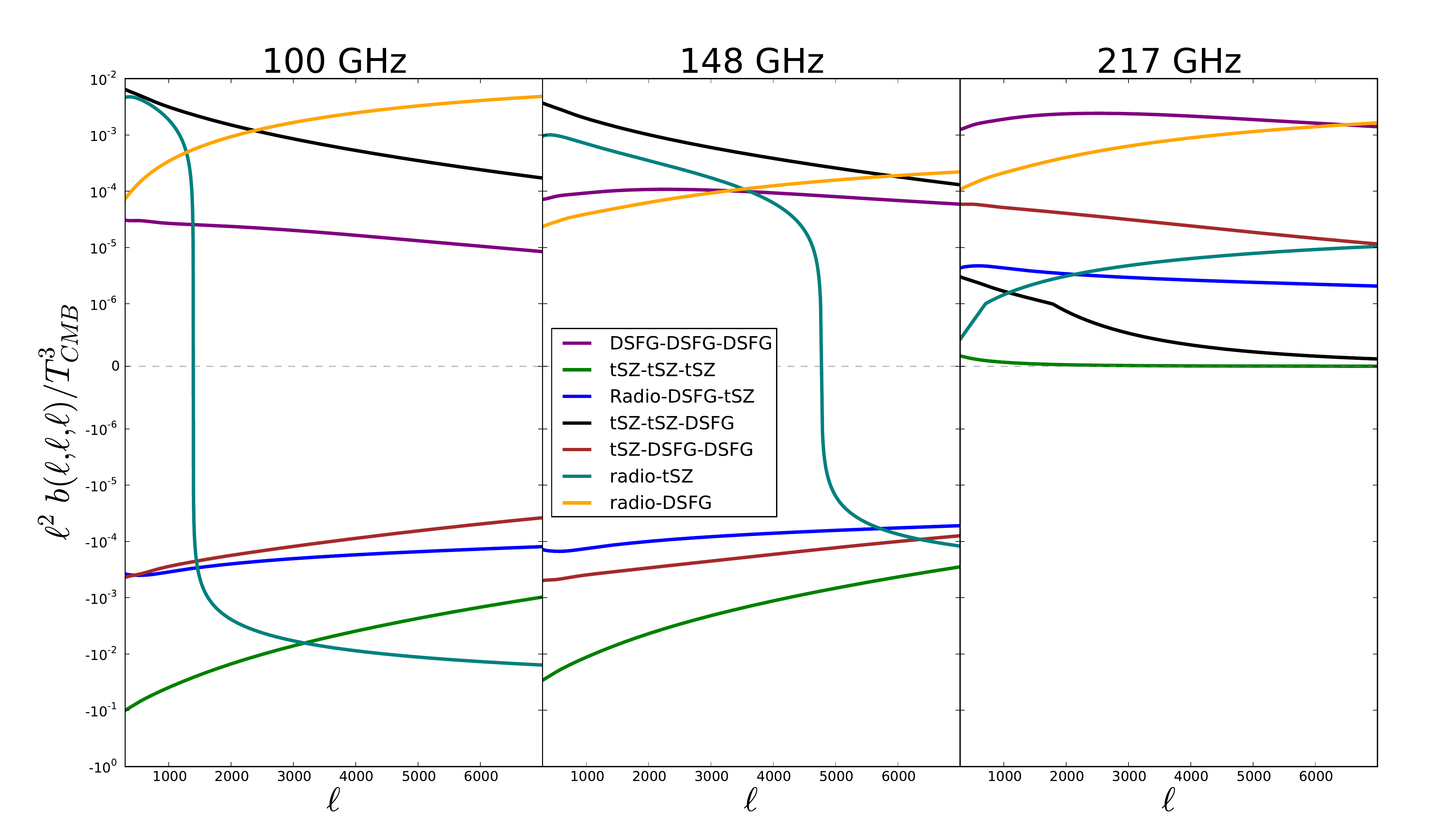}
\caption{The one-halo bispectrum templates with $l_1=l_2=l_3$ for the tSZ, DSFGs and radio galaxies for three frequencies: 100, 148 and 217 GHz. The DSFG term contains the two and three galaxy term but not the one galaxy term; the one galaxy term is a Poisson term with trivial scale dependence and so is excluded. Similarly we exclude the radio point source term. The radio-tSZ is composed of two physical terms: the radio-tSZ-tSZ and the radio-radio-tSZ terms. The feature in the radio-tSZ template arises as the two contributions have different signs below 220 GHz and are dominant on different scales; the radio-tSZ-tSZ term is dominant on the largest scales and is positive at all frequencies and the radio-radio-tSZ term is dominant on the smallest scales and is negative at frequencies below 220 GHz. Similarly the radio-DSFG term is composed of two physical terms: the radio-radio-DSFG and the radio-DSFG-DSFG terms. The scale is linear between $10^{-6}$ and  $10^{6}$ and logarithmic outside this range. This leads to kinks in the curves that cross the log-linear transition.}
\label{fig:templateSummary}
\end{figure}

\begin{figure}
 \centering
 \includegraphics[width=.5\textwidth]{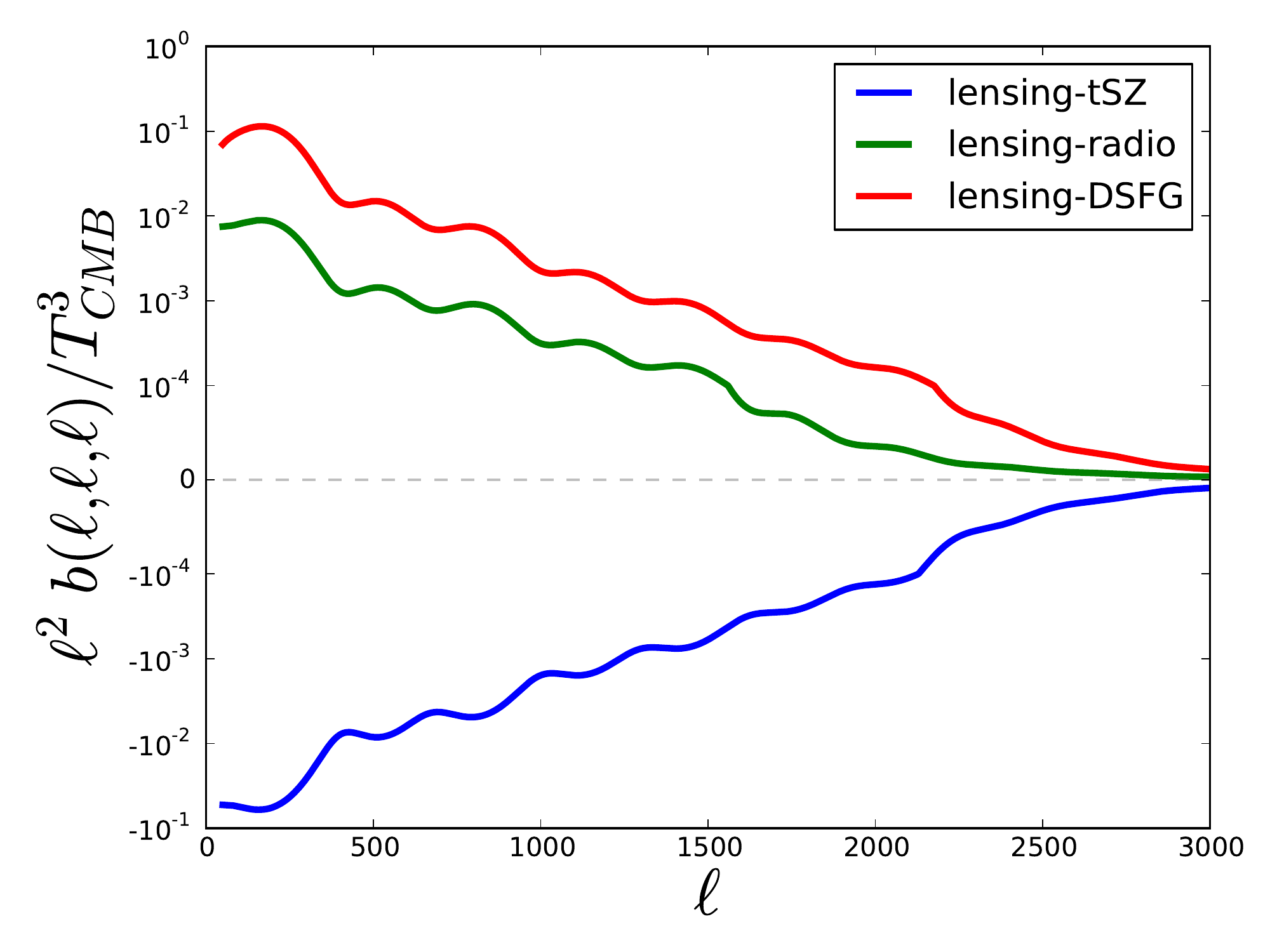}
 \caption{The lensing cross secondary source bispectrum templates with $l_1=l_2=l_3$ at 148 GHz. We exclude the ISW-lensing template here as it has been well studied before. The scale is linear between $10^{-4}$ and  $10^{4}$ and logarithmic otherwise. }
\label{fig:templateSummaryLens}
\end{figure}

In this work we consider non-Gaussianity from lensing-induced anisotropies, the tSZ effect, DSFGs and radio galaxies. The kSZ effect is too small to be constrained with the data used in this work; the form of its bispectrum template is described in Appendix \ref{app:kSZeffect} for completeness. In this section we present a brief overview of the structure of the bispectrum templates that are used in this analysis and provide the details of these calculations in Appendix \ref{app:bispectrumTemplates}. 

\subsection{Poisson point source templates}
Unclustered point sources in our maps follow Poisson statistics. Their reduced bispectrum is a flux weighted count of the number of point sources and has the form
\begin{equation}
b^{\mathrm{Point\, Sources}}_{\ell_1,\ell_2,\ell_3}=k_{\nu}^{3}  \int_{0}^{S_{c}}\mathrm{d}S_{\nu}\, {S_{\nu}}^3 \frac{\mathrm{d}n}{\mathrm{d}S\mathrm{d}\Omega},
\end{equation}
where $S_\nu$ is the flux density, $\mathrm{d}n/\mathrm{d}S\mathrm{d}\Omega$ is the number count per steradian per flux interval, S$_c$ is the flux cut and $k_\nu$ is the conversion from flux to map temperature. The flux cut is uniform across the survey area and different cuts are applied for the different frequency map as described in section \ref{section:DataSets}. Radio galaxies and dusty star forming galaxies (DSFGs) both have a Poisson point source contribution and the details of their source counts is given in Appendicies \ref{section:radioGals} and \ref{section:haloModel}. 
\subsection{Thermal Sunyaev Zel'dovich, dusty star forming galaxy and radio galaxy templates}
We construct templates for non-Gaussianity from auto and cross correlations from the tSZ effect, DSFG and radio galaxies with the halo model. Our analysis is focused on the smallest scales and so we only include the one halo term. Here we briefly overview the general prescription for calculating one halo bispectrum templates. Our discussion is based on that of \cite{hill2013b} and \cite{Lacasa2014}.

In the halo model it is assumed that all the matter is contained in dark matter halos. We then assume that each quantity of interest is distributed within each cluster with profile $X^i(\bm{x},M)$, where $X^i$ could be the pressure of the halo gas or the distribution of radio galaxies and $M$ is the cluster virial mass. Note that we use the index $i$ to label different physical effects and not different halos, as is common in the halo model literature. We also make the assumption that these profiles are all spherically symmetric. The reduced bispectrum is given by
\begin{align}
b^{i,j,k}_{\ell_1,\ell_2,\ell_3}=&\int \mathrm{d}\chi_i \mathrm{d}\chi_j \mathrm{d}\chi_k a(\chi_i)a(\chi_j)a(\chi_k) \times \nonumber \\ & \int \mathrm{d}x x^2\prod\limits_{a}\mathrm{d}k_a k_a^2 J_{\ell_a+\frac{1}{2}}(k_a \chi_a) J_{\ell_a+\frac{1}{2}}(k_a x)\frac{1}{k_a \sqrt{x \chi_a}}  \int\mathrm{d}\,\ln M\,\frac{\mathrm{d}n}{\mathrm{d}\ln M} \tilde{X}^i({k_i},M)  \tilde{X}^j({k_j},M) \tilde{X}^k({k_k},M),
\end{align}
where $k_i=|\bm{k}_i |$, $\chi$ is the comoving distance, $a$ is the scale factor, $ \mathrm{d}n/ \mathrm{d}\ln M$ is the mass function and $ \tilde{X}^i({k_i},M)$ is the Fourier transform of the line of sight projected profile $X(\bm{x},M)$. Using the Limber approximation we have the final form (see Appendix \ref{sec:FullSkyBispectrumCal} for the details)
\begin{align}
b^{i,j,k}_{\ell_1,\ell_2,\ell_3}= &\int \mathrm{d}\chi \frac{a(\chi)a(\chi)a(\chi)}{\chi^4}  \int\mathrm{d}\ln M\,\frac{\mathrm{d}n}{\mathrm{d}\ln M} \tilde{X}^i(k_i,M)  \tilde{X}^j(k_j,M) \tilde{X}^k(k_k,M) \nonumber \\ 
=& \int \mathrm{d}z \frac{\mathrm{d}^2V}{\mathrm{d}z\mathrm{d}\Omega} \frac{a(\chi)a(\chi)a(\chi)}{\chi^6}  \int\mathrm{d}\ln M\,\frac{\mathrm{d}n}{\mathrm{d}\ln M} \tilde{X}^i\left(\frac{\ell_1+\frac{1}{2}}{\chi(z)},M\right)  \tilde{X}^j \left(\frac{\ell_2+1\frac{1}{2}}{\chi(z)},M\right)  \tilde{X}^k \left(\frac{\ell_3+\frac{1}{2}}{\chi(z)},M\right). 
\end{align}

With this formalism we construct eight templates to characterize the non-Gaussianity from auto and cross correlations among the tSZ effect, DSFG and radio galaxies. The equatorial slice for these bispectra is plotted for three different frequencies in figure \ref{fig:templateSummary}. The tSZ-tSZ-tSZ template describes non-Gaussianty from the tSZ effect only; the DSFG template describes the Poisson and clustered components of the DSFGs. The tSZ-tSZ-DSFG and tSZ-DSFG-DSFG templates model the cross correlations between the tSZ effect and DSFGs. The radio galaxy template describes the Poisson contributions of the radio galaxies. The radio-tSZ template contains two terms, the radio-radio-tSZ and radio-tSZ-tSZ terms, and arises from cross correlations between radio galaxies and the tSZ effect. The radio-DSFG template, which similarly contains both the radio-radio-DSFG and radio-DSFG-DSFG terms, describes the cross correlations between the DSFGs and radio galaxies. Finally the radio-DSFG-tSZ template describes the cross correlations among the three effects. Extending the results shown in \cite{Lacasa2014} we find that all the one-halo bispectrum terms are largely insensitive to the configuration but display strong scale dependence. Motivated by this we plot only equatorial slices, where $\ell_1=\ell_2=\ell_3$, of our templates to show the scale dependence. 

\subsection{Lensing cross secondary sources templates}\label{subsec:lensingCrossMain}
Bispectra from lensing-induced anisotropies have been well studied \citep{goldberg1999,goldberg1999b,wHu2000,lewis2011}. We briefly summarize the origin of this non-Gaussianity here. We can decompose the anisotropies into contributions from the early universe and late-time sources:
\begin{equation}
\Delta T(\bm{n})=\Delta T^P(\bm{n}+\nabla \phi)+\Delta T^s(\bm{n}),
\end{equation}
where $\Delta T^P$ is the unlensed CMB fluctuations, $\phi$ is the lensing potential, and $\Delta T^S$ encodes the contributions from late time sources such as radio galaxies or clusters via the tSZ effect. If we expand in the lensing potential to first order we find
\begin{equation}
\Delta T(\bm{n})=\Delta T^P(\bm{n}) + \nabla\phi \cdot \nabla \Delta T^P(\bm{n}) + \Delta T^s(\bm{n}).
\end{equation}
From this we see that we can get non-vanishing bispectra arising from the terms $\left\langle\Delta T^P(\bm{n}) \nabla \phi \cdot \nabla \Delta T^P(\bm{n})\Delta T^s(\bm{n})\right\rangle $ as the secondary sources are correlated with the structures that cause the lensing. More precisely these lead to the following reduced bispectrum
\begin{align} \label{eq:lensingBase}
b^{\mathrm{lensing-sec.}}_{\ell_1,\ell_2,\ell_3} =-\bm{\ell}_1\cdot\bm{\ell}_2C_{\ell_1}^{TT}C_{\ell_2}^{\phi S} + 5\, \mathrm{ permutations},
\end{align}
where $C^{\phi S}$ is the cross-correlation between the lensing potential and the secondary source. We consider four lensing bispectrum templates arising from cross-correlations with the tSZ effect, radio galaxies, DSFG and the integrated Sachs Wolfe effect.  In figure \ref{fig:templateSummaryLens} we show the equatorial slice through the bispectrum for several of the lensing templates considered in this work. The shape of the lensing templates is largely insensitive to frequency so we display only a single frequency here. The shape of the ISW-lensing template has been well studied before \citep{lewis2011} and so is not shown here. The other lensing templates have a similar spatial dependence so we plot only the equatorial configuration here.

\section{Data Sets}\label{section:DataSets} 
\begin{figure}
\centering
 \includegraphics[width=.75\textwidth]{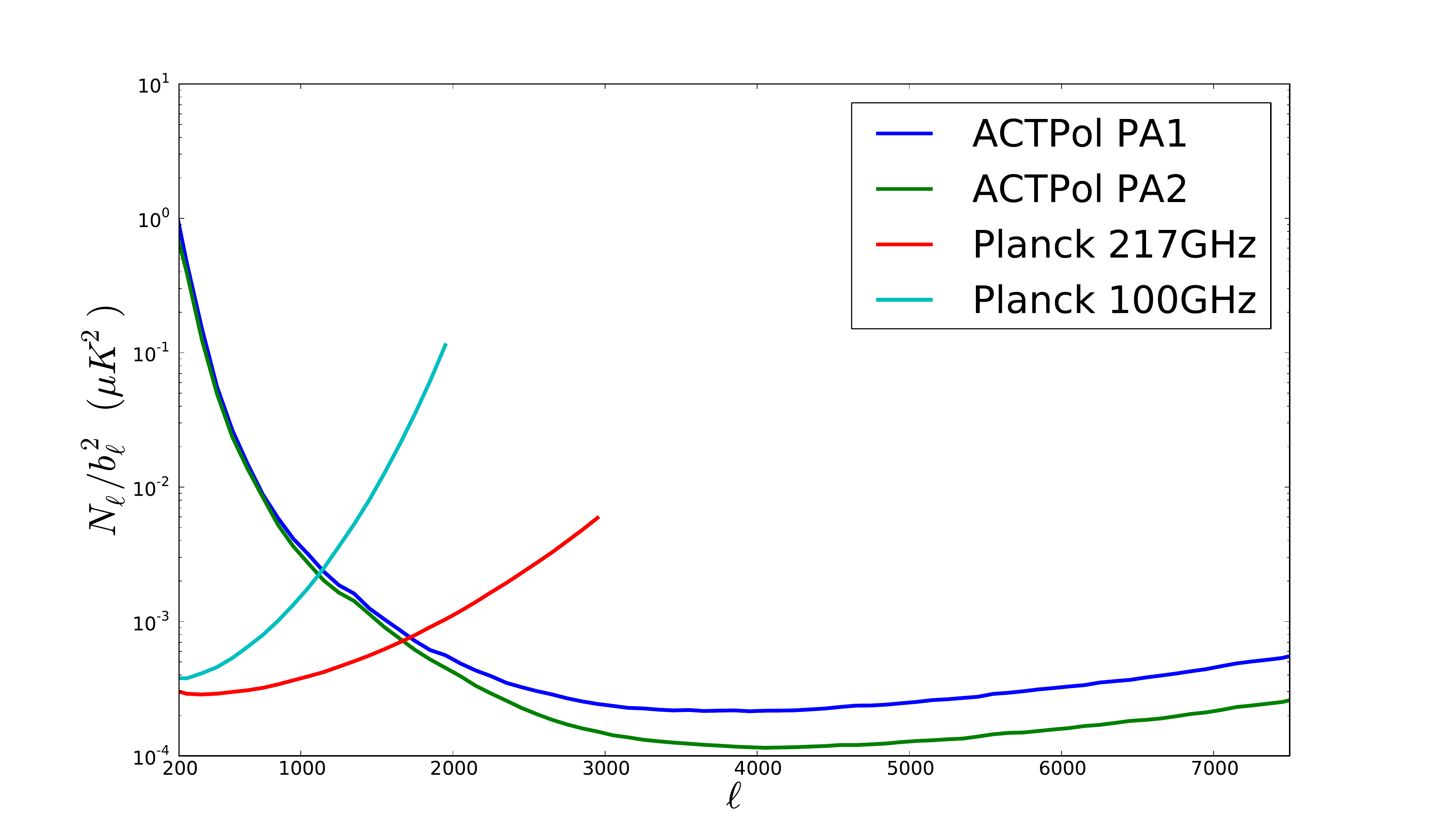}
\caption{The noise power spectra divided by the beam function squared for the four maps used in this analysis. The beams for the \textit{Planck} 100 GHz and 217 GHz maps are treated as zero for $\ell>2000$ and $\ell>3000$ respectively.}
\label{fig:NoiseComp}
\end{figure}

The analysis in this paper is focused on using the small-scale information from the ACTPol experiment. Several non-Gaussian sources, such as the Poisson contributions from DSFG and radio galaxies, have similar template shapes and cannot be distinguished with the current 148GHz ACTPol maps. The different spectral behavior of the sources enables their differentiation with multi-frequency data. For this purpose we use data from the \textit{Planck Experiment}. Through combining ACTPol and \textit{Planck} $100$ GHz and $217$ GHz data we can break many of the degeneracies of single frequency measurements and thus distinguish more sources of non-Gaussianity. Most of our constraining power will come from bispectrum configurations involving one  \textit{Planck} map and two ACTPol maps. Figure \ref{fig:NoiseComp} shows that our estimator will be most sensitive to scales of $1000<\ell<2000$ for the  \textit{Planck} maps and $2000<\ell<5000$ for the ACTPol maps.

\subsection{ACTPol data sets}\label{section:DataSetsACT}
We use data in the `D56' field, a patch of sky on the equator with coordinates $-$7.2$^{\circ}$ < dec < 4$^{\circ}$ and 352$^{\circ}$ < RA < 41$^{\circ}$. This is part of the data described in \cite{Louis2017}; in particular we use only the wide field and not the deep fields. In this work we use the data from the two arrays (called PA1 and PA2 hereafter) which observed the sky at 148 GHz. The ACT experiment has a full width at half maximum (FWHM) of $1.4^\prime$ at 148 GHz. As was shown in \cite{Louis2017} non-white atmospheric noise, from atmospheric temperature brightness fluctuations, dominates the largest scales. On the smallest scales the noise is approximately white at  $31$ $\mu$K-arcmin and $25$ $\mu$K-arcmin for PA1 and PA2 respectively. We mask all the point sources whose fluxes were measured to be above $30$ mJy with discs of radius $5^\prime$ and perform no masking of clusters or of galactic dust. 

\subsection{Planck Data sets}\label{section:DataSetsPlanck}
We use only the \textit{Planck} $100$ and $217$ GHz maps as the lower frequencies have too limited sensitivity to the small-scales (due to the $> 5^\prime$ beam) and the higher frequencies are obscured by dust in this region. Instead of using the full information of the Planck maps we remain in the flat-sky regime and use only the portion of the \textit{Planck} data that overlaps with the ACTPol D56 region. The ACTPol and Planck maps we used as input do not come with the same pixelization or coordinate system. ACTPol uses equatorial coordinates in equal-area coordinates pixelization (CEA), while \textit{Planck} uses HEALPix-pixelized galactic coordinates. Before cross-correlating the maps, we reprojected \textit{Planck} onto the ACTPol pixels by first expanding it in spherical harmonic coefficients and rotating these coefficients from galactic to equatorial coordinates using \textit{healpy}, and then evaluating these coefficients on the ACTPol pixels using \textit{libsharp} \citep{libsharp}.

We mask point sources that were detected in the ACTPol $148$ GHz map with fluxes above $269$ mJy and $152$ mJy for the $100$ GHz and $217$ GHz maps respectively with discs of radii $12.5^\prime$ and $6.5^\prime$. These masking levels are \textit{Planck}'s 90$\%$ completeness limit for these frequencies \citep{planck2014-a35}. We chose a point source mask based on sources detected in the ACTPol maps (rather than using the \textit{Planck} point source mask) to ensure we had a constant flux cut across the map. We perform no masking of clusters or of galactic dust. The \textit{Planck} experiment's beam FWHM are 9.7$^\prime$, and $5.0^\prime$ at $100$ GHz and $217$ GHz respectively. The small scale \textit{Planck} noise is white with levels of 77.4 $\mu$K-arcmin and 46.8 $\mu$K-arcmin at $100$ GHz and $217$ GHz respectively \citep{planck2014-a09}. To avoid issues of nearly singular matrices during our inverse covariance filtering we add uncorrelated noise to the \textit{Planck} maps with a power spectrum equal to $10\%$ of the primary CMB (with the appropriate beam and window functions). Without this the largest scale modes of the \textit{Planck} maps would be very highly correlated and this can lead to  numerical instabilities when performing inverse covariance operations.

\section{Analysis Pipeline}\label{section:Pipeline} 
Full inverse-variance weighting of these maps is a computationally costly process and whilst methods exists for performing this \citep{elsner2013,smith2009}, the \textit{Planck} team has shown \citep{planck2013-p09a} that near optimality can be obtained using approximate methods that assume the covariance matrix is diagonal. In this work we will use the approximate method described in \cite{planck2013-p09a}.

First we apply a point source mask that masks the brightest point sources as stated in sections \ref{section:DataSetsACT} and \ref{section:DataSetsPlanck}. We then use a similar method to \cite{planck2013-p09a}, in which we fill in masked point source pixels with the average of their neighbors and iterate until the solution converges. Convergence is typically attained within 1000 iterations and 2000 iterations are used for our final analysis. This infilling procedure reduces mode coupling and allows the approximation of a diagonal covariance matrix without loss of optimality. 

Next we apply our real space mask. The real space mask is constructed from two components. The first is the square root of the smoothed hit counts map ($H_s(\bm{n})$). The hits map was smoothed with a Gaussian of FWHM 9.5$^\prime$ to limit leakage. The second component is a smoothed top hat mask $F(\bm{n})$. To obtain  $F(\bm{n})$, we mask all pixels with fewer than $5500$ counts in either of the ACTPol arrays and then convolve this mask with a cosine squared of width 25$^\prime$. This mask serves two purposes: it down-weights noisy and infrequently observed regions of the sky, thereby reducing the influence of inhomogeneous noise, and it serves as an edge taper that reduces mode coupling. The total mask $M(\bm{n})$ is given by
\begin{align}
M(\bm{n})\propto \frac{H_s(\bm{n})}{H_s(\bm{n})+15000}F(\bm{n}).
\end{align}
On the scale of our smoothed hits-map mask, the power in the map for heavily observed regions (hit counts $\gtrsim15,000$) is dominated by the CMB. In these regions of the maps our final mask is constructed to be approximately uniform to prevent overweighting these regions.

The FFT methods we use assume periodic boundary conditions. To prevent spurious correlations we zero-pad our maps by a factor of two before performing a Fourier transform on the maps. Finally, we Fourier transform our maps and then apply a $k$-space mask to our maps. For the ACTPol maps, we mask modes with $\ell<500$  as these modes are dominated by atmospheric $1/f$ noise. We then remove all modes with $| \ell_x| <90 $ and $| \ell_y| <90 $, as these are dominated by ground contamination \citep{das2011}. For the \textit{Planck} maps, we filter modes with $\ell<200$. This is done for two reasons; first the flat-sky approximation becomes less accurate at low $\ell$ and second the two and three halo terms cannot be ignored if the lowest $\ell$ modes are used.

Once processed as above, we use the maps to estimate the amplitudes of the theoretical templates described in section \ref{section:templates}. We use the KSW implementation of equation \ref{eq:estFlatSky} to calculate the template amplitudes and equation \ref{eq:estJointFit} to jointly fit our templates. The Gaussian errors are calculated using equation \ref{eq:estGausCov}. Our estimator jointly fits the amplitudes of all the non-Gaussian templates and uses all the possible auto and cross bispectra from the four maps. We use an $\ell_{\rm{max}}=7500$. This cut off was chosen as there is little signal at small scales (higher $\ell$) and it is computational expensive to include the smaller scales. The validation of the analysis pipeline is described in Appendix \ref{app:CodeValidation}.

\section{Template measurements with ACTPol and \textit{Planck}}\label{section:Results} 
\begin{figure}
\includegraphics[width=0.95\textwidth]{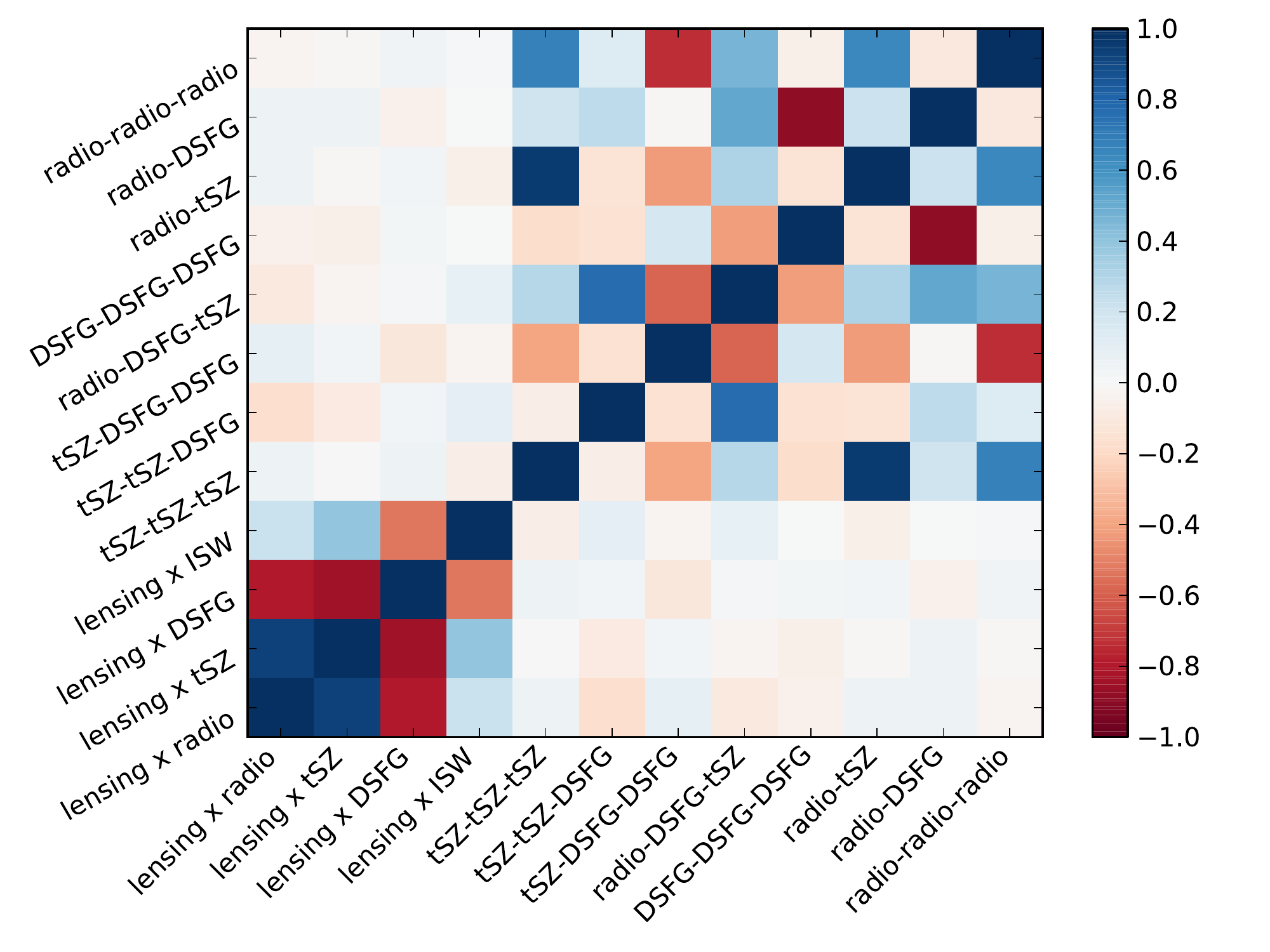}
\caption{The covariance matrix of template amplitudes only accounting for the Gaussian component to the six point function. The strong covariances between several templates highlights the difficult of disentangling all of the sources of non-Gaussianity.}
\label{fig:GausErrorCov}
\end{figure}

In table \ref{table:ResultsGaussian} we present the measured template amplitudes and the Gaussian errors for these templates. The ``joint fit'' results come from simultaneously fitting all of the templates. The Gaussian error characterizes how well we can measure the different types of non- Gaussianity given the finite number of noisy modes in our data set. Our measured amplitudes with the Gaussian error describe the level of non-Gaussianity and its significance in this data set. We find significant amplitudes for the tSZ-tSZ-tSZ, tSZ-tSZ-DSFG, radio galaxy, and radio-DSFG-tSZ templates. When simultaneously measuring multiple templates, the Gaussian covariance represents how well the estimator can differentiate between the types of non-Gaussianity. In figure \ref{fig:GausErrorCov} we present the Gaussian covariance matrix for our measurements; it can be seen that, despite the use of multifrequency information, we still have several highly correlated terms. In particular, we see strong covariances between the radio-DSFG term and the DSFG only term; between the radio-tSZ term and the tSZ only term; and between the tSZ-DSFG-DSFG term and the radio galaxy term. These strong covariances, as well as the others in the dataset, make our measurements particularly sensitive to our models, since inaccurate theoretical models can `leak' the signal of one template into another, tightly correlated template.
 \begin{table}
 \centering
\begin{tabular}{|l |c |c |}
\hline

\multirow{2}{*}{Type} &\multicolumn{2}{c|}{Measured $A_{i}$}\\
\cline{2-3}&Gaussian Errors & Full non-Gaussian Errors\\ \hline
lensing x radio& $-0.31 \pm 6.26 $ & $-0.31 \pm 6.37$ \\
lensing x tSZ& $1.74 \pm 1.55 $ & $1.74 \pm 1.60 $  \\
lensing x DSFG& $0.43 \pm 0.41 $  & $0.43 \pm 0.45 $ \\
lensing x ISW& $47.86 \pm 29.22 $& $47.86 \pm 28.81 $ \\
tSZ-tSZ-tSZ& $0.80 \pm 0.25 $ & $0.80 \pm 0.65 $  \\
tSZ-tSZ-DSFG& $1.20 \pm 0.29 $ & $1.20 \pm 0.80 $ \\
tSZ-DSFG-DSFG& $-0.96 \pm 0.47 $ & $-0.96 \pm 0.78 $ \\
radio-DSFG-tSZ& $6.03 \pm 1.31 $ & $6.03 \pm 1.83 $ \\
DSFG-DSFG-DSFG& $1.65 \pm 0.44 $ & $1.65 \pm 0.45 $ \\
radio-tSZ& $1.20 \pm 0.82 $ & $1.20 \pm 1.05 $  \\
radio-DSFG& $-0.45 \pm 1.40 $& $-0.45 \pm 1.65 $ \\
radio-radio-radio& $0.99 \pm 0.08 $ & $0.99 \pm 0.13 $  \\
\hline
\end{tabular}
\caption{Template amplitudes obtained from a joint fit using 100 GHz and 217 GHz Planck maps with the ACTPol PA1 and PA2 148 GHz maps and $\ell_{max} = 7500$. Point sources below $269$, $152$, $30$ mJy for frequencies $100,217,148$ GHz respectively were masked and no masking of clusters was performed. The amplitudes, $A_i$, are overall scalings of our templates and are dimensionless. The Gaussian errors only indicate the significance of the non-Gaussianity in these data sets. The non-Gaussian error results include the one-halo contributions to the cosmic variance and these errors reflect the statistical uncertainty in the template amplitudes.  The $\chi^2$ for the fit to our model is $\chi^2 = 43.6$ for 12 degrees of freedom.}
\label{table:ResultsGaussian}
\end{table}

\begin{figure}
\includegraphics[width=0.95\textwidth]{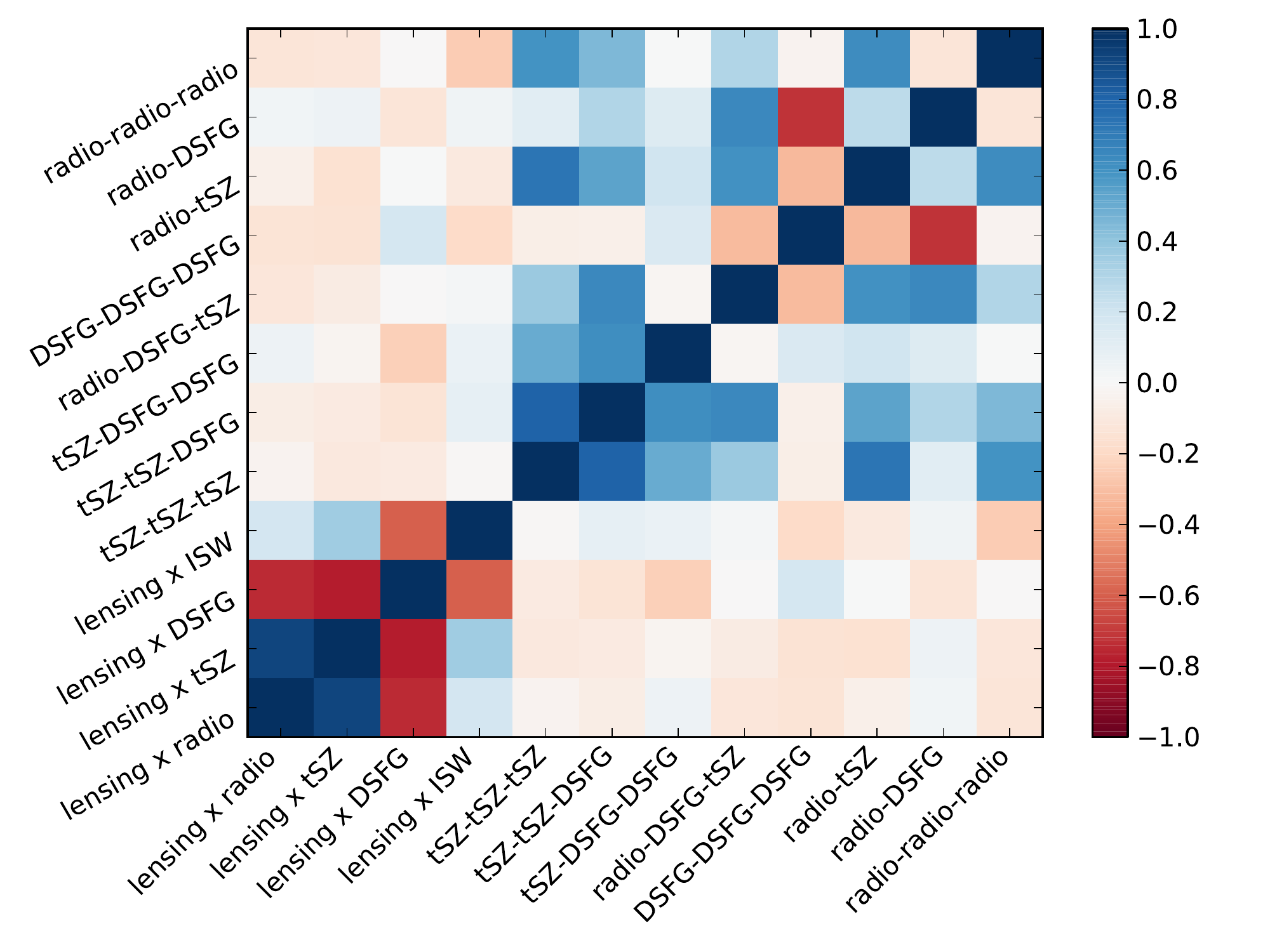}
\caption{The full one-halo covariance matrix for our measured estimators. The strong covariances present in this figure, but absent in Fig \ref{fig:GausErrorCov} arise due to the physical covariances of these templates.}
\label{fig:FullErrorCov}
\end{figure}

\begin{figure}
\includegraphics[width=0.95\textwidth]{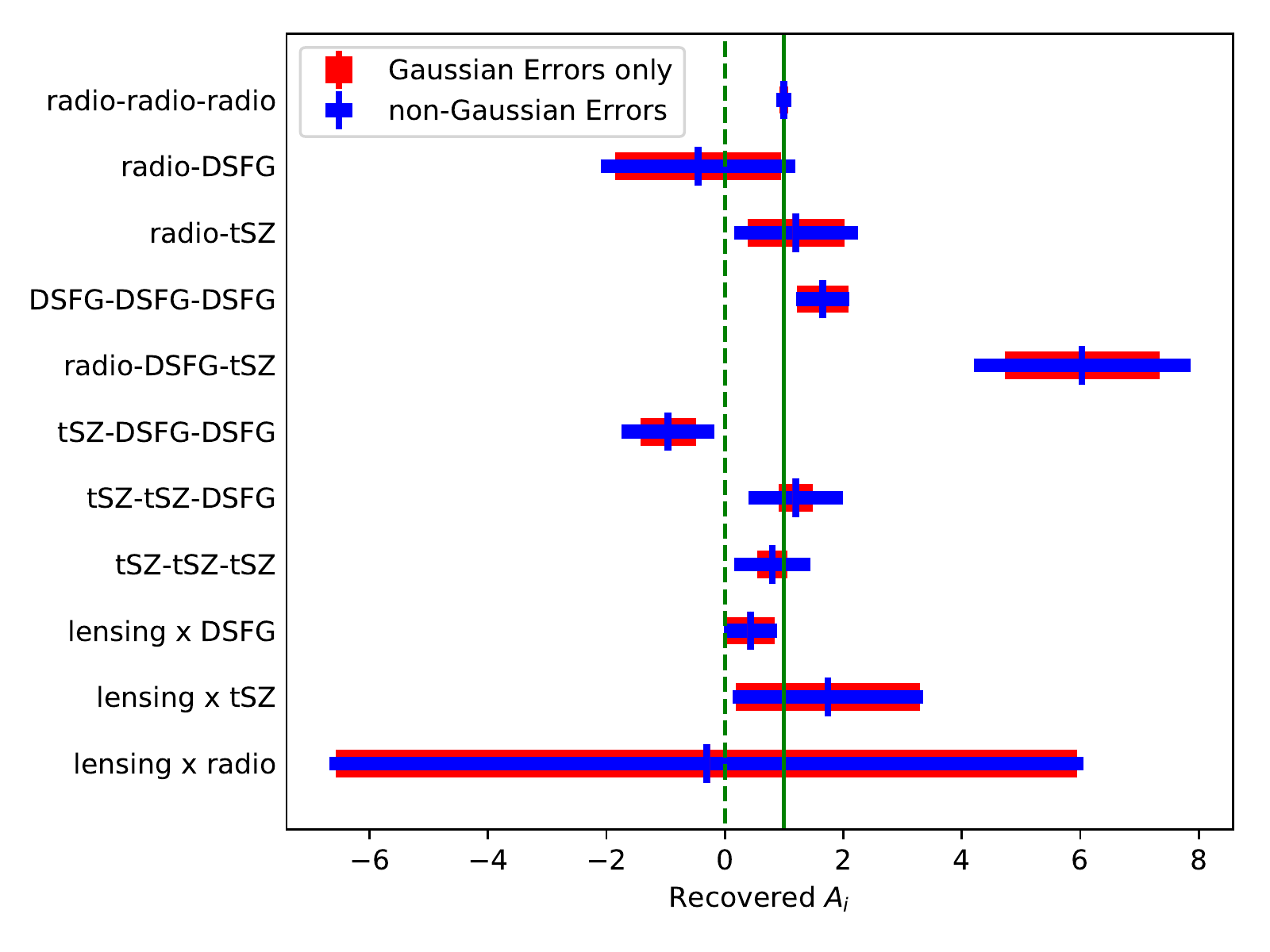}
\caption{A graphical representation of our joint fit results from table \ref{table:ResultsGaussian} excluding the lensing-ISW constraint in order to restrict the scales. The solid green line is the values predicted by our model and the dashed green line is the null value.}
\label{fig:whiskerPlot}
\end{figure}

As we have strong detections of these templates it is important to verify whether the Gaussian errors (that are valid in the limit of weak non-Gaussianity) are still dominant and, if not, to include the non-Gaussian contribution to the cosmic variance. Our method for calculating the six point function is described in Appendix \ref{app:TemplateErrors}. In table \ref{table:ResultsGaussian} we present our constraints on the amplitudes with the errors updated to include the non-Gaussian contributions. Note that for both the Gaussian and non-Gaussian results we used the same estimator with the same normalisation and thus the only difference between the two results is the error bars.
It can be seen that the errors for some of the terms are effectively doubled by the inclusion of the six point terms. Figure \ref{fig:whiskerPlot} shows our results with the Gaussian only errors and with the full non-Gaussian errors. In figure \ref{fig:FullErrorCov} we present the covariance matrix of the templates when we include the full six point function. The full covariance deviates significantly from the Gaussian-only covariance matrix further indicating the necessity of including the non-Gaussian components. The source of the non-Gaussian covariances differs fundamentally from those in the Gaussian case: the Gaussian covariances arise due to the similarity of the templates and the difficulty of differentiating between them with our data sets, whereas the non-Gaussian covariances arise from the physical correlations among the sources of these templates. For example, the large covariance that exists between the tSZ-tSZ-tSZ and tSZ-tSZ-DSFG template primarily arises from the Poissonian fluctuations of the number of dark matter halos, as an increase (or decrease) in the number of dark matter halos will increase (decrease) the tSZ-tSZ-tSZ bispectrum and tSZ-tSZ-DSFG bispectrum in an identical and correlated manner. In contrast, the large covariance between the lensing-tSZ and lensing-ISW templates arises as we are unable to differentiate between the contributions of these two sources of non-Gaussianity with our data sets.

The different physical origin of the Gaussian and non-Gaussian errors means that there is a difference in the interpretation of the results in columns 2 and 3 of table \ref{table:ResultsGaussian}. The results in column 2 give the detection significance of the templates and the amplitude of the template in our patch of sky. Thus we have a $3.2 \sigma$ detection of the tSZ-tSZ-tSZ template in our patch of sky. The results in column 3 give our constraint on the global population amplitude of these templates, we find that we can constrain the amplitude of the tSZ-tSZ-tSZ globally to be $0.80\pm 0.65$. This distinction means that if another experiment was performed in our patch of sky we would expect a result consistent with the results from column 2, however if the same experiment was performed on a different patch of sky we would expect results consistent with column 3.


\section{Discussion, conclusions and future directions}\label{section:conclusions}
We have applied the KSW estimator to simultaneously measure a wide range of sources of non-Gaussianity in the CMB. These first measurements with ACTPol extend the similar work done by \cite{Crawford2014} with the SPT experiment by examining cross-correlations between the sources. In future work we will have the sensitivity to constrain the bispectrum of the kinetic Sunyaev Zel'dovich effect (see Appendix \ref{app:kSZeffect} for current sensitivities) and thereby fully constrain the bispectrum of known CMB secondary sources.

\begin{table}
\centering
 \begin{tabular}{|l |l |l |l |l |}
\hline
 & Komatsu et al. & Calabrese et al. & Crawford et al. & Our model \\ \hline
 
WMAP Q band radio galaxies & $6.0 \pm 1.3 \times 10^{-5}$ &$13.7\pm 1.1\times 10^{-5} $ &$ - $&$ 5.5 \pm 0.7\times 10^{-5} $ \\ \hline 
WMAP V band radio galaxies & $0.43 \pm 0.31\times 10^{-5} $ &$2.3 \pm 0.5 \times 10^{-5}$ &$ - $&$ 0.49\pm 0.06\times 10^{-5} $\\ \hline
WMAP W band radio galaxies & $0.12 \pm 0.24 \times 10^{-5}$ &$1.0\pm 0.5 \times 10^{-5}$ &$ - $&$0.053 \pm 0.007 \times 10^{-5} $ \\ \hline 
220 GHz DSFG (Poisson and Clustered) & $ -$ &$  -$ &$0.917\pm0.14 \times 10^{-9} $&$ 1.41 \pm 0.38 \times10^{-9} $\\ \hline 
\end{tabular}
\caption{A comparison between various different bispectrum measurements with all results given in $\mathrm{\mu K^3 sr^2}$. The Komatsu et al. column contains the results from \protect\cite{Komatsu2009}, Calabrese et al. column contains the results from \protect\cite{Calabrese2010} and Crawford et al. column is from \protect\cite{Crawford2014}. The final column shows what our model, with amplitudes fitted from the ACTPol and Planck data as presented here, would predict for these measurements. To compare our DSFG result to the Crawford et al. result we evaluate our model at $\ell=3000$, $220$ GHz and at their flux cut of $6.4$ mJy. The factor of two difference between our results and those from \protect\cite{Calabrese2010} is discussed in the text.}
\label{tab:ResultComparison}
\end{table}

There have been several previous measurements of non-Gaussianity from secondary anisotropies. In table \ref{tab:ResultComparison} we show the measurements of the bispectrum from the DSFG and radio galaxies from previous work. We compare our results to measurements of the bispectrum amplitude of radio sources from \cite{Komatsu2009} and \cite{Calabrese2010} for the WMAP Q, V and W bands (40, 60 and 90 GHz respectively)  and to measurements of the DSFGs bispectrum from \cite{Crawford2014}. We also show the bispectrum contribution that we would predict from our model as constrained by the measurements. For the WMAP results we assume that the only contribution is from the radio galaxies and use a flux cut of 0.5 Jy at 22 GHz. In both of the analyses of the WMAP data, the authors mask all the point sources detected in \cite{Wright2009} as well radio galaxies identified in external surveys with fluxes greater than 0.5 Jy at 22 GHz  \cite[as described in][]{Hinshaw2007}. The comparison to the WMAP Q and V bands is done by extrapolating our model beyond the frequencies measured in this work.  To compare our measurements of DSFGs with \cite{Crawford2014} we have combined their measurements of the Poisson and clustered components of the DSFGs and compared this to our model's prediction for the DSFG contribution at $\ell=2000$ and 220 GHz. The radio Poisson contribution at this frequency with a $6.4$ mJy mask is negligible. We see that our model has good agreement with the measurements in \cite{Komatsu2009}, but differ from \cite{Calabrese2010} by a factor of two in the Q band and a factor of 20 in W band. A similar level of disagreement was seen between the \cite{Komatsu2009} and the \cite{Calabrese2010} results, and was discussed in \cite{Calabrese2010}. The origin of this disparity is not understood. We see good agreement with the DSFG measurements, though the errors are large; a more interesting comparison will be possible with future data sets. Our measurements of the tSZ bispectrum and lensing-ISW bispectrum are consistent with previous work \citep{Crawford2014,planck2014-a26,wilson2012,Calabrese2010}, but with large errors. 

From our tSZ bispectrum template amplitude we can compute a constraint on $\sigma_8 \Omega_m^{0.17}$, a combination of the amplitude of matter fluctuations smoothed on $8h^{-1}$ Mpc scales and the matter density $\Omega_m$. Our analysis follows that of \cite{Crawford2014} and is described in detail in Appendix \ref{app:sig8}. The resulting constraint is
\begin{equation}
\sigma_8 \Omega_m^{0.17}=0.65^{+0.05}_{-0.06}.
\end{equation}
Whilst we jointly fit the amplitude of the tSZ effect with the other templates, the large degeneracies seen in section \ref{section:Results} and the generally poor fit of our templates, which is discussed more in the following paragraph, mean that this result is a demonstration of the constraining power of the tSZ bispectrum, rather than a robust constraint. As such we do not discuss the cosmological interpretations of our $\sigma_8 \Omega_m^{0.17}$ constraint.

We now consider the ability of the data set used in this work to discriminate among the dozen bispectrum templates in our model.  The model is not a good fit to the data, with $\chi^2 = 43.6$ for 12 degrees of freedom.  This poor fit is likely due to the strong degeneracies seen in the Gaussian covariance matrix ( figure \ref{fig:GausErrorCov}) and deficiencies in the halo model templates. These degeneracies mean that small changes to the covariance matrix lead to large shifts in our results. Physical fluctuations, for example those caused by fluctuations in the number of massive dark matter halos in the patch, or deficiencies in our modeling could mean that our model covariance differs from the true Gaussian covariance of our data and could cause leakage of one template into another. The interpretation of the chi-squared is further complicated as the distribution of the template amplitudes is non-Gaussian. The full non-Gaussian probability distribution for the template amplitudes is difficult to calculate and will be the subject of future work.

There are two further possible causes of the poor fit. Firstly our best fit $\sigma_8 \Omega_m^{0.17}$ value of $ 0.65^{+0.05}_{-0.06}$ is less than the \textit{Planck} best fit value, of 0.67, used to generate our templates. A lower value of  $\sigma_8 \Omega_m^{0.17}$ would reduce the amplitude of our templates and it would also slightly alter the shape of our templates, as the number of most massive halos would be more significantly reduced than the number of lower mass halos. A consistent fitting of the cosmological parameters from bispectrum measurements is a subject of on-going work. Secondly our model of the DSFGs is potentially over simplistic, in particularly we assume no mass dependence in the emissivity as has been included in recent models e.g.  \citet{Viero2013,Shang2012,Bethermin2013,Wu2017}. These models including quenching of star forming galaxies in massive halos and this reduces the one halo term. In future work we will explore the effect of these models on our bispectrum measurements.

All these points suggest that the theoretical uncertainty of our results could be significant and is likely the limiting factor in our analysis; as such, we stress that our results are a proof of principle for this approach, rather than robust measurements of these secondary anisotropies. 

The issues raised above will be addressed with future measurements. Multi-frequency small-scale information, particularly at low frequencies, which will be available from Advanced ACT, SPT3G, Simons Observatory and CMB-S4 \citep{cmbs4ScienceBook,Henderson2016,Benson2014}, should break the strong degeneracies. This will significantly reduce the dependence on the modeling and allow the use of advanced statistics, such as Skew-C$_\ell$s \citep{Munshi2010}, to assess the accuracy of our templates, and to study the astrophysics of these sources. In particular, it will be interesting to see if the high amplitudes of the radio-tSZ-DSFG terms are caused by template degeneracies or are due to radio galaxies occupying lower mass clusters at redshift $\sim 1$ than is assumed in our model. If it is the later then it suggests that radio galaxies, DSFGs and the tSZ effect can be found in the same halos and understanding the correlations among these terms will be important for searches for the kSZ signal.

The calculation of the six point function indicated that non-Gaussian contributions for such measurements are already important and cannot be neglected in future more sensitive measurements. The contribution of the non-Gaussian component to the variance has the potential to limit the power of the bispectrum to measure astrophysics and cosmology through its reduction on the information. However there are many measures that can be employed to reduce this error. For example, masking the largest clusters will significantly reduce the importance of these terms as the six point function receives the largest contribution from low redshift and massive clusters (even more so than the bispectrum itself). An even better approach, which was done in \citet{Hurier2017}, would be to jointly analyse the bispectrum and cluster counts; such an approach would allow a reduction of the errors without loss of the information from the largest clusters.

\section{Acknowledgements}
This work was supported by the U.S. National Science Foundation through awards AST-1440226, AST-0965625 and AST-0408698 for the ACT project, as well as awards PHY-1214379 and PHY-0855887. Funding was also provided by Princeton University, the University of Pennsylvania, and a Canada Foundation for Innovation (CFI) award to UBC. ACT operates in the Parque Astron\'omico Atacama in northern Chile under the auspices of the Comisi\'on Nacional de Investigaci\'on Cient\'ifica y Tecnol\'ogica de Chile (CONICYT). Computations were performed on the GPC supercomputer at the SciNet HPC Consortium and on the hippo cluster at the University of KwaZulu-Natal. SciNet is funded by the CFI under the auspices of Compute Canada, the Government of Ontario, the Ontario Research
Fund -- Research Excellence; and the University of Toronto. The development of multichroic detectors and lenses was supported by NASA grants NNX13AE56G and NNX14AB58G. WRC was partially funded by NSF grant AST-1311756. EC is supported by an STFC Rutherford Fellowship. AvE was supported by the Vincent and Beatrice Tremaine Fellowship. LM is funded by CONICYT FONDECYT grant 3170846. JCH gratefully acknowledges support from the Friends of the Institute for Advanced Study.

\appendix
\section{The flat-sky -full-sky bispectrum correspondence}\label{app:appendixFlatToFull}
In this section we follow the power spectrum derivation of \citet{loverde2008} and derive a relation between the full-sky and flat-sky bispectrum.
\subsection{Full-sky}\label{sec:FullSkyBispectrumCal}
 The 2-D three-point function of three fields is

\begin{equation}
\langle A(\bm{n}_1)B(\bm{n}_2)C(\bm{n}_3)\rangle=\left\langle \int\mathrm{d}r_1 \mathrm{d}r_2 \mathrm{d}r_3 F_{A}(r_1)A^{\mathrm{3D}}(r_1\bm{n}_1)F_{B}(r_2)B^{\mathrm{3D}}(r_2\bm{n}_2)F_{C}(r_3)C^{\mathrm{3D}}(r_3\bm{n}_3)\right \rangle,
\end{equation}
where $F_{X}(\bm{x})$ is the projection kernel and in harmonic space we have
\begin{equation}
\langle A_{\ell_1,m_1} B_{\ell_2,m_2} C_{\ell_3,m_3}\rangle=\int \mathrm{d}\Omega_1\mathrm{d}\Omega_2\mathrm{d}\Omega_3\left\langle A(\bm{n}_1)B(\bm{n}_2)C(\bm{n}_3) Y^*_{\ell,m}(\bm{n}_1)Y^*_{\ell',m'}(\bm{n}_2)Y^*_{\ell'',m''}(\bm{n}_3)\right\rangle.
\end{equation}
To proceed we decompose each 3D field into its Fourier components and then use the plane wave expansion:
\begin{equation}
e^{i\bm{k}\cdot\bm{x}}= 4\pi\sum\limits_{\ell,m} i^{\ell}j_{\ell}(kr)Y^*_{\ell,m}(\bm{\hat{k}})Y^*_{\ell,m}(\bm{\hat{n}}).
\end{equation}
Performing these steps and evaluating the angular integrals leaves
\begin{equation}
\langle A_{\ell_1,m_1}B_{\ell_2,m_2} C_{\ell_3,m_3}\rangle=\int \mathrm{d}r_1 \mathrm{d}r_2 \mathrm{d}r_3 F_{A}(r_1)F_{B}(r_2)F_{C}(r_3)\int\prod\limits_{i}\frac{\mathrm{d}\bm{k}^3_i}{(2\pi)^3} 4\pi i^{\ell_i} j_{\ell_i}(k_i r_i)Y^*_{\ell_i,m_i}(\bm{\hat{k}})\langle \tilde{A}^{\mathrm{3D}}(\bm{k}_1) \tilde{B}^{\mathrm{3D}}(\bm{k}_2) \tilde{C}^{\mathrm{3D}}(\bm{k}_3) \rangle.
\end{equation}
We then assume translational and rotational invariance, i.e.
\begin{equation}
\langle \tilde{A}^{\mathrm{3D}}(\bm{k}_1) \tilde{B}^{\mathrm{3D}}(\bm{k}_2) \tilde{C}^{\mathrm{3D}}(\bm{k}_3) \rangle = (2\pi)^3\delta^{(3)}(\bm{k}_1+\bm{k}_2+\bm{k}_3)B(k_1,k_2,k_3)
\end{equation}
so
\begin{align}
\langle A_{\ell_1,m_1} B_{\ell_2,m_2} C_{\ell_3,m_3}\rangle=&\int \mathrm{d}r_1 \mathrm{d}r_2 \mathrm{d}r_3 F_{A}(r_1)F_{B}(r_2)F_{C}(r_3)\times \nonumber \\ & \int\prod\limits_{i}\frac{\mathrm{d}\bm{k}^3_i}{(2\pi)^3} 4\pi i^{\ell_i} j_{\ell_i}(k_i r_i)Y^*_{\ell_i,m_i}(\bm{\hat{k}})B(k_1,k_2,k_3) \int\mathrm{d}\bm{x} e^{i(\bm{k}_1+\bm{k}_2+\bm{k}_3)\cdot\bm{x}}.
\end{align}
Again using the plane wave expansion and evaluating the angular integrals gives
\begin{align}
\langle A_{\ell_1,m_1} B_{\ell_2,m_2} C_{\ell_3,m_3}\rangle=& \mathcal{G}^{\,\,\ell_1\; \ell_2\; \ell_3}_{m_1 m_2 m_3 } \int \mathrm{d}r_1 \mathrm{d}r_2 \mathrm{d}r_3 F_{A}(r_1)F_{B}(r_2)F_{C}(r_3) \times \nonumber \\ & \int \mathrm{d}x x^2\prod\limits_{i}\frac{\mathrm{d}k_i (4\pi k_i)^2}{(2\pi)^3} j_{\ell_i}(k_i r_i) j_{\ell_i}(k_i x) B(k_1,k_2,k_3).
\end{align}
where:
\begin{equation}\label{eq:GauntIntegral}
\mathcal{G}^{\,\,\ell_1\; \ell_2\; \ell_3}_{m_1 m_2 m_3 } =\int\mathrm{d}\Omega^2 Y_{\ell_1,m_1}(\bm{n})Y_{\ell_2,m_2}(\bm{n})Y_{\ell_3,m_3}(\bm{n})
\end{equation} is the Gaunt integral. Note that the bispectrum can now be expressed in terms of the reduced bispectrum of the full-sky, $b^{\mathrm{FULL}}_{\ell_1,\ell_2,\ell_3}$,
\begin{equation}
\langle A_{\ell_1,m_1} B_{\ell',m'} C_{\ell'',m''}\rangle=\mathcal{G}^{\,\,\ell_1\; \ell_2\; \ell_3}_{m_1 m_2 m_3 } b^{\mathrm{FULL}}_{\ell_1,\ell_2,\ell_3}.
\end{equation}We will focus on the reduced bispectrum. Next we replace the spherical Bessel functions, $j_{\ell}(x)$, with Bessel functions of the first kind, $J_{\ell+\frac{1}{2}}(x)=j_{\ell}(x)\sqrt{\frac{2x}{\pi}} $ to obtain
\begin{equation}
b^{\mathrm{FULL}}_{\ell_1,\ell_2,\ell_3}=\int \mathrm{d}r_1 \mathrm{d}r_2 \mathrm{d}r_3 F_{A}(r_1)F_{B}(r_2)F_{C}(r_3)\int \mathrm{d}x x^2\prod\limits_{i}\mathrm{d}k_i k_i^2 J_{\ell_i+\frac{1}{2}}(k_i r_i) J_{\ell_i+\frac{1}{2}}(k_i x)\frac{1}{k_i \sqrt{x r_i}} B(k_1,k_2,k_3).
\end{equation}
For notational simplicity we will define
\begin{equation}
f_i(r)=\frac{F_{A}(r)}{\sqrt{r}}.
\end{equation}
We can now use the relation \citep{loverde2008}
\begin{align}
\int\mathrm{d}x J_{\nu}(x)f(x)=f(\nu)-\frac{1}{2}f''(\nu) -\frac{\nu}{6}f'''(\nu)+....\nonumber\\
\int\mathrm{d}x J_{\nu}(kx)f(x)=\frac{f\left(\frac{\nu}{k}\right)}{k}-\frac{1}{2k^3}f''\left(\frac{\nu}{k}\right)-\frac{\nu}{6k^4}f'''\left(\frac{\nu}{k}\right)+....
\end{align}
to get
\begin{align}
b^{\mathrm{FULL}}_{\ell_1,\ell_2,\ell_3}=&\int \mathrm{d}x x^2\prod\limits_{i}\mathrm{d}k_i J_{\ell_i+\frac{1}{2}}(k_i x)\frac{1}{ \sqrt{x}} \left(f_{A}(r_1)-\frac{f''_{A}(r_1)}{2 k_1^2}-\frac{\nu_1}{6k_1^3}f'''_{A}(r_1)+....\right) \nonumber \\ & \left(f_{B}(r_2)-\frac{f''_{B}(r_2)}{2 k_2^2}-\frac{\nu_2}{6k_2^3}f'''_{B}(r_2)+....\right)  \left(f_{C}(r_3)-\frac{f''_{C}(r_3)}{2 k_3^2}-\frac{\nu_3}{6k_3^3}f'''_{C}(r_3)+....\right)B(k_1,k_2,k_3)
\end{align}
with $\nu_i=\ell_i+\frac{1}{2}=k_i r_i$. Applying that relation again and retaining only the lowest terms in $\nu_i$ leads to
\begin{align}\label{eq:fullSkyExpanded}
&b^{\mathrm{FULL}}_{\ell_1,\ell_2,\ell_3}=\int \frac{\mathrm{d}x}{x^4}F_A(x)F_B(x)F_C(x) B\left(\frac{\nu_1}{x},\frac{\nu_2}{x},\frac{\nu_3}{x}\right) \times \nonumber \\ & \left(1 -\sum\limits_{i}\frac{1}{6\nu_i^2} \left[ \left( \frac{\mathrm{d}\ln (f_iB)}{\mathrm{d}\ln k_i}\right)^3+\left(\frac{\mathrm{d}\ln f_i}{\mathrm{d}\ln x}\right)^3+\frac{\mathrm{d}^3\ln(f_i B)}{\mathrm{d}\ln ^3k_i}+\frac{\mathrm{d}^3\ln f_i}{\mathrm{d}\ln ^3x}+3\frac{\mathrm{d}^2\ln(f_i B)}{\mathrm{d}\ln ^2k_i}\frac{\mathrm{d}\ln (f_i B)}{\mathrm{d}\ln k_i}+3\frac{\mathrm{d}^2\ln f_i}{\mathrm{d}\ln ^2x}\frac{\mathrm{d}\ln f_i}{\mathrm{d}\ln x} -\frac{\mathrm{d}\ln B}{\mathrm{d}\ln k_i}\right]+...O\left(\nu^{-4}\right)\right) .
\end{align}
Note that we now have $\nu_i=\ell_i+\frac{1}{2}=k_i r_i = k_i x$.
 
\subsection{Flat-sky}
In the flat-sky regime we have
\begin{equation}
\langle A(\bm{x}^{\bot}_1)B(\bm{x}^{\bot}_2)C(\bm{x}^{\bot}_3)\rangle=\left\langle \int\mathrm{d}x^\parallel_1 \mathrm{d}x^\parallel_2 \mathrm{d}x^\parallel_3 F_{A}(x^\parallel_1)A^{\mathrm{3D}}(x^\parallel_1,\bm{x}^{\bot}_1)F_{B}(x^\parallel_2)B^{\mathrm{3D}}(x^\parallel_2,\bm{x}^{\bot}_2)F_{C}(x^\parallel_3)C^{\mathrm{3D}}(x^\parallel_3,\bm{x}^{\bot}_3) \right\rangle
\end{equation}
where $x^\parallel$ is the coordinate along the line of sight and $\bm{x}^{\bot}$ is perpendicular to the line of sight. Expanding in flat-sky Fourier modes gives
\begin{equation}
\langle\bar{A}(\bm{k}^{\bot}_1)\bar{B}(\bm{k}^{\bot}_2)\bar{C}(\bm{k}^{\bot}_3)\rangle=\prod_i\int\mathrm{d}x_i^\parallel \frac{\mathrm{d}k_i^\parallel}{2\pi} F_i(x^\parallel_i)e^{ik^\parallel_i x^\parallel_i} \left\langle \tilde{A}^{\mathrm{3D}}(\bm{k}^{\bot}_1,k^\parallel_1)\tilde{B}^{\mathrm{3D}}(\bm{k}^{\bot}_2,k^\parallel_2)\tilde{C}^{\mathrm{3D}}(\bm{k}^{\bot}_3,k^\parallel_3) \right\rangle.
\end{equation}
Assuming translational and rotational invariance we can then write 
\begin{equation}
\langle \tilde{A}^{\mathrm{3D}}(\bm{k}^{\bot}_1,k^\parallel_1)\tilde{B}^{\mathrm{3D}}(\bm{k}^{\bot}_2,k^\parallel_2)\tilde{C}^{\mathrm{3D}}(\bm{k}^{\bot}_3,k^\parallel_3) = (2\pi)^3\delta^{(3)}(\bm{k}_1+\bm{k}_2+\bm{k}_3)B(k_1,k_2,k_3)
\end{equation}
so
\begin{align}
\langle \bar{A}(\bm{k}^{\bot}_1)\bar{B}(\bm{k}^{\bot}_2)\bar{C}(\bm{k}^{\bot}_3)\rangle=\prod_i\int\mathrm{d}x_i^\parallel \frac{\mathrm{d}k_i^\parallel}{2\pi} F_i(x^\parallel_i)e^{ik^\parallel_i x^\parallel_i} (2\pi)^3\delta^{(3)}(\bm{k}_1+\bm{k}_2+\bm{k}_3)B(k_1,k_2,k_3).
\end{align}
This can be written in terms of the flat-sky bispectrum:
\begin{align}
\langle\bar{A}(\bm{k}^{\bot}_1)\bar{B}(\bm{k}^{\bot}_2)\bar{C}(\bm{k}^{\bot}_3)\rangle=(2\pi)^2\delta^{(2)}(\bm{k}^{\bot}_1+\bm{k}^{\bot}_2+\bm{k}^{\bot}_3)b^{\mathrm{FLAT}}(k^{\bot}_1,k^{\bot}_2,k^{\bot}_3)
\end{align}
with
\begin{align}
b^{\mathrm{FLAT}}(k^{\bot}_1,k^{\bot}_2,k^{\bot}_3)=\prod_i\int\mathrm{d}x_i^\parallel \frac{\mathrm{d}k_i^\parallel}{2\pi} F_i(x^\parallel_i)e^{ik^\parallel_i x^\parallel_i} (2\pi)\delta(k^\parallel_1 +k^\parallel_2 +k^\parallel_3 )B\left(\sqrt{{k^\parallel_1}^2+{k^{\bot}_1}^2},\sqrt{{k^\parallel_2}^2+{k^{\bot}_2}^2},\sqrt{{k^\parallel_3}^2+{k^{\bot}_3}^2}\right).
\end{align}
Expanding $B\left(\sqrt{{k^\parallel_1}^2+{k^{\bot}_1}^2},\sqrt{{k^\parallel_2}^2+{k^{\bot}_2}^2},\sqrt{{k^\parallel_3}^2+{k^{\bot}_3}^2}\right)$ about $k_i^\parallel$ we obtain
\begin{align}
b^{\mathrm{FLAT}}(k^{\bot}_1,k^{\bot}_2,k^{\bot}_3)=\int\mathrm{d}x F_A(x)F_B(x)F_C(x)B(k_1,k_2,k_3)\left(1+\frac{1}{2(k_1x)^2}\frac{\mathrm{d}\ln F_A}{\mathrm{d}\ln x}\frac{\mathrm{d}(\ln F_B\ln F_C)}{\mathrm{d}\ln x}\frac{\mathrm{d}\ln B}{\mathrm{d}\ln  k_1} \nonumber \right.\\ \left. + \frac{1}{2(k_2x)^2}\frac{\mathrm{d}\ln F_B}{\mathrm{d}\ln x}\frac{\mathrm{d}(\ln F_A\ln F_C)}{\mathrm{d}\ln x}\frac{\mathrm{d}\ln B}{\mathrm{d}\ln  k_2}+\frac{1}{2(k_3x)^2}\frac{\mathrm{d}(\ln F_A\ln F_B)}{\mathrm{d}\ln x}\frac{\mathrm{d}\ln F_B}{\mathrm{d}\ln x}\frac{\mathrm{d}\ln B}{\mathrm{d}\ln  k_3}\right).
\end{align}
Expanding $1/x^4 B(\frac{\nu_1}{x},\frac{\nu_2}{x},\frac{\nu_3}{x})$ in equation \ref{eq:fullSkyExpanded} around the peak, $\bar{r}$, of $F_A(x)F_B(x)F_C(x)$ we find $k^4b^{\mathrm{FLAT}}(k^{\bot}_1,k^{\bot}_2,k^{\bot}_3)\approx \ell^4 b^{\mathrm{FULL}}_{\ell_1,\ell_2,\ell_3}$ with $l_i+1/2=k_i \bar{r}$. Note that as in the power spectrum case, the convergence of these series depend both on $\ell+1/2$ and on the projection kernels, if the projection kernels are peaked at different $r_i$ then more terms will be required in this series.

\section{Detailed Bispectrum Templates}\label{app:bispectrumTemplates}

In this appendix we briefly summarize the origin and bispectrum structure of the various significant secondary sources of non-Gaussianity. While the kSZ effect is too small to be constrained with the data used in this work we discuss the form of its bispectrum for completeness. 

In this section we define a deconvolved reduced bispectrum, $\tilde{b}$, where we have factored out the beam and pixelization function, as
\begin{align}
b^{X_1,X_2,X_3}_{\ell_1,\ell_2,\ell_3} \equiv \mathrm{w}^{X_1}_{\ell_1}\mathrm{w}^{X_2}_{\ell_3}\mathrm{w}^{X_2}_{\ell_3}\tilde{b}^{X_1,X_2,X_3}_{\ell_1,\ell_2,\ell_3}
\end{align}
and for notational convenience we suppress the map indicies $\{X_1,X_2,X_3\}$ on the reduced bispectrum. 

\subsection{Radio galaxies}\label{section:radioGals}
Radio galaxies are point sources with strong synchrotron emission and include AGN, flat-spectrum radio quasars (FSRQs) and BL Lacertae type objects (BL Lacs). It has been shown \citep{Toffolatti1998,gonzalez2005} that at CMB frequencies radio galaxies are unclustered. Thus radio galaxies follow Poisson statistics and they have a power spectrum of
\begin{align}
C^{\mathrm{radio}}_{\ell}=k^2_\nu \sum\limits_i\int_{0}^{S_{c}} \mathrm{d}S_\nu^{(i)}\, {S_\nu^{(i)}}^2 \frac{\mathrm{d}n^{(i)}}{\mathrm{d}S\mathrm{d}\Omega}
\end{align}
and a reduce bispectrum of
\begin{equation}
\tilde{b}^{\mathrm{radio-radio-radio}}_{\ell_1,\ell_2,\ell_3}=k_{\nu}^{3} \sum\limits_i  \int_{0}^{S_{c}}\mathrm{d}S^{(i)}_{\nu}\, {S^{(i)}_{\nu}}^3 \frac{\mathrm{d}n^{(i)}}{\mathrm{d}S\mathrm{d}\Omega},
\end{equation}
where we sum over the different populations of radio galaxies, $S_\nu^{(i)}$ is the flux density, $\mathrm{d}n^{(i)} /\mathrm{d}S\mathrm{d}\Omega$ is the number count per steradian per flux interval, S$_c$ is the flux cut and $k_{\nu}$ is the conversion from flux to map temperature. We compute the theoretical level and frequency dependence using the model described in \cite{deZotti2010}. The model has three components, each described by their own number counts and with flux densities $S\propto \nu^{-\alpha}$. The components are flat-spectrum radio quasars and BL Lacs with $ \alpha_{\mathrm{FSRQ}}= \alpha_{\mathrm{BLLac}} = 0.1$, and a population of steep spectrum sources, arising from AGNs, with $\alpha_{\mathrm{steep}} = 0.8$. We extend \cite{deZotti2010} by using their model for differential source counts at 100 GHz and steepening the spectral indices of the three components in agreement with the results from \cite{tucci2011}. 

The bispectrum is weighted more towards the most luminous galaxies than the power spectrum and, given the relatively slow decrease of the source counts as a function of flux (${\mathrm{d}n}/{\mathrm{d}S\mathrm{d}\Omega} \propto S^{-1.5}$) the bispectrum of radio galaxies is dominated by the brightest sources and thus is sensitive to the level of point source masking.
\subsection{ Thermal Sunyaev Zel'dovich}\label{tSZsection}
Light propagating from the surface of last scattering to the present day can be up-scattered by hot gas in a process known as the thermal Sunyaev Zel'dovich effect (tSZ). The magnitude of this effect depends upon the integrated line-of-sight electron pressure, therefore this effect is dominated by galaxy groups and clusters where there is a large amount of hot intercluster gas. The observable signature depends on the frequency of observation; it can be seen as a decrement below frequencies of $ 220$ GHz and an excess at higher frequencies. We briefly review the tSZ bispectrum as developed in \cite{bhattacharya2012}. It should be noted that we only include effects from the one-halo term as previous work \citep{Komatsu1999} demonstrated that for the angular scales of interest in the power spectrum the two halo term is negligible. We have verified that this extends to the bispectrum. In the non-relativistic approximation the temperature fluctuation at angular position $\bm{\theta}$ from the center of a cluster of mass $M$ at redshift $z$ is proportional to the integral along the line-of-sight of the electron pressure:
\begin{align}
\frac{\Delta T(\bm{\theta},M,z)}{T} &=f(x_\nu)y(\bm{\theta},M,z) \nonumber \\
&=f(x_\nu) \frac{ \sigma_T}{m_e c^2} \int_{\mathrm{LOS}} \mathrm{d}r P_e(\sqrt{r^2+D_A(z)^2 |\bm{\theta}|^2},M,z)
\end{align}
where $f (x_\nu) = x_\nu (\coth(x_\nu /2) - 4)$ is the spectral function of the SZ effect, $x_\nu={k_b T_{\mathrm{CMB}}}/{\mathrm{h} \nu}$ with $\nu$ the observing frequency, $D_A(z)$ is the angular diameter distance, and $P_e$ is the electron pressure. The Fourier transform of the projected SZ profile is \citep{Komatsu2002}:
\begin{equation}
\tilde{y}(\ell,M,z)=\frac{4\pi r_s}{\ell_s^2} \frac{\sigma_T}{m_e c^2} \int\limits_{0}^{\infty} \mathrm{d}x x^2 P_e(M,Z,x) \frac{\sin({x \ell }/{\ell_s})}{x \ell/\ell_s},
\end{equation}
where $x=r/r_s$ with $r_s$ a characteristic radius of the cluster which we take to be $r_s=r_{\mathrm{vir}}/c_{\mathrm{NFW}} $ following \cite{navarro1997}, $\ell_s=r_s/D_A(z)$. As is consistent with the literature we have included the projection ($a$) and Limber terms ($1/\chi^2$) in our definition of $\tilde{y}(\ell,M,z)$.  The power spectrum can be calculated by summing the square of the Fourier transform of the projected SZ-profile weighted by the mass function giving:
\begin{align}
 C^{\mathrm{tSZ}}_{\ell}=&f(x_{\nu_1})f(x_{\nu_2})\int_0^\infty \mathrm{d} z \frac{\mathrm{d}^2V}{\mathrm{d}z\mathrm{d}\Omega} \int_0^\infty \mathrm{d} \ln M \frac{\mathrm{d} n}{\mathrm{d} \ln M} \tilde{y}(\ell,M,z) \tilde{y}(\ell,M,z) ,
\end{align}
where $ \frac{\mathrm{d} n}{\mathrm{d} \ln M}$ is the halo mass function \citep[for which we used][]{tinker2008} and $ \frac{\mathrm{d} V}{\mathrm{d} z \mathrm{d}\Omega} $ is the comoving volume per steradian. In this work we use cluster pressure profiles obtained from simulations \citep{Battaglia2012b} and we verified our results are not significantly changed when using observationally constrained profiles \citep{arnaud2010,planck2012-V}. Similarly the one halo reduced bispectrum is obtained by summing the cube of the Fourier transform of the projected SZ-profile weighted by the mass function giving:
\begin{align} 
\tilde{b}^{\mathrm{tSZ-tSZ-tSZ}}_{\ell_1,\ell_2,\ell_3} = f(x_{\nu_1})f(x_{\nu_2})f(x_{\nu_3})\int_0^\infty \mathrm{d} z \frac{\mathrm{d}^2V}{\mathrm{d}z\mathrm{d}\Omega} \int_0^\infty \mathrm{d} \ln M \frac{\mathrm{d} n}{\mathrm{d} \ln M} \tilde{y}(\ell_1,M,z) \tilde{y}(\ell_2,M,z) \tilde{y}(\ell_3,M,z).
\end{align}

\subsection{Dusty star-forming Galaxies}\label{section:haloModel}
Dusty star-forming galaxies (DSFGs) are high redshift (z $\sim 2-3$), dusty, star-forming galaxies; the UV radiation emitted by the newly formed massive stars is absorbed by the dust and re-emitted at longer wavelengths.
In CMB maps these objects appear as point sources which are domi- nant at high frequencies but still important for the frequencies considered here \citep{Hauser2001}. We model these sources with two components: a Poisson component and a clustered component. We use the halo model \citep{cooray2002} as a unified way to calculate both the theoretical DSFG template and the cross-correlations with other physical effects. In this we follow the works of \cite{Lacasa2014}, \cite{addison2012} and \cite{xia2012} and we combine the halo model with halo occupation statistics to populate dark matter halos with DSFGs \citep{zheng2005}. We use the assumption that the properties of the DSFGs depend only upon redshift and use the parametric model by \cite{bethermin2012} to model the source properties. This model incorporates two populations of galaxies, main sequence and starburst, with redshift evolution. This parametric model was fit to differential counts of mm objects from \textit{Herschel} and \textit{Spitzer} data \citep{bethermin2010,berta2011b,Glenn2010}. The two and three halo terms, which have been measured by the \textit{Planck} satellite \citep{planck2013-pip56}, are only important on the largest scales \citep{addison2012,Lacasa2014}. As we have very limit sensitivity to scales with $\ell<1000$, we only consider the one-halo and Poisson terms that dominate the small scales. This model makes the simplifying assumption that the galaxy emissivity only varies as a function of redshift. Recent work has developed models that include mass dependence on the emissivity \citep{Viero2013,Shang2012,Bethermin2013,Wu2017}. In future work we will explore the effect of mass dependence on the bispectrum, but in this work we focus on the simpler, mass independent, case.

For a detailed description of this bispectrum we refer the reader to \cite{Lacasa2014} and here we summarize their results.
We define the n$^{th}$ order differential source intensity as :
\begin{equation}
\frac{\mathrm{d}I^{(n)}(\nu_1,\nu_2...,\nu_n)}{\mathrm{d}V}=\int\limits_{0}^{S_{cut}} dS_{\nu_1}S_{\nu_1}S_{\nu_2}...S_{\nu_n} \frac{\mathrm{d}^2 n(S_{\nu_1},z)}{\mathrm{d}V\mathrm{d}S},
\end{equation}
$S_{\nu_i}$ is the observer frame flux at frequency $\nu_i$, $S_{cut}$ is the flux above which all sources are masked and ${\mathrm{d}^2 n(S_{\nu_1},z)}/{\mathrm{d}V\mathrm{d}S}$ is the differential count and is obtained from the \cite{bethermin2012} model. Note that this can be related to the generalized differential source emissivity $j^{(n)}(\nu_1,\nu_2...,\nu_n)$ \citep{knox2001} used in \cite{Lacasa2014} by
\begin{equation} 
\frac{\mathrm{d}I^{(n)}(\nu_1,\nu_2...,\nu_n)}{\mathrm{d}V}= \frac{a^n}{\chi^{2n}}j^{(n)}(\nu_1,\nu_2...,\nu_n);
\end{equation}
in other words we absorbed the projection operator (a) and the Limber terms($1/\chi^2$) into our differential source count. 
Through the halo occupation model prescription we have
\begin{equation}
\langle N_{\mathrm{gal}}\rangle=\langle N_{\mathrm{cen}}\rangle+\langle N_{\mathrm{sat}}\rangle
\end{equation}
with
\begin{align}\label{eq:nCentralDef}
\langle N_{\mathrm{cen}}\rangle &=\frac{1}{2}\left[1+\mathrm{erf} \left(\frac{\log(M)-\log(M_{\mathrm{min}})}{\sigma_{\log(M)}}\right)\right]
\end{align}
and
\begin{align}\label{eq:nSatDef}
\langle N_{\mathrm{sat}}\rangle &=\frac{1}{2} \left[ 1+\mathrm{erf} \left(\frac{\log(M)-\log(2M_{\mathrm{min}})}{\sigma_{\log(M)}}\right)\right]\left(\frac{M}{M_{\mathrm{sat}}}\right)^{\alpha_{\mathrm{sat}}},
\end{align}
where $M$ is the halo mass, and $M_{\mathrm{min}}$, $\sigma_{\log(M)}$, $M_{\mathrm{sat}}$ and $ \alpha_{\mathrm{sat}}$ parameterize the occupation of the halos. $M_{\mathrm{min}}$ gives the mass at which a halo has a $50\%$ probability of having a central galaxy and $\sigma_{\log(M)}$ describes the width of the transition from no central galaxy to one central galaxy. $M_{\mathrm{sat}}$ and $ \alpha_{\mathrm{sat}}$ describe the number of satellite galaxies. The parameters for the HOD model were fit to \textit{Planck} power spectra in \cite{penin2014} and we use these values in this work. Finally we assume that the distribution of the DSFGs follows that of the dark matter profile in the halo. Thus the Fourier transform of the 3D DSFG spatial distribution, $\tilde{u}_{\mathrm{DSFGs}}$ is
\begin{align}
\tilde{u}_{\mathrm{DSFG}}(\ell,M,z)=\tilde{u}_{\mathrm{dm}}(\ell,M,z)=\int\limits_0^{r_{vir}}\mathrm{d}r \frac{\sin((\ell+1/2)r/\chi)}{(\ell+1/2)r/\chi}\frac{\rho(r| M,z)}{M}
\end{align}
where $\tilde{u}_{\mathrm{dm}}(\ell,M,z)$ is the Fourier transformed of the normalised halo profile.  Bringing all these pieces together we can write down the power spectrum of the DSFG. There are two components a one galaxy term, called the Poisson component
\begin{equation}
C^{\mathrm{DSFG-Pois}}_{\ell} = \int_0^\infty \mathrm{d} z \frac{\mathrm{d}^2V}{\mathrm{d}z\mathrm{d}\Omega} \frac{\mathrm{d}I^{(2),\mathrm{DSFG}}(\nu_1,\nu_2)}{\mathrm{dV} }.
\end{equation}
and two galaxy, or clustered component,
\begin{equation}
C^{\rm{DSFG-Clustered}}= \int_0^\infty \mathrm{d} z \frac{\mathrm{d}^2V}{\mathrm{d}z\mathrm{d}\Omega} \frac{\mathrm{d}I^{(1),\mathrm{DSFG}}(\nu_1)}{\mathrm{d}V}\frac{\mathrm{d}I^{(1),\mathrm{DSFG}}(\nu_2)}{\mathrm{d}V} \int_0^\infty \mathrm{d} \ln M \frac{\mathrm{d} n}{\mathrm{d} \ln M} \tilde{u}_{\mathrm{dm}}(\ell_1,M,z) \tilde{u}_{\mathrm{dm}}(\bm{\ell}_2,M,z) \frac{\langle N_{\mathrm{gal}}(N_{\mathrm{gal}}-1)\rangle}{\bar{n}_{\mathrm{gal}}^2}.
\end{equation}

For the bispectrum there are three DSFG components. The three galaxy term
\begin{align}
\tilde{b}^{\mathrm{DSFG} - 3 \mathrm{\mathrm{gal}}}_{\ell_1,\ell_2,\ell_3}=& \int_0^\infty \mathrm{d} z \frac{\mathrm{d}^2V}{\mathrm{d}z\mathrm{d}\Omega} \frac{\mathrm{d}I^{(1),\mathrm{DSFG}}(\nu_1)}{\mathrm{d}V}\frac{\mathrm{d}I^{(1),\mathrm{DSFG}}(\nu_2)}{\mathrm{d}V}\frac{\mathrm{d}I^{(1),\mathrm{DSFG}}(\nu_3)}{\mathrm{d}V} \\ & \int_0^\infty \mathrm{d} \ln M \frac{\mathrm{d} n}{\mathrm{d} \ln M} \tilde{u}_{\mathrm{dm}}(\ell_1,M,z)\tilde{u}_{\mathrm{dm}}(\ell_2,M,z)\tilde{u}_{\mathrm{dm}}(\ell_3,M,z)\frac{\langle N_{\mathrm{gal}}(N_{\mathrm{gal}}-1)(N_{\mathrm{gal}}-2)\rangle}{\bar{n}_{\mathrm{gal}}^3},
\end{align}
the two galaxy term,
\begin{align}\label{eq:2galDSFG}
\tilde{b}^{\mathrm{DSFG} - 2 \mathrm{\mathrm{gal}}}_{\ell_1,\ell_2,\ell_3} =& \int_0^\infty \mathrm{d} z \frac{\mathrm{d}^2V}{\mathrm{d}z\mathrm{d}\Omega} \frac{\mathrm{d}I^{(2),\mathrm{DSFG}}(\nu_1,\nu_2)}{\mathrm{d}V}\frac{\mathrm{d}I^{(1),\mathrm{DSFG}}(\nu_3)}{\mathrm{d}V} \int_0^\infty \mathrm{d} \ln M \frac{\mathrm{d} n}{\mathrm{d} \ln M} \tilde{u}_{\mathrm{dm}}(\ell_1,M,z) \tilde{u}_{\mathrm{dm}}(|\bm{\ell}_2+\bm{\ell}_3|,M,z) \frac{\langle N_{\mathrm{gal}}(N_{\mathrm{gal}}-1)\rangle}{\bar{n}_{\mathrm{gal}}^2}  \nonumber \\ &+ \text{ 2 cyclic permutations},
\end{align}
 and the one galaxy term :
\begin{align}
\tilde{b}^{\mathrm{DSFG} -1 \mathrm{\mathrm{gal}}}_{\ell_1,\ell_2,\ell_3} =& \int_0^\infty \mathrm{d} z \frac{\mathrm{d}^2V}{\mathrm{d}z\mathrm{d}\Omega} \frac{\mathrm{d}I^{(3),\mathrm{DSFG}}(\nu_1,\nu_2,\nu_3)}{\mathrm{dV} }.
\end{align}
Typically the first two terms are called the clustered component and the last term is called the Poisson component. Note that note that $\ell_1 = | \bm{\ell}_2+\bm{\ell}_3 |$ and thus the integrand in eq. \ref{eq:2galDSFG} is actually separable. In this work we fix the relative amplitude of these three contributions and fit for one overall amplitude, hereafter called the DSFG-DSFG-DSFG component. The source counts for the Poisson component decrease rapidly as a function of flux and as such it is largely insensitive to the cut level used. We also express the $N_{\mathrm{gal}}$ averages in terms of $N_{\mathrm{cen}}$ and $N_{\mathrm{sat}}$. Assuming that $N_{\mathrm{sat}}$ follows a Poisson distribution, we obtain the following (see Appendix \ref{app:NorderHODmoments}):
\begin{align}
{\langle N_{\mathrm{gal}}(N_{\mathrm{gal}}-1)\rangle}=2\langle N_{\mathrm{sat}} \rangle +\langle N_{\mathrm{sat}}\rangle^2 
\end{align}
and
\begin{align}
{\langle N_{\mathrm{gal}}(N_{\mathrm{gal}}-1)(N_{\mathrm{gal}}-2)\rangle}=3 \langle N_{\mathrm{sat}} \rangle^2 +\langle N_{\mathrm{sat}}\rangle^3.
\end{align}
The halo formalism allows the cross-correlation between the DSFGs and the tSZ effect to be easily calculated. The tSZ-DSFG power spectrum is given by
\begin{align}
C^{\mathrm{tSZ-DSFG}}_{\ell}=&f(x_{\nu_1})\int_0^\infty \mathrm{d} z \frac{\mathrm{d}^2V}{\mathrm{d}z\mathrm{d}\Omega}\frac{\mathrm{d}I^{(1),\mathrm{DSFG}}(\nu_2)}{\mathrm{d}V} \left[ \int_0^\infty \mathrm{d} \ln M \frac{\mathrm{d} n}{\mathrm{d} \ln M} \tilde{y}(\ell,M,z) \tilde{u}(\ell,M,z)\langle \frac{N_{\mathrm{DSFG}}(M)}{\bar{n}_{\mathrm{DSFG}}}\rangle \nonumber \right.\\
 & \left. +\int_0^\infty \mathrm{d} \ln M \frac{\mathrm{d} n}{\mathrm{d} \ln M}b(M)\, \tilde{y}(\ell,M,z)\int_0^\infty \mathrm{d} \ln M' \frac{\mathrm{d} n}{\mathrm{d} \ln M}b(M')\, \tilde{u}(\ell,M',z)\langle \frac{N_{\mathrm{DSFG}}(M')}{\bar{n}_{\mathrm{DSFG}}}\rangle P_{\mathrm{lin}}\left(\frac{\ell+1/2}{\chi(z)},z\right) \right].
\end{align}
There are two bispectra arsing from cross correlations between the tSZ effect and DSFGs:
\begin{align}
\tilde{b}^{\mathrm{tSZ-DSFG-DSFG}}_{\ell_1,\ell_2,\ell_3} =& \int_0^\infty \mathrm{d} z \frac{\mathrm{d}^2V}{\mathrm{d}z\mathrm{d}\Omega} \left[f(x_{\nu_1})\frac{\mathrm{d}I^{(1),\mathrm{DSFG}}(\nu_2)}{\mathrm{d}V}\frac{\mathrm{d}I^{(1),\mathrm{DSFG}}(\nu_3)}{\mathrm{d}V} \int_0^\infty \mathrm{d} \ln M \frac{\mathrm{d} n}{\mathrm{d} \ln M}\tilde{y}(\ell_1,M,z)  \tilde{u}_{\mathrm{dm}}(\ell_2,M,z)\tilde{u}_{\mathrm{dm}}(\ell_3,M,z)\right. \nonumber \\ &\left. \frac{\langle N_{\mathrm{gal}}(N_{\mathrm{gal}}-1)\rangle}{\bar{n}_{\mathrm{gal}}^2} 
 +f(x_{\nu_1})\frac{\mathrm{d}I^{(2),\mathrm{DSFG}}(\nu_2,\nu_3)}{\mathrm{d}V} \int_0^\infty \mathrm{d} \ln M \frac{\mathrm{d} n}{\mathrm{d} \ln M} \tilde{y}(\ell_1,M,z)\tilde{u}_{\mathrm{dm}}(|\bm{\ell}_2+\bm{\ell}_3|,M,z) \frac{\langle N_{\mathrm{gal}}\rangle}{\bar{n}_{\mathrm{gal}}} \right]  \nonumber \\ &+ \text{ 2 cyclic permutations}
\end{align}
and
\begin{align}       
\tilde{b}^{\mathrm{tSZ-tSZ-DSFG}}_{\ell_1,\ell_2,\ell_3} = \int_0^\infty \mathrm{d} z \frac{\mathrm{d}^2V}{\mathrm{d}z\mathrm{d}\Omega} f(x_{\nu_1} )f(x_{\nu_2}) \frac{\mathrm{d}I^{(1),\mathrm{DSFG}}(\nu_3)}{\mathrm{d}V} \int_0^\infty \mathrm{d} \ln M \frac{\mathrm{d} n}{\mathrm{d} \ln M} \tilde{y}(\ell_1,M,z)\tilde{y}(\ell_2,M,z) \tilde{u}_{\mathrm{dm}}(\ell_3,M,z)\frac{\langle N_{\mathrm{gal}}\rangle}{\bar{n}_{\mathrm{gal}}}  \nonumber \\+ \text{ 2 cyclic permutations}.
\end{align}

\subsection{Cross-bispectra among radio galaxies, the tSZ effect and DSFGs} \label{section:radioCrossTemplates}
Whilst the radio galaxies show little clustering, recent work by \cite{Gralla2013,Gupta2017} has found evidence for a correlation between radio galaxies and tSZ sources, since radio galaxies occupy the same clusters that source the tSZ effect. Building on this we consider the cross-bispectra that arise from correlations among the tSZ effect, DSFGs and radio galaxies. There are five possible bispectra: radio-tSZ-tSZ, radio-radio-tSZ, radio-DSFG-DSFG, radio-radio-DSFG and radio-DSFG-tSZ. In this work we fix the relative amplitude of the radio-tSZ-tSZ and radio-radio-tSZ terms, hereafter this joint term is called the radio-tSZ template, and the relative amplitude of the radio-DSFG-DSFG, radio-radio-DSFG terms, hereafter the joint term is called the radio-DSFG template. These terms were combined as they cannot be distinguished with our current data sets. This is because the radio terms are expected to the smallest bispectra, and radio terms are most important at low frequencies, which can not be throughly studied with the \textit{Planck} $100$ GHz data used in this analysis. The tSZ-DSFG templates discussed in section \ref{section:haloModel} were not combined as they are expected to be larger than the radio templates and the \textit{Planck} FWHM at $217$ GHz is smaller, allowing the small scales to be more accurately measured.

To estimate the radio halo occupation, we use the model from \cite{Wake2008}, which is also reexamined in \cite{smolcic2011}. In \cite{Wake2008}, the authors constrain HOD parameters with data from a set of combined surveys (see \cite{Sadler2007} for more details). They are unable to constrain the parameters associated with satellite galaxies, implying most halos host only one radio galaxy, and as such we use only a simplified HOD model:
\begin{equation}
N_{\mathrm{radio}}=e^{-\frac{M^*}{M}},
\end{equation}
where $M$ is the halo mass and $M^*=9.65 \times10^{13}h^{-1}M_\odot$ was as measured in \cite{Wake2008}. We combine this model with the halo model and force the radio galaxies to lie in the center of the cluster as was preferred in \cite{smolcic2011} and \cite{Wake2008} so that $\tilde{r}(\ell,M,z)=1$.  Combining this with the results of the previous section we can write down the power spectrum for radio, tSZ and DSFG cross correlations
\begin{align}
C^{\mathrm{radio-tSZ}}_{\ell}=&f(x_{\nu_1})\int_0^\infty \mathrm{d} z \frac{\mathrm{d}^2V}{\mathrm{d}z\mathrm{d}\Omega}\frac{\mathrm{d}I^{(1),\mathrm{radio}}(\nu_2)}{\mathrm{d}V} \left[ \int_0^\infty \mathrm{d} \ln M \frac{\mathrm{d} n}{\mathrm{d} \ln M} \tilde{y}(\ell,M,z) \tilde{r}(\ell,M,z)\langle \frac{N_{\mathrm{radio}}(M)}{\bar{n}_{\mathrm{radio}}}\rangle \right. \nonumber \\
 &\left.+\int_0^\infty \mathrm{d} \ln M \frac{\mathrm{d} n}{\mathrm{d} \ln M}b(M)\, \tilde{y}(\ell,M,z)\int_0^\infty \mathrm{d} \ln M' \frac{\mathrm{d} n}{\mathrm{d} \ln M}b(M')\, \tilde{r}(\ell,M',z)\langle \frac{N_{\mathrm{radio}}(M')}{\bar{n}_{\mathrm{radio}}}\rangle P_{\mathrm{lin}}\left(\frac{\ell+1/2}{\chi(z)},z\right)\right],
\end{align}
and
\begin{align}
C^{\mathrm{radio-DSFG}}_{\ell}=& \int_0^\infty \mathrm{d} z \frac{\mathrm{d}^2V}{\mathrm{d}z\mathrm{d}\Omega}\frac{\mathrm{d}I^{(1),\mathrm{DSFG}}(\nu_1)}{\mathrm{d}V}\frac{\mathrm{d}I^{(1),\mathrm{radio}}(\nu_2)}{\mathrm{d}V} \left[ \int_0^\infty \mathrm{d} \ln M \frac{\mathrm{d} n}{\mathrm{d} \ln M} \tilde{u}_{\mathrm{dm}}(\ell,M,z) \tilde{r}(\ell,M,z)\langle \frac{N_{\mathrm{DSFG}}(M)}{\bar{n}_{\mathrm{DSFG}}} \frac{N_{\mathrm{radio}}(M)}{\bar{n}_{\mathrm{radio}}}\rangle \right. \nonumber \\ 
& \left. +\int_0^\infty \mathrm{d} \ln M \frac{\mathrm{d} n}{\mathrm{d} \ln M}b(M)\, \tilde{u}_{\mathrm{dm}}(\ell,M,z) \langle \frac{N_{\mathrm{DSFG}}(M)}{\bar{n}_{\mathrm{DSFG}}} \rangle\int_0^\infty \mathrm{d} \ln M' \frac{\mathrm{d} n}{\mathrm{d} \ln M}b(M')\, \tilde{r}(\ell,M',z)\langle \frac{N_{\mathrm{radio}}(M')}{\bar{n}_{\mathrm{radio}}}\rangle P_{\mathrm{lin}}\left(\frac{\ell+1/2}{\chi(z)},z\right)\right].
\end{align}
At the power spectrum level it is very challenging to separate the contributions from tSZ-DSFG, radio-tSZ and radio-DSFG terms. The strength of the cross-correlations between these sources is still poorly understood. Small-scale power spectrum measurements provide constraints on these terms \citep{george2015,dunkley2013} whilst the tSZ-DSFG cross-correlation at large scales has been explored in \cite{planck2014-a29}. Our model predicts a correlation coefficient between the DSFGs and the tSZ effect of $0.28$ at $\ell=3000$. This is higher than was seen in \cite{george2015} who found $0.113^{+0.057}_{-0.051}$, but consistent with other previous work \citep{addison2012}. 

We can then write down the template for the radio-tSZ term
\begin{align}
\tilde{b}^{\mathrm{radio-tSZ}}_{\ell_1,\ell_2,\ell_3} =& f(x_{\nu_2})f(x_{\nu_3})\int_0^\infty \mathrm{d} z \frac{\mathrm{d}^2V}{\mathrm{d}z\mathrm{d}\Omega}{\mathrm{d}z}\frac{\mathrm{d}I^{(1),\mathrm{radio}}(\nu_1)}{\mathrm{d}V} \int_0^\infty \mathrm{d} \ln M \frac{\mathrm{d} n}{\mathrm{d} \ln M}\tilde{r}(\ell_1,M,z) \tilde{y}(\ell_2,M,z)\tilde{y}(\ell_3,M,z)  \langle \frac{N_{\mathrm{radio}}(M)}{\bar{n}_{\mathrm{radio}}}\rangle \nonumber \\
& +f(x_{\nu_3}) \int_0^\infty \mathrm{d} z \frac{\mathrm{d}^2V}{\mathrm{d}z\mathrm{d}\Omega} \frac{\mathrm{d}I^{(2),\mathrm{radio}}(\nu_1,\nu_2)}{\mathrm{d}V} \int_0^\infty \mathrm{d} \ln M \frac{\mathrm{d} n}{\mathrm{d} \ln M}\tilde{r}(|\bm{\ell}_1+\bm{\ell}_2|,M,z)\tilde{y}(\ell_3,M,z) \langle \frac{N_{\mathrm{radio}}(M)}{\bar{n}_{\mathrm{radio}}}\rangle  \nonumber \\ &+ \text{ 2 cyclic permutations}.
\end{align}
Similarly, the radio-DSFG template is:
\begin{align}
\tilde{b}^{\mathrm{radio-DSFG}}_{\ell_1,\ell_2,\ell_3} =& \int_0^\infty \mathrm{d} z \frac{\mathrm{d}^2V}{\mathrm{d}z\mathrm{d}\Omega}{\mathrm{d}z} \frac{\mathrm{d}I^{(1),\mathrm{radio}}(\nu_1)}{\mathrm{d}V}\frac{\mathrm{d}I^{(1),\mathrm{DSFG}}(\nu_2)}{\mathrm{d}V}\frac{\mathrm{d}I^{(1),\mathrm{DSFG}}(\nu_3)}{\mathrm{d}V} \nonumber\\ & \times \int_0^\infty \mathrm{d} \ln M \frac{\mathrm{d} n}{\mathrm{d} \ln M}\tilde{r}(\ell_1,M,z) \tilde{u}_{\mathrm{dm}}(\ell_2,M,z)\tilde{u}_{\mathrm{dm}}(\ell_3,M,z) \langle \frac{N_{\mathrm{DSFG}}(M)(N_{\mathrm{DSFG}}(M)-1)}{\bar{n}^2_{\mathrm{DSFG}}} \rangle \langle \frac{N_{\mathrm{radio}}(M)}{\bar{n}_{\mathrm{radio}}}\rangle \nonumber \\
& + \int_0^\infty \mathrm{d} z \frac{\mathrm{d}^2V}{\mathrm{d}z\mathrm{d}\Omega} \frac{\mathrm{d}I^{(1),\mathrm{radio}}(\nu_1)}{\mathrm{d}V}\frac{\mathrm{d}I^{(2),\mathrm{DSFG}}(\nu_2,\nu_3)}{\mathrm{d}V} \int_0^\infty \mathrm{d} \ln M \frac{\mathrm{d} n}{\mathrm{d} \ln M} \tilde{r}(\bm{\ell}_1,M,z) \tilde{u}_{\mathrm{dm}}(|\bm{\ell}_2+\bm{\ell}_3|,M,z)\langle \frac{N_{\mathrm{DSFG}}(M)}{\bar{n}_{\mathrm{DSFG}}} \frac{N_{\mathrm{radio}}(M)}{\bar{n}_{\mathrm{radio}}}\rangle \nonumber \\
& +\int_0^\infty \mathrm{d} z \frac{\mathrm{d}^2V}{\mathrm{d}z\mathrm{d}\Omega} \frac{\mathrm{d}I^{(2),\mathrm{radio}}(\nu_1,\nu_2)}{\mathrm{d}V} \frac{\mathrm{d}I^{(1),\mathrm{DSFG}}(\nu_3)}{\mathrm{d}V} \int_0^\infty \mathrm{d} \ln M \frac{\mathrm{d} n}{\mathrm{d} \ln M}\tilde{r}(|\bm{\ell}_1+\bm{\ell}_2|,M,z) \tilde{u}_{\mathrm{dm}}(\ell_3,M,z)\langle \frac{N_{\mathrm{DSFG}}(M)}{\bar{n}_{\mathrm{DSFG}}} \frac{N_{\mathrm{radio}}(M)}{\bar{n}_{\mathrm{radio}}}\rangle  \nonumber \\ &+ \text{ 2 cyclic permutations},
\end{align}
and the final template, the radio-DSFG-tSZ term, is
\begin{align}
\tilde{b}^{\mathrm{radio-DSFG-tSZ}}_{\ell_1,\ell_2,\ell_3} =& f(x_{\nu_3}) \int_0^\infty \mathrm{d} z \frac{\mathrm{d}^2V}{\mathrm{d}z\mathrm{d}\Omega}\frac{\mathrm{d}I^{(1),\mathrm{radio}}(\nu_1)}{\mathrm{d}V} \frac{\mathrm{d}I^{(1),\mathrm{DSFG}}(\nu_2)}{\mathrm{d}V} \int_0^\infty \mathrm{d} \ln M \frac{\mathrm{d} n}{\mathrm{d} \ln M}\tilde{r}({\ell_1})\tilde{u}(\ell_2)\tilde{y}(\ell_3) \langle \frac{N_{\mathrm{DSFG}}(M)}{\bar{n}_{\mathrm{DSFG}}} \frac{N_{\mathrm{radio}}(M)}{\bar{n}_{\mathrm{radio}}}\rangle  \nonumber \\ &+ \text{ 5 permutations}.
\end{align}
To calculate these templates we need to know the joint radio-DSFG halo occupation model. In this work we assume they are independent so that $\langle N_{\mathrm{DSFG}}(M) N_{\mathrm{radio}}(M)\rangle=\langle N_{\mathrm{DSFG}}(M)\rangle \langle N_{\mathrm{radio}}(M)\rangle$. This assumption could overestimate the size of the correlated terms as radio galaxies often correspond to strong AGN that could quench the star formation. An exploration of this is left to future work.  

\subsection{Cross-bispectra between lensing and secondary sources}
As was described in section \ref{subsec:lensingCrossMain} we can decompose the temperature anisotropies, having expanded the effect of lensing to first order in $\phi$, as
\begin{equation}
\Delta T(\bm{n})=\Delta T^P(\bm{n}) + \nabla\phi \cdot \nabla \Delta T^P(\bm{n}) + \Delta T^s(\bm{n}).
\end{equation}
and this leads to a bispectrum with the form
\begin{align} \label{eq:lensingBaseApp}
\tilde{b}^{\mathrm{lensing-sec.}}_{\ell_1,\ell_2,\ell_3} =-\bm{\ell}_1\cdot\bm{\ell}_2C_{\ell_1}^{TT}C_{\ell_2}^{\phi S} + 5\, \mathrm{ permutations}.
\end{align}
\cite{lewis2011} showed that the effect of higher order terms in $\phi$ lead to $\sim 10\%$ corrections to the perturbative result in equation \ref{eq:lensingBaseApp} and they showed these higher order terms could accurately be approximated by replacing $C^{TT}$ with the lensed temperature power spectrum. Unlike the other bispectra listed here, this source of non-Gaussianity has a significant contribution from polarized maps through the lensing term. In this work we only analyze temperature maps so do not not considered the polarized contribution in this work. 

In the following subsections we describe our models of the cross-correlations between the lensing potential and the secondary sources. The two halo term has a larger contribution to the power spectrum than in the three-point function so we include its contribution in these terms.

\subsubsection{Thermal Sunyaev Zel'dovich lensing cross-correlation} 
The cross-correlation between the lensing potential and thermal SZ effect is calculated in \cite{hill2014} and \cite{Battaglia2015} and has two contributions, the one and two halo terms are
\begin{align}
C^{\phi-t\mathrm{SZ}}_{\ell}=&\int \mathrm{d}z\frac{\mathrm{d}^2{V}}{\mathrm{d}z\mathrm{d}\Omega} \int\mathrm{d}M\frac{\mathrm{d}n(M,z)}{\mathrm{d}\ln M}\tilde{y}(\ell,M,z)\tilde{\phi}(\ell,M,z)\nonumber\\
&+\int \mathrm{d}z\frac{\mathrm{d}^2{V}}{\mathrm{d}z\mathrm{d}\Omega} \int\mathrm{d}M\frac{\mathrm{d}n(M,z)}{\mathrm{d}\ln M} b(M)\, \tilde{y}(\ell,M,z) \int\mathrm{d}M'\frac{\mathrm{d}n(M',z)}{\mathrm{d}\ln M'}b(M')\,\tilde{\phi}(\ell,M',z) P_{\mathrm{lin}}\left(\frac{\ell+1/2}{\chi(z)},z\right) 
\end{align}
where $\tilde{\phi}$ is the lensing potential profile for a cluster
\begin{equation}
\tilde{\phi}(\ell,M,z)=\frac{2}{\ell(\ell+1)}\frac{4 \pi r_{s,\phi}}{l_{s,\phi}^2} \int \mathrm{d}x_{\phi} x^2_{\phi} \frac{\sin((\ell+1/2)x_{\phi}/l_{s,\phi})}{(\ell+1/2)x_{\phi}/l_{s,\phi}}\frac{\rho (x_{\phi}r_{s,\phi},M,z)}{\Sigma_{crit}(z)},
\end{equation}              
$r_{s,\phi}$ is a characteristic scale radius for the cluster, in our case the NFW scale radius, $\rho$ is the cluster density, given by the NFW profile, $\Sigma_{crit}$ is the critical surface density for lensing at redshift z
\begin{equation}
\Sigma^{-1}_{crit}(z)=\frac{4\pi G \chi(z)(\chi_*-\chi(z))}{c^2\chi_*(1+z)},
\end{equation}
where $\chi_*$ is the comoving distance to the surface of last scattering, and the other terms are as in section \ref{tSZsection}.

\subsubsection{DSFGs lensing cross-correlation} 
The DSFG-lensing cross-correlation has been detected \citep[e.g.][]{Omori2017,vanEngelen2015,planck2013-p13}. We use the halo model for the DSFGs as described in section \ref{section:haloModel} to model the lensing-DSFGs cross-correlation. Through this formalism we obtain the following relations
\begin{align}
&C^{\phi-\mathrm{DSFG}}_{\ell}=\int \mathrm{d}z\frac{\mathrm{d}^2{V}}{\mathrm{d}z\mathrm{d}\Omega} \frac{\mathrm{d}I^{(1),\mathrm{DSFG}}(\nu_1)}{\mathrm{d}V} \int\mathrm{d}M\frac{\mathrm{d}n(M,z)}{\mathrm{d}\ln M}\tilde{u}_{\mathrm{dm}}(\ell,M,z)\tilde{\phi}(\ell,M,z)\langle \frac{N_{\mathrm{gal}}(M)}{\bar{n}_{\mathrm{gal}}}\rangle \nonumber \\
&+\int \mathrm{d}z\frac{\mathrm{d}^2{V}}{\mathrm{d}z\mathrm{d}\Omega}\frac{\mathrm{d}I^{(1),\mathrm{DSFG}}(\nu_1)}{\mathrm{d}V} \int\mathrm{d}M\frac{\mathrm{d}n(M,z)}{\mathrm{d}\ln M}b(M)\,\tilde{u}_{\mathrm{dm}}(\ell,M,z) \langle \frac{N_{\mathrm{gal}}(M)}{\bar{n}_{\mathrm{gal}}} \rangle\int\mathrm{d}M'\frac{\mathrm{d}n(M',z)}{\mathrm{d}\ln M'}b(M')\,\tilde{\phi}(\ell,M',z) P_{lin}\left(\frac{\ell+1/2}{\chi(z)},z\right).
\end{align}

\subsubsection{Radio--Lensing cross-correlation}
Given the discussion above in section \ref{section:radioCrossTemplates} we should expect a cross-correlation between the $\phi$ field and radio sources. This will have the same structure as the DSFG-lensing cross-correlation:
\begin{align}
&C^{\phi-\mathrm{radio}}_{\ell}=\int \mathrm{d}z\frac{\mathrm{d}^2{V}}{\mathrm{d}z\mathrm{d}\Omega} \frac{\mathrm{d}I^{(1),\mathrm{radio}}(\nu_1)}{\mathrm{d}V} \int\mathrm{d}M\frac{\mathrm{d}n(M,z)}{\mathrm{d}\ln M}\tilde{u}_{\mathrm{dm}}(\ell,M,z)\tilde{\phi} (\ell,M,z)\langle \frac{N_{\mathrm{radio}}(M)}{\bar{n}_{\mathrm{radio}}}\rangle \nonumber \\
&+\int \mathrm{d}z\frac{\mathrm{d}^2{V}}{\mathrm{d}z\mathrm{d}\Omega}\frac{\mathrm{d}I^{(1),\mathrm{radio}}(\nu_1)}{\mathrm{d}V} \int\mathrm{d}M\frac{\mathrm{d}n(M,z)}{\mathrm{d}\ln M}b(M)\,\tilde{u}_{\mathrm{dm}}(\ell,M,z) \langle \frac{N_{\mathrm{radio}}(M)}{\bar{n}_{\mathrm{radio}}} \rangle\int\mathrm{d}M'\frac{\mathrm{d}n(M',z)}{\mathrm{d}\ln M'}b(M')\,\tilde{\phi}(\ell,M',z) P_{lin}\left(\frac{\ell+1/2}{\chi(z)},z\right).
\end{align}

\subsubsection{ISW-Lensing cross-correlation}
As was noted in \cite{verde2002} and seen in \cite{planck2013-p14,planck2014-a26}, secondary CMB anisotropies caused by the late time decay of the gravitational potential, the integrated Sachs Wolfe effect \citep{sachs1967}, and their growth through non-linear structure \citep{Rees1968} are correlated with CMB lensing leading to bispectrum with the above form. This bispectrum is primarily important on the largest scales and so we expect very limited sensitivity to this bispectrum with ACTPol. We use CAMB to calculate the lensing-ISW cross-correlation, $C^{ISW-\phi}$ and combine this with equation \ref{eq:lensingBaseApp} to form a bispectrum template.

\subsection{The kinetic Sunyaev Zel'dovich effect}\label{app:kSZeffect}
The kinetic Sunyaev Zel'dovich effect (kSZ) arises due to CMB photons being scattered from moving gas and this motion imparts a Doppler shift to the CMB photons. This effect can be either positive or negative (depending on the direction of the motion of gas) and so in the bispectrum analysis only kSZ squared terms are non vanishing. In this section we outline an approximate template for the kSZ bispectrum and discuss its measurability with our data set. This approach to measure the kSZ bispectrum is complimentary to other methods such as those discussed in \cite{Hill2016,Ferraro2016,Dore2004,DeDeo2005}.

We use the halo model to calculate simplified one-halo templates for cross-correlations of the kSZ squared with other tracers. We stress that these are only approximate templates and that there will be significant contributions from two and three halo terms, which are not considered here. The temperature shift caused by the kSZ effect is given by \citep{Ostriker1986,Sunyaev1972}
\begin{equation}
\frac{\Delta T(\bm{n})}{T}= - \sigma_T \int \frac{\mathrm{d}{\chi}}{1+z} e^{-\tau(z)} n_e(\chi \bm{n},\chi) \frac{ \bm{v}_e}{c} \cdot \bm{n}
\end{equation}
where $\tau(z)$ is the optical depth, $n_e$ is the electron number density and $\bm{v}_e$ is the electron velocity. In our one-halo approximation the deconvolved reduced bispectrum arising from the kSZ cross secondary is
\begin{equation}
\tilde{b}^{kSZ-kSZ-X}_{\ell_1,\ell_2,\ell_3} =\sigma_T^2 \int_0^\infty \mathrm{d} z \frac{\mathrm{d}^2V}{\mathrm{d}z\mathrm{d}\Omega} \int_0^\infty \mathrm{d} \ln M \frac{\mathrm{d} n}{\mathrm{d} \ln M} \tilde{n}_e(M,z,\ell_1) \tilde{n}_e(M,z,\ell_2) \tilde{X}(M,z,\ell_3)\frac{\sigma^2_{vc}(M)}{3}e^{-2\tau(z)} 
\end{equation}
where $ \tilde{n}_e(M,z,\ell_1)$ is the projected line of sight integral of the electron number density, $\sigma_{vc}(M)$ is the cluster velocity dispersion and $\tilde{X}(M,z,\ell_3)$ is $f(\nu) \tilde{y}(M,z,\ell)$ for the kSZ-kSZ-tSZ bispectrum and $\frac{\mathrm{d}I}{\mathrm{d}z} \tilde{u}(M,z,\ell)$ for the kSZ-kSZ-DSFGs bispectrum.
 We approximate the cluster velocity dispersion as the velocity of the halo given by peaks theory as \citep{Bardeen1986} 
\begin{equation}
\sigma_{vc}(M)=\sigma_v(M)\sqrt{1-\frac{\sigma_0^4}{\sigma_1^2 \sigma_{-1}^2}},
\end{equation}
where:
\begin{align}
\sigma_v&=(faH)^2\sigma_{-1}, \\
\sigma_{j}&=\frac{1}{2\pi}\int \mathrm{d}k P(k)W(Rk)k^{2j+2},
\end{align}
and $f={\mathrm{d}\ln(g)}/{\mathrm{d}\ln(a)}$ is the derivative of the growth function. The optical depth is given by the integral of the mean ionized electron number density to the halo
\begin{equation}
\tau(z)=\sigma_T \int\frac{\mathrm{d}{\chi}}{1+z}\bar{n}_e(z).
\end{equation}
To close these equations we need the electron number density profile and, motivated by the results in \cite{schaan2016}, we consider a Gaussian distribution.  The Gaussian distribution describes the profile of the electrons in the halos. The width of the Gaussian is given by halo's the virial radius and we normalize the profile by fixing the baryon mass to be given by $f_b M_{dm}$ and then using the mean molecular weight and the mean molecular weight per free electron to convert to the total number of electrons. The choice of the Gaussian is purely phenomenological. A physical model for the distribution should be related to the tSZ effect as the same electrons are involved in both processes.

 We then performed a fisher forecast to assess their measurability. In table \ref{table:kSZResults} we present the expected signal to noise from fitting just the three kSZ templates separately, fitting a combined kSZ template and jointly fitting these templates with the templates considered in section \ref{section:templates}. When we fit the combined template we assume that the relative amplitude of the three different kSZ templates is fixed and that only a possible overall scaling can exist. This is a simplification but the results show that even with such simplifications we have no constraining power current for measuring the kSZ effect; as such we exclude this template from our analysis. It should be noted the combined template is not constrained better than the individual templates, despite being a sum of the other contributions, due to the opposite signs of the tSZ-kSZ-kSZ and DSFG-kSZ-kSZ terms. It should be noted that if we could break the degeneracies between these templates and the other secondary source templates, such that the joint fit and single fit errors are similar, then future experiments should be able to start constraining these terms. The degeneracy between these templates will be reduced for experiments with multifrequency small-scale measurements.

\begin{table}
\centering
\begin{tabular}{|l |l |l |}
\hline
\multirow{2}{*}{Type} &\multicolumn{2}{c|}{ S/N}\\
\cline{2-3}&Fit alone & Fit with other templates\\ \hline
kSZ-kSZ-tSZ& $0.45 $ & $0.019$ \\
kSZ-kSZ-DSFG& $0.31 $ & $ 38.1 $ \\
kSZ-kSZ-radio& $ 0.15 $ & $ 0.026 $ \\
Combined kSZ Template & $0.30$& $0.032$ \\
\hline
\end{tabular}
\caption{The expected signal to noise the kinetic Sunyaev Zel'dovich templates. We see that whilst the sensitivities for the individual templates is close to having constraining power, when the templates are jointly fit the similarity of the templates to our data sets leaves no constraining power.}
\label{table:kSZResults}
\end{table}

\section{Pipeline Validation Tests}\label{app:CodeValidation}
\subsection{Validation Templates}
In this section we describe the suite of tests that we have preformed to verify the accuracy and efficiency of our analysis.  To validate our pipeline we used three types of bispectra that can be easily simulated; these are Poisson noise, pseudo-local and primordial local non-Gaussianity. We did not investigate the amplitudes of these templates in the real data and only used them for validation. These types were chosen due to their ease of simulation, which can be challenging for arbitrary non-Gaussianity. 

To cross check the analysis pipeline described in section \ref{section:Pipeline} (hereafter pipeline 1) we implemented a second pipeline based on \cite{Das2009} (hereafter pipeline 2) which is described in section \ref{sec:pipeline2} . For most of these analyses we limit the $\ell$ range of our analysis to be $\ell<4000$, as Pipeline 2 is restricted to these scales.

\subsection{Non-Gaussian Simulations}\label{sec:nonGausSims}
In order to validate our pipelines we need to verify that we can accurately measure non-Gaussianity in the maps. In general we use the method described in \cite{Smith2011}, hereafter called the indirect method, to generate maps with arbitrary bispectra. We briefly summarize this method here. First we generate a Gaussian map (as in \ref{sec:estLinearAndNorm}) and then perform the following operation to its Fourier coefficients ($ \mathrm{a}_{\bm{\ell}}^{L}$):
\begin{equation}\label{eq:nonGausIndirectSim}
{a^X_{\bm{\ell}}}^{NL}= {a^X_{\bm{\ell}}}^{L} + \frac{1}{6}\sum\limits_{X_1,X_2} \int \mathrm{d}^2\ell_1 \mathrm{d}^2\ell_2 4 \pi^2 \delta^{(2)}(\bm{\ell} +\bm{\ell}_1+\bm{\ell}_2) b^{X,X_1,X_2}_{\bm{\ell},\bm{\ell}_1,\bm{\ell}_2} {C_{\bm{\ell}_1}^{-1}}^{X_1}a^{L} {C_{\bm{\ell}_2}^{-1}}^{X_2}a^{L}
\end{equation}
where $b_{\bm{\ell},\bm{\ell}_i,\bm{\ell}_j}$ is the reduced bispectrum of the non-Gaussianity that you wish to generate. This method is useful for testing our pipelines although higher order correlations are not accurately reproduced. The method of generating this non-Gaussianity is similar to the method used to calculate the normalization of our estimator, and so this method would not be able to identify errors that jointly affect this method and the normalization calculation.
 
As a cross check of this method and for further robustness of our analysis pipeline we generate several types of non-Gaussianity via direct methods. For this, we consider the three templates Poisson, pseudo-local and local non-Gaussianity.
\subsubsection{Poisson Noise}\label{sec:poisNoise}
The Poisson noise bispectrum has a trivial structure:
\begin{align}
\tilde{b}^{\mathrm{const}}_{\ell_1,\ell_2,\ell_3} =\mathrm{Const}.
\end{align}
This template is the same as the radio point source term or the DSFG poisson component, except we have removed any frequency dependence to simplify the template. Poisson non-Gaussianity is simulated by adding Poisson noise to the maps.
 
\subsubsection{Pseudo-Local bispectrum}\label{sec:pseudoLocal}
This template is inspired by the primordial local bispectrum and it is generated by adding $A_{\mathrm{pseudo-local}}/T_{CMB}^2(\Delta T(\bm{n})-\bar{\Delta T})^2$ to the map. This has a bispectrum of the following form:
\begin{align}
\tilde{b}^{\mathrm{pseudo-local}}_{\ell_1,\ell_2,\ell_3} =2 C(\ell_1)C(\ell_2) +3 \, \mathrm{permutations}.
\end{align}
This type of non-Gaussianity is not physical. We use it to validate our pipelines due to the simplicity of simulating it and as it is orthogonal to the other test templates.

\subsubsection{Primordial Local non-Gaussianity}
 Primordial local non-Gaussianity is a physical  primordial type of non-Gaussianity, so-called as it generated by the real space interactions of fields and is thus local. The KSW estimator \citep{komatsu2005} used in this work was original used to constrain this type of non-Gaussianity. In on-going work we are using the ACTPol data to primordial non-Gaussianity, including local non-Gaussianity, and so it is important to check our pipeline on primordial templates as well. Physically it arises when two short wavelength modes interact with a longer wavelength mode. The local primordial three point structure is given by \citep{Falk1993,Gangui1994,Verde2000,Wang2000,komastu2001}
\begin{align}
B^{\mathrm{local}}_{\Phi}(k_1,k_2,k_3)&= 2 f_{local}^{\mathrm{NL}}( P_{\Phi}(k_1)P_{\Phi}(k_2) + P_{\Phi}(k_1)P_{\Phi}(k_3)+ P_{\Phi}(k_2)P_{\Phi}(k_3)) \nonumber \\
&=2A^2 f_{\mathrm{local}}^{NL}\left(\frac{1}{k_1^{(4-n_s)}k_2^{(4-n_s)}}+ \frac{1}{k_1^{(4-n_s)}k_3^{(4-n_s)}}+\frac{1}{k_2^{(4-n_s)}k_3^{(4-n_s)}}\right),
\end{align}
where $n_s$ is the primordial spectral tilt and A is the amplitude of fluctuations. From this we obtain the reduced bipsectrum as \begin{equation}
\tilde{b}^{\mathrm{prim}}_{\ell_1,\ell_2,\ell_3} =\frac{8}{\pi^3}\int \mathrm{d}r\,r^2\prod\limits_i \left( \int \mathrm{d}k_i\,k_i^2 g^{X_i}_{\ell}(k_i)j_{\ell}(k_i r) \right)B_{\Phi}(k_1,k_2,k_3),
\end{equation}
 where $g^{X_i}_{\ell}(k_i)$ is the transfer function that can be computed with a Boltzmann code such as CAMB \citep{LewisCAMB}. 

Local non-Gaussianity can be generated by first generating a realization of primordial Gaussian fluctuations,$(\Phi(\bm{x})$ and then modifying the primordial fluctuations in the following manner:
\begin{equation}
\Phi^{NL}(\bm{x})=\Phi(\bm{x}) + A_{\mathrm{local}}\left(\Phi^2(\bm{x})-\langle \Phi^2(\bm{x}) \rangle\right).
\end{equation} 
We then apply the transfer functions to evolve these fluctuations to the present. We use algorithms described in \cite{elsner2009} and \cite{liguori2003} to efficiently generate non-Gaussian maps at high resolution. We use their method to generate full-sky maps and then cut out a patch that is appropriate for our fields. This method exploits the fact that the $a_{lm}$ are related to primordial fluctuations in the following manner
\begin{equation}\label{eq:localSimGen}
a_{lm}=\int \mathrm{d} r r^2 \Phi_{lm}(r) \alpha_{l}(r),
\end{equation}
where $\alpha_l (r)$ are the real space transfer function and $ \Phi_{lm}(r)$ are harmonic transforms of the real space primordial potential fluctuations. We draw $ \Phi_{lm}(r)$ from a Gaussian distribution, with the appropriate radial correlations, perform a spherical harmonic transform of these, square them, subtract the variance and transform back. Then we use equation \ref{eq:localSimGen} to evolve these perturbations and attain the late time $a_{lm}$. This quantity is then added back to the original map scaled to the appropiate f$_{\mathrm{NL}}$ to add local non Gaussianity. We then transform these to a HEALpix map \citep{gorski2005} and cut out a patch to use in our flat-sky analysis. We refer the reader to \cite{liguori2003} for further details on this method. 

\subsection{Pre-whitening Analysis Pipeline}\label{sec:pipeline2}
This pipeline is based on \cite{Das2009} and uses the same real space and Fourier masks as described in section \ref{section:Pipeline}. We pre-whiten the maps by using a disc differencing method; this involves convolving two copies of the maps with discs of radius $R=1.5'$ and $3R$ and then differencing them. We then add back a fraction of the original maps to these maps and then convolve the maps with a Gaussian. This pre-whitening reduces the mode coupling when we apply the point source and the mask (as above). When we move to Fourier space we deconvolve these effects to attain the desired maps. This method is limited to probing a reduced range of $\ell$ space as deconvolving the disc-differencing involves a sinc function, and the signal near the zero of the sinc function cannot be accurately recovered. For this reason we restrict the use of this method to testing our pipelines at $\ell<4000$.

\subsection{Fisher Information Comparisons}
It is important to assess how the variance of our estimators compares to the theoretical limits. A Fisher analysis shows that the expected variance of our estimator should be (in the diagonal covariance case)
\begin{equation}
\sigma^2(\mathrm{A_{sec}}^{\alpha}) =F^{-1}_{\alpha,\alpha}\nonumber \\
 \end{equation}
 \begin{equation}
  F_{\alpha,\beta}=\frac{\mathrm{f^{(3)}_{\mathrm{sky}}f^{(3)}_{\mathrm{sky}}}}{6} \int \prod\limits_{i} \mathrm{d}^2\bm{\ell}_i (2\pi)^2 \delta(\bm{\ell}_1+ \bm{\ell}_2+\bm{\ell}_3) \frac{b_{\ell_1,\ell_2,\ell_3} b_{\ell_1,\ell_2,\ell_3}}{C_{\bm{\ell}_1}C_{\bm{\ell}_2}C_{\bm{\ell}_3}}.
 \end{equation}
 We generate sets of 100 Gaussian simulations of maps as would be measured by the ACTPol PA2, including beam effects and anisotropic noise as described section \ref{sec:estLinearAndNorm}. In table \ref{tab:stdDis} we show the expected standard deviation of our estimators against the measured standard deviation. As mentioned above pipeline $2$ only works for $\ell<4000$ and so when we consider $\ell>4000$ we limit our analysis to pipeline $1$. We see generally good agreement between the different methods, with pipeline 2 having slightly higher variance. There is close agreement between our estimator variances with the Fisher estimations, which implies that our methods are close to satisfying the Cram\'er-Rao bound.

\begin{table}
\centering
 \begin{tabular}{|l |l |l |l |l |}
\hline
Type & Max $\ell$ &$\sigma_{\mathrm{Fisher}}$ &$\sigma_{\mathrm{measured}}$ Pipeline 1 &$\sigma_{\mathrm{measured}} $ Pipeline 2 \\ \hline
 
Poisson Noise & $ 3000$ &$ 8.2$ &$ 8.0$&$ 8.8$ \\ \hline 
Poisson Noise & $ 7500$ &$ 0.19$ &$ 0.21$&$ -$\\ \hline
Pseudo Local& $ 3000$ &$ 65$ &$ 68$&$ 87$ \\ \hline 
Local & $ 3000$ &$  430$ &$ 420$&$ 450$\\ \hline 
tSZ-tSZ-tSZ & $ 3000$ &$  0.23$ &$ 0.23 $&$ 0.22$\\ \hline 
tSZ-tSZ-tSZ & $ 7500$ &$  0.085$ &$ 0.086 $&$ -$\\ \hline 
\end{tabular}
\caption{A comparison between the expected and measured standard deviations of our estimator when applied to simulated Gaussian maps. The consistency of our estimator is a check of the optimality of our implementation, especially our approximate inverse covariance filtering.}
\label{tab:stdDis}
\end{table}
\subsection{Simulated non-Gaussian Maps}
It is important to verify that our estimators are able to accurately recover known levels of non-Gaussianity. To do this we first consider the three types that we can directly simulate: the constant, pseudo-local and local form. We compared direct non-Gaussian simulations to non-Gaussianity generated by the method described in \cite{Smith2011} and section \ref{sec:nonGausSims}. By confirming the accuracy of the \cite{Smith2011} method, we can then use it to verify the accuracy of estimators on the remaining templates, which are much more challenging to simulate directly. 

\begin{figure}
\subfloat[Constant non-Gaussianity with $\ell_{\mathrm{max}}=7500$ .]{
    \centering
    \includegraphics[width=0.5\textwidth]{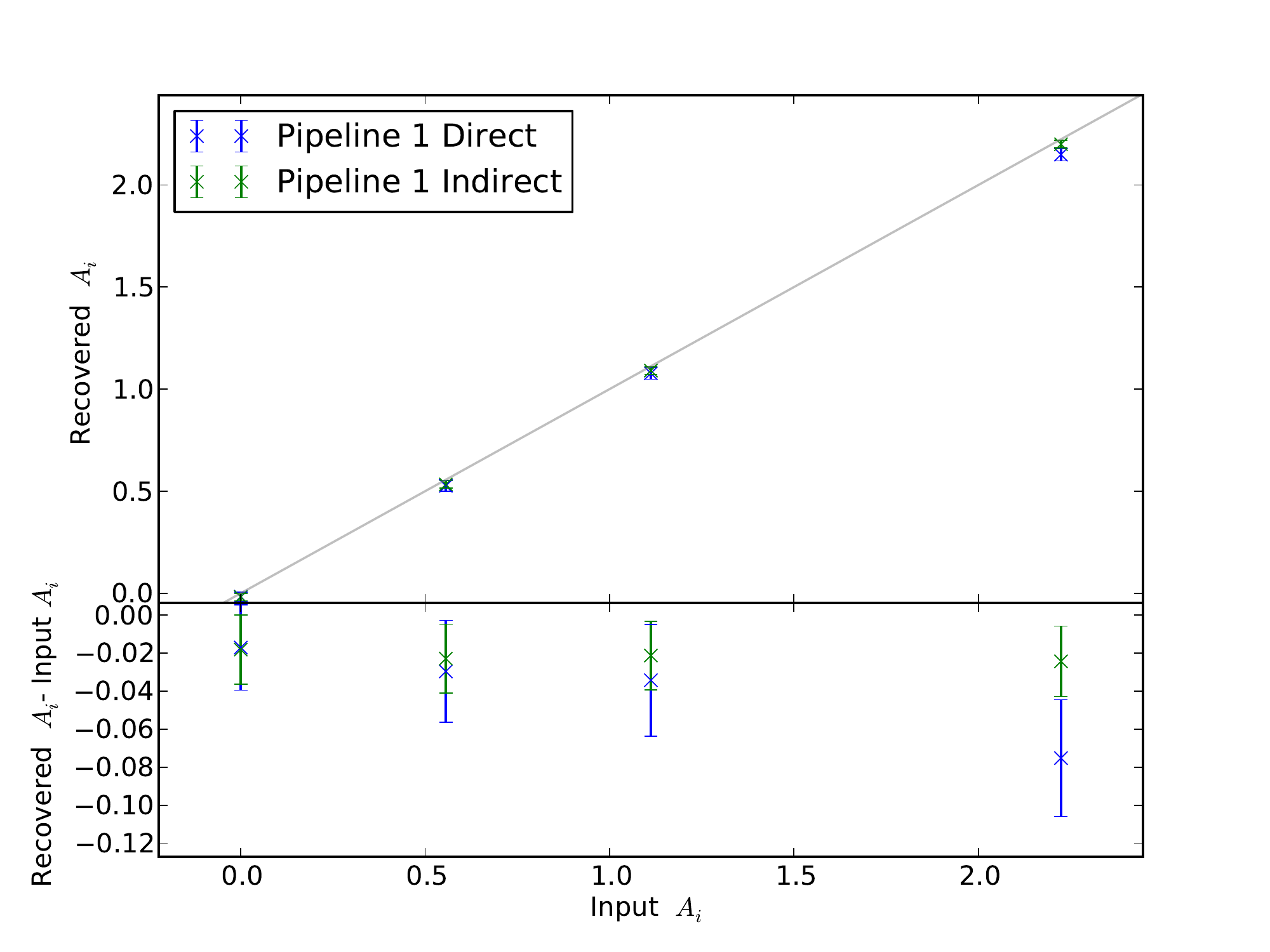}
    \label{fig:fnl_simCons}}\qquad
\subfloat[Constant non-Gaussianity with $\ell_{\mathrm{max}}=3000$.]{
    \centering
    \includegraphics[width=0.5\textwidth]{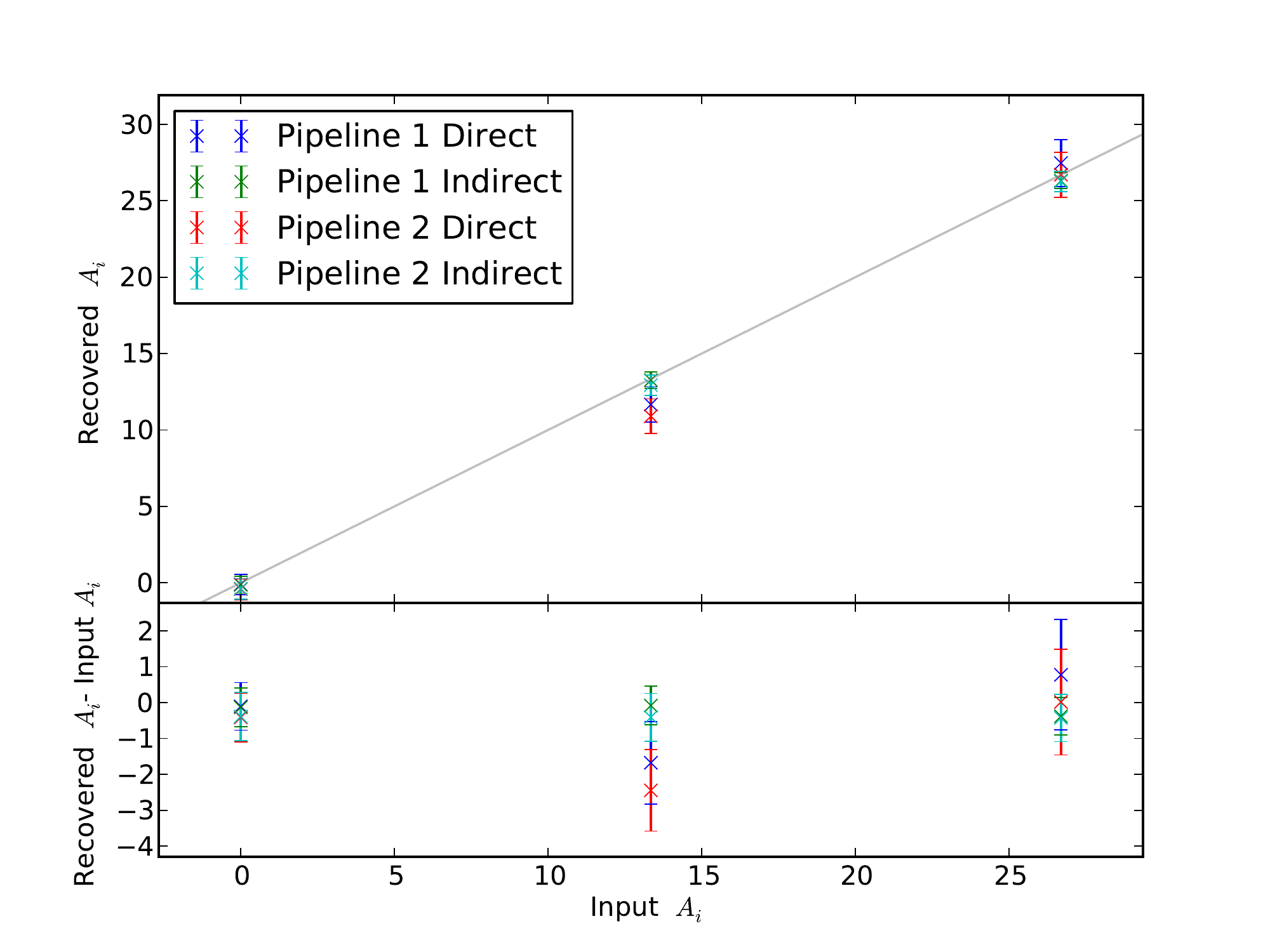}
    \label{fig:fnl_simLocal}}
    \qquad
\subfloat[Local non-Gaussianity with $\ell_{\mathrm{max}}=3000$.]{
    \centering
    \includegraphics[width=0.5\textwidth]{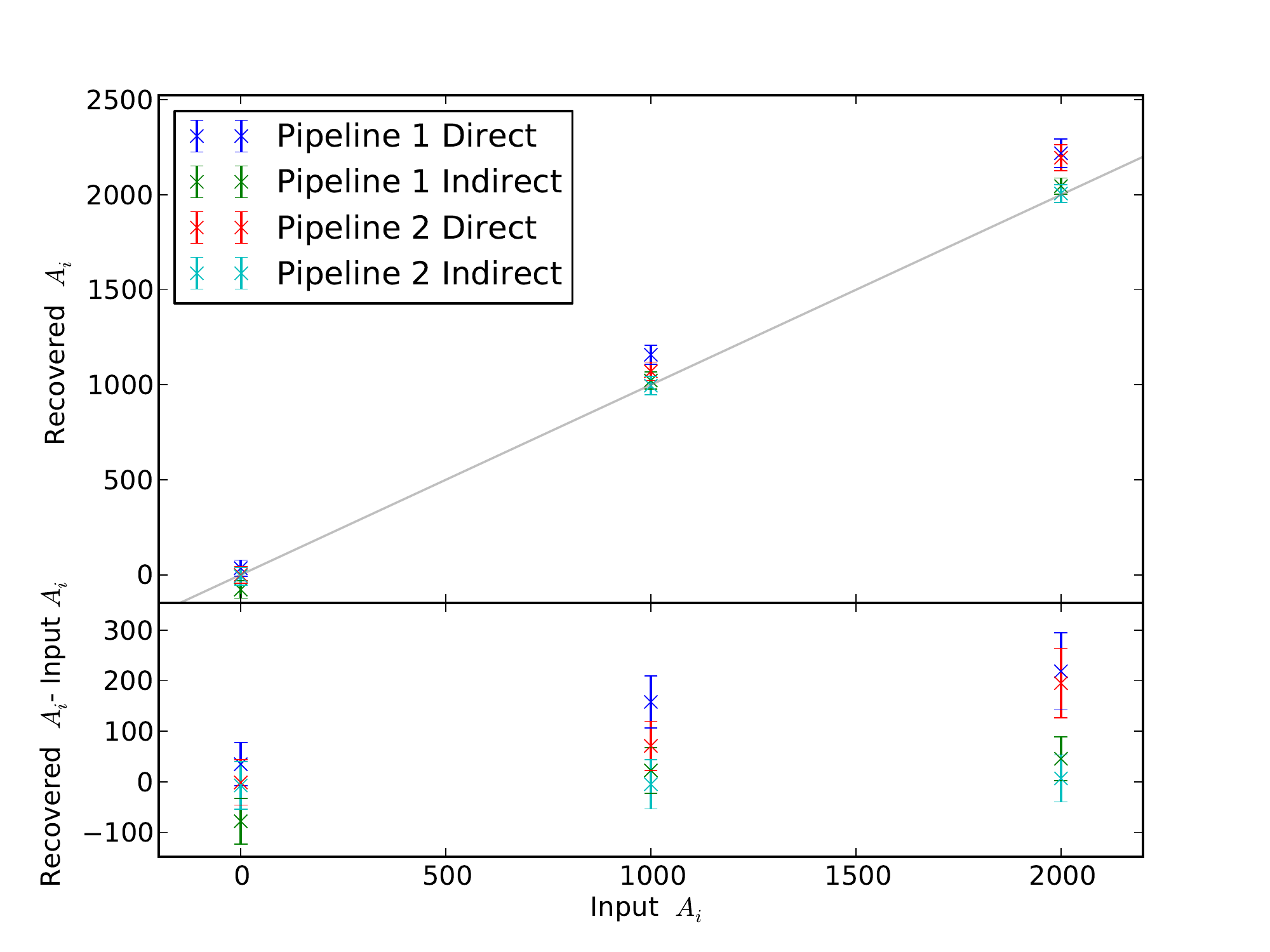}
    \label{fig:fnl_simLocal}}\qquad
\subfloat[Pseudo local non-Gaussianity with $\ell_{\mathrm{max}}=3000$.]{
    \centering
    \includegraphics[width=0.5\textwidth]{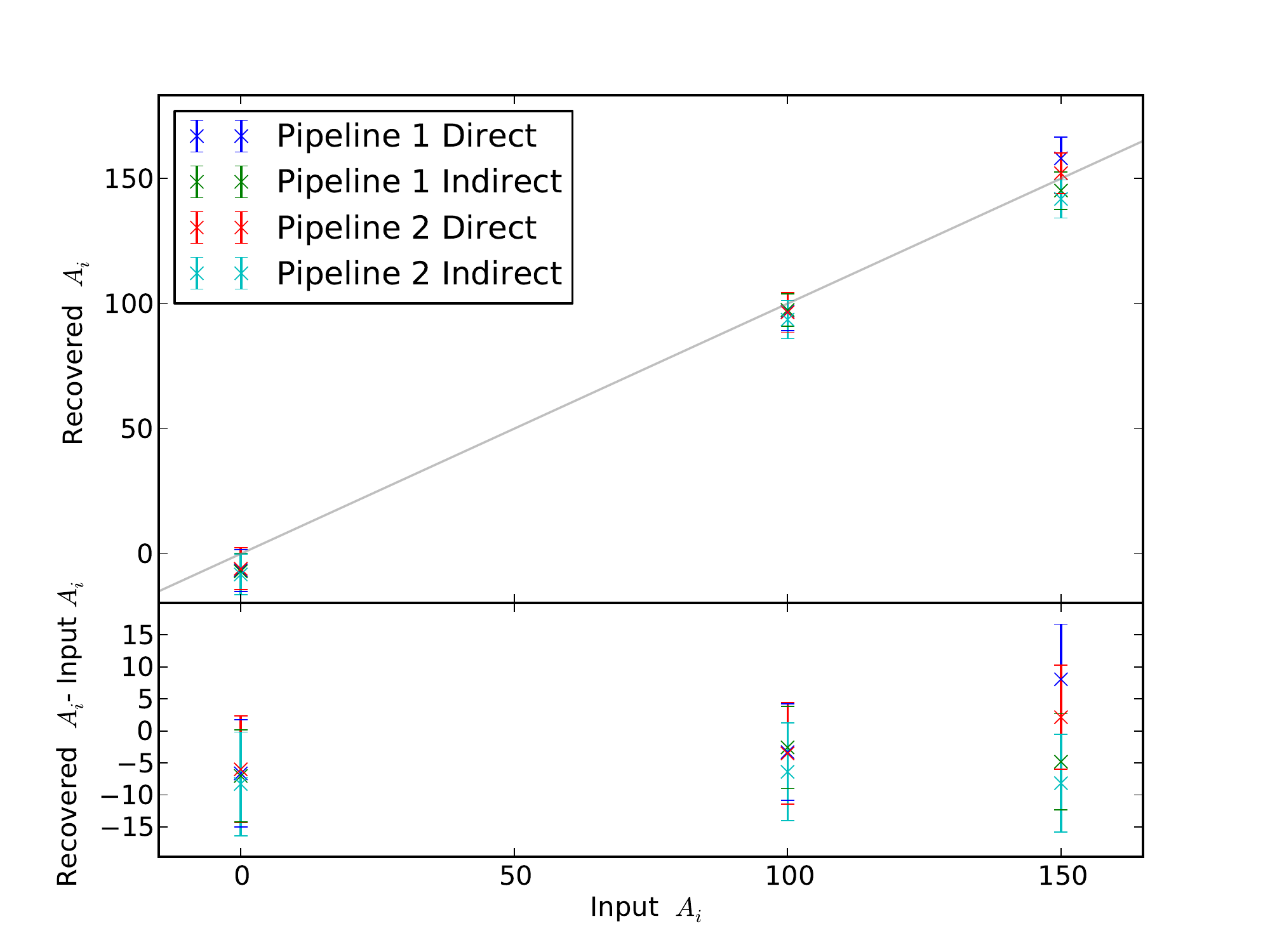}
    \label{fig:fnl_simPseudoTrim}}
\caption{A set of graphs demonstrating how our pipelines can recover an input level of non-Gaussianity, input A$_{i}$, for a set of different templates with different max $\ell$. The amplitudes, $A_i$, are overall scalings of our templates and so are dimensionless. We compare our two methods of simulating non-Gaussian templates, the ``direct'' method and the ``indirect'' method, as described in section \ref{sec:nonGausSims}. Having verified that our estimators are unbiased on this set of templates we then use the ``indirect" method to validate our full range of templates, which cannot easily be verified by direct simulation methods.}
\label{fig:simNonG}
\end{figure}

In figure \ref{fig:simNonG} we compare the accuracy of our simulation methods for several different types of non-Gaussianity with various different cut offs obtained from sets of $100$ simulations of PA2 observations. We see that there is generally good agreement between the different methods and pipelines. However we see that the residuals are predominantly slightly negative, this indicates a slight bias in our analysis at the level of $\sim 1\% $. We believe this arises from ignoring the linear term in our normalization. For the measurements reported here this effect is not significant but for future precision measurements this should be accounted for. It should be noted that the variances of the two non-Gaussian simulation methods for large non-Gaussianity is different. This is because the method in \cite{Smith2011} accurately reproduces the power spectrum and bispectrum but does not reproduce the higher order statistics, which results in an inaccurate variance for the bispectrum (as the variance of the bispectrum is a six point function). 

\begin{figure} 
\subfloat[Local non-Gaussianity with $\ell_{\mathrm{max}}=3000$.]{
  \centering
    \includegraphics[width=.49\textwidth]{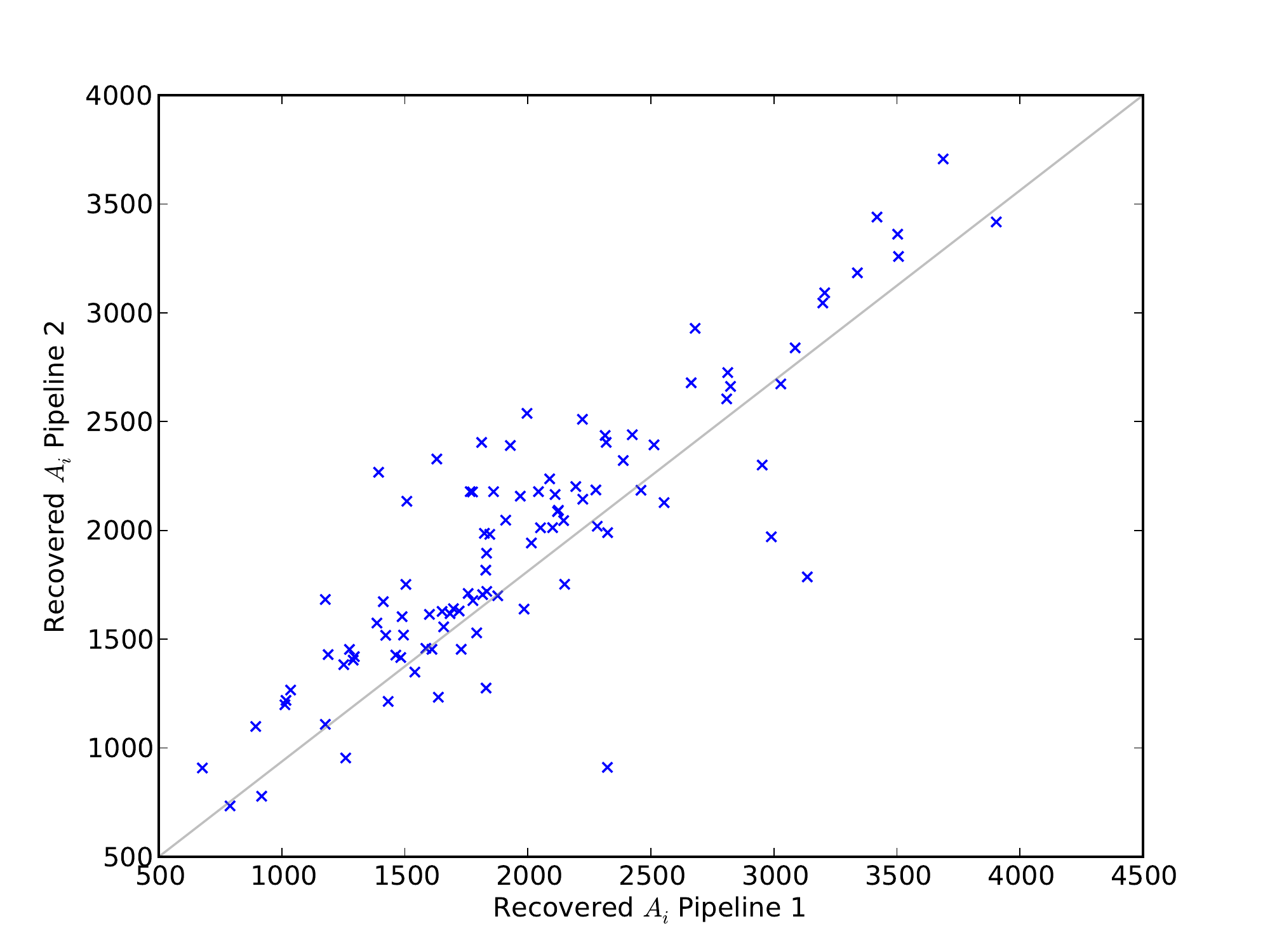}
    \label{fig:fnl_mapBymapLocal}}
    \qquad
\subfloat[Poisson non-Gaussianity with $\ell_{\mathrm{max}}=3000$,]{
    \centering
    \includegraphics[width=.49\textwidth]{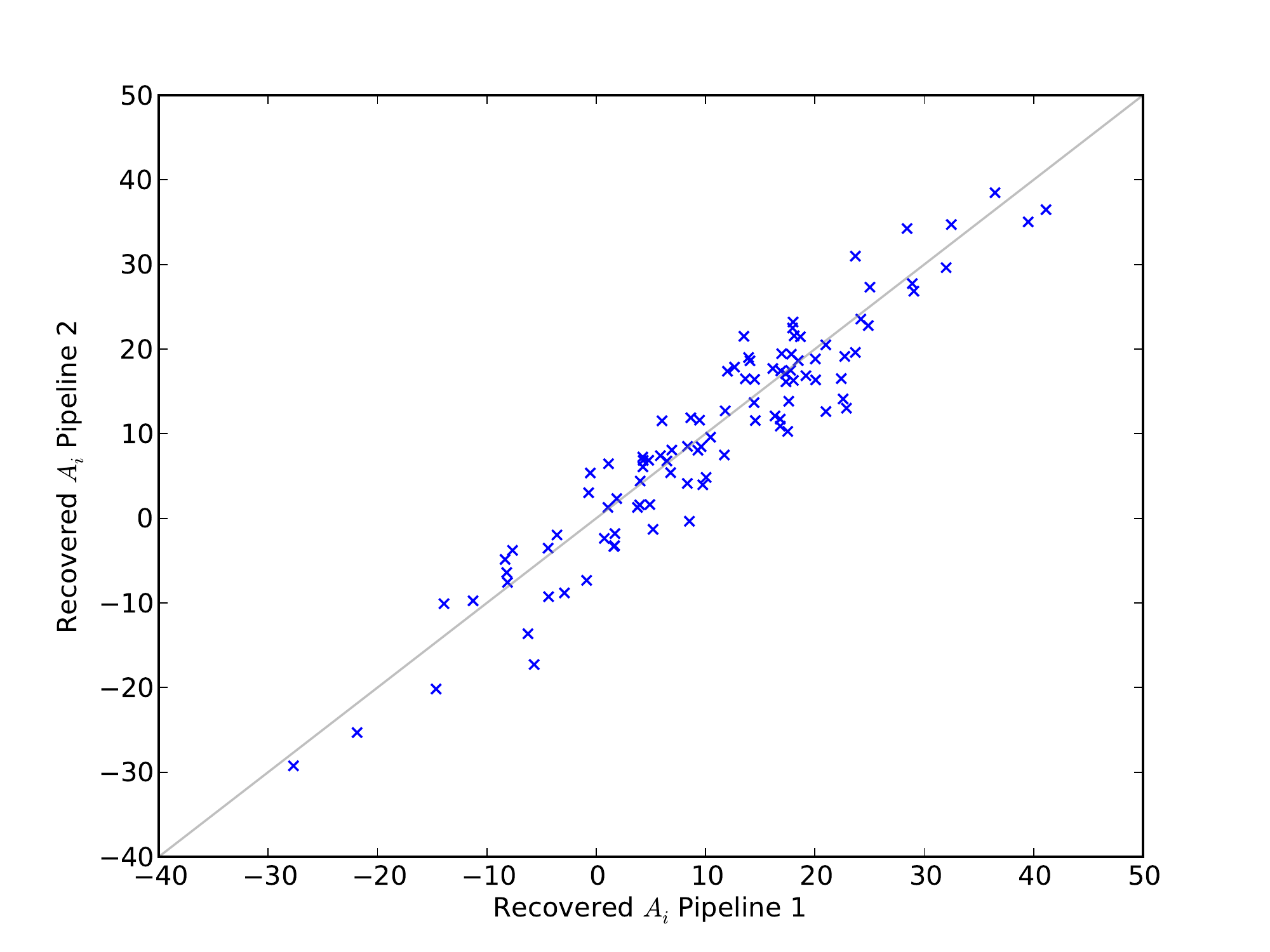}
    \label{fig:fnl_mapBymapLocal}}
\caption{A map-by-map comparison of our estimators applied to maps with local type and Poisson noise non-Gaussianity simulated with the direct method. The input level of non-Gaussianity for local type this is $A_{\mathrm{local}}=2000$ and for Poisson noise this is $A_{\mathrm{Poisson}}=10.6$. It can be seen the two pipelines agree well on a map by map basis, increasing the confidence in the accuracy of our pipelines.}
\label{fig:MapCom}
\end{figure}

Another useful diagnostic is to ensure that our pipelines measure similar levels of non-Gaussianity on individual maps, as well as for the statistical ensemble. In figure \ref{fig:MapCom} we view the recovered amplitudes, A$_{\mathrm{i}}$, for our local and constant estimator on a map-by-map basis for our two analysis pipelines. We see that in general there is reasonable agreement between the two methods. The scatter seen in figure \ref{fig:MapCom} is larger than the Gaussian error of table \ref{tab:stdDis}; this is thought to arise from the non-Gaussian distribution, and enlarged variance, of the estimator when applied to maps with non-zero signal \citep[see][for more details]{SmithT2011,Creminelli2007,Liguori2007}. Together, these tests give us confidence that our estimators are unbiased and efficient. 

\section{Template Errors} \label{app:TemplateErrors}
When the non-Gaussianity is weak the variance on the template is given by the inverse of the normalization term. However as we have detections of these templates, and further as we wish to estimate the template amplitude rather than just the significance of the deviation from null, it is important to either verify that these errors are still dominant or, in the case that they are subdominant, calculate the correct error bars. For simplicity we will consider the case when we only have a single map, but the generalization is simple. From our estimator the full covariance is given as: 
\begin{align}\label{eq:sixPointTerms}
\langle( \hat{A}_i-{\hat{A}}^{\rm{mean}}_i)(\hat{A}_j-{\hat{A}}^{\rm{mean}}_j)\rangle = &\sum\limits_{i',j'}\mathcal{N}^{-1}_{i,i'}\mathcal{N}^{-1}_{j,j'}\sum\limits_{\bm{\ell}_k,t_i} \delta^{(2)}(\bm{\ell}_1+\bm{\ell}_2+\bm{\ell}_3) \delta^{(2)}(\bm{\ell}_4+\bm{\ell}_5+\bm{\ell}_6) b^{i'}_{\ell_1,\ell_2,\ell_3}
b^{j'}_{\ell_4,\ell_5,\ell_6} C^{-1}_{\bm{\ell}_1}C^{-1}_{\bm{\ell}_2}C^{-1}_{\bm{\ell}_3}C^{-1}_{\bm{\ell}_4}C^{-1}_{\bm{\ell}_5}C^{-1}_{\bm{\ell}_6} \nonumber \\ 
& \left( C_{\bm{\ell}_1}C_{\bm{\ell}_2}C_{\bm{\ell}_3}  \delta(\bm{\ell}_1+\bm{\ell}_4)\delta(\bm{\ell}_2+\bm{\ell}_5)\delta(\bm{\ell}_3+\bm{\ell}_6) +\delta(\bm{\ell}_1+\bm{\ell}_4)b^{t_i}_{\ell_1,\ell_2,\ell_3}b^{t_j}_{\ell_4,\ell_5,\ell_6}+\delta(\bm{\ell}_1+\bm{\ell}_4) C_{\bm{\ell}_1}T_{\ell_2,\ell_3,\ell_5,\ell_6} \right. \nonumber \\ 
& \left. +S_{\ell_1,\ell_2,\ell_3,\ell_4,\ell_5,\ell_6} +V^{SSV}_{\ell_1,\ell_2,\ell_3,\ell_4,\ell_5,\ell_6} +\mathrm{ permutations} \right)
\end{align}
where $S$ is the fully connected six point function, $T$ is the trispectrum and $V^{SSV}$ is the halo sample variance or the super sample variance \citep{hamilton2006}. The first term is the `Gaussian' term described in equation \ref{eq:estGausCov}.  To calculate these contributions we use the halo model and keep only the one-halo term. This results in the following form for the trispectrum:
\begin{align}
T_{\ell_2,\ell_3,\ell_5,\ell_6}=\sum\limits_{i,j,k,l} \int_0^\infty \mathrm{d} z \frac{\mathrm{d}^2V}{\mathrm{d}z\mathrm{d}\Omega} \int_0^\infty \mathrm{d} \ln M \frac{\mathrm{d} n}{\mathrm{d} \ln M} X^i_{\ell_2}X^j_{\ell_3}X^k_{\ell_5}X^l_{\ell_6}
\end{align}
where $X^i_{\ell}$ is the Fourier transform of the integrated line of sight halo property, {\it i.e.} $X^1$ could be the Compton Y parameter and $X^2$ could be the DSFG profile etc. We sum over all the contributions to the trispectrum from the tSZ effect, DSFGs and radio galaxies (including both the Poisson and clustered terms). Similarly for the six point function:
\begin{align}\label{eq:6pnt_6connected}
S_{\ell_1,\ell_2,\ell_3,\ell_4,\ell_5,\ell_6}=\sum\limits_{i,j,k,l,m,n} \int_0^\infty \mathrm{d} z \frac{\mathrm{d}^2V}{\mathrm{d}z\mathrm{d}\Omega} \int_0^\infty \mathrm{d}\ln M \frac{\mathrm{d} n}{\mathrm{d} \ln M}X^i_{\ell_1} X^j_{\ell_2}X^k_{\ell_3}X^l_{\ell_4}X^m_{\ell_5}X^n_{\ell_6}
\end{align}
where we again sum over all of the contributions listed above. The super sample variance term can be though of additional variance induced by the presence of modes longer than the survey volume. For example, if the region is located near a peak of such a long mode then more regions will have density fluctuations greater than the level required for collapse, resulting in more halos. We refer the reader to \cite{takada2013} and \cite{schaan2014} for a more detailed description of the super sample variance term. As is shown in \cite{kayo2013} and \cite{schaan2014} the one-halo term is most important for the bispectrum and has the following form:
\begin{align}
V^{SSV}_{\ell_1,\ell_2,\ell_3,\ell_4,\ell_5,\ell_6} =\sum\limits_{i,j,k,l,m,n} \int_0^\infty \mathrm{d} z \, \sigma^2_s(V_s) \frac{\mathrm{d}^2V}{\mathrm{d}z\mathrm{d}\Omega} \int_0^\infty \mathrm{d} \ln M \frac{\mathrm{d} n}{\mathrm{d} \ln M}b (m)X^i_{\ell_1} X^j_{\ell_2}X^k_{\ell_3} \int_0^\infty \mathrm{d} \ln M \frac{\mathrm{d} n}{\mathrm{d} \ln M}b(m)X^l_{\ell_4}X^m_{\ell_5}X^n_{\ell_6}
\end{align}
where $b(m)$ is the linear bias and $\sigma_s(V_s)$ encapsulates the finite survey volume by:
\begin{align}
\sigma^2_s(V_s)=\frac{1}{4\pi^2}\int \mathrm{d}^2\bm{k}| W(\bm{k})|^2 P_{lin}(\bm{k},z)
\end{align}
where $ W(\bm{k})$ is the $3$D Fourier transform of the $3$D survey window function \citep{Krause2017}.

With these pieces, equation \ref{eq:sixPointTerms} could be directly calculated by using the halo model to calculate $b_{\ell_1,\ell_2,\ell_3}$, $b_{\ell_4,\ell_5,\ell_6}$, $S_{\ell_1,\ell_2,\ell_3,\ell_4,\ell_5,\ell_6}$ etc.  However such a direct calculation is very challenging and so in this work we use an ensemble average method to calculate these terms. In particular we need to exploit some form of factorization as the six point matrix $S_{\ell_1,\ell_2,\ell_3,\ell_4,\ell_5,\ell_6}$ for modern surveys is extremely large (O$(10^{40})$ elements for our data) and so cannot be calculated directly. Instead we use methods inspired by the non-Gaussian simulation methods described in \cite{Smith2011} and section \ref{sec:nonGausSims} .
As is done in \cite{Smith2011} and above in equation \ref{eq:nonGausIndirectSim} we first calculate :
\begin{equation}\label{eq:nonGausIndirectSim2}
{a_{\bm{\ell}}}^{NL}= \frac{1}{2}\sum\limits_{X_1,X_2} \int \mathrm{d}^2\ell_1 \mathrm{d}^2\ell_2 4 \pi^2 \delta^{(2)}(\bm{\ell} +\bm{\ell}_1+\bm{\ell}_2) b_{\bm{\ell},\bm{\ell}_1,\bm{\ell}_2} {C_{\bm{\ell}_1}^{-1}}a {C_{\bm{\ell}_2}^{-1}}a
\end{equation}
This has the property that $\langle a_{\bm{\ell}_1}a_{\bm{\ell}_2}a^{NL}_{\bm{\ell}_3}\rangle=4\pi^2\delta^{(2)}(\bm{\ell}_1+\bm{\ell}_2+\bm{\ell}_3)b_{\ell_1,\ell_2,\ell_3}$. We calculate this for two different Gaussian realizations, $a^{(1)},a^{(2)}$ and then calculate the following quantity:
\begin{align}\label{eq:ensmbleSixPointT1}
\int \prod_{i=1}^6 \mathrm{d}^2{\ell} S_{\ell_1,\ell_2,\ell_3,\ell_4,\ell_5,\ell_6}  {C_{\bm{\ell}_1}^{-1}}a^{(1)} {C_{\bm{\ell}_2}^{-1}}a^{(1)} {C_{\bm{\ell}_3}^{-1}}{a^{NL}}^{(1)} {C_{\bm{\ell}_4}^{-1}}a^{(2)} {C_{\bm{\ell}_5}^{-1}}a^{(2)} {C_{\bm{\ell}_6}^{-1}}{a^{NL}}^{(2)}
\end{align}
and similar terms for the other components of the six point function. As with the calculation of the estimator, we compute this by utilizing the separability of the integrand of the double integral. This means for each redshift and mass in our numerical calculation of S (or the other six point terms) we factorize the above quantity and efficiently calculate the sums by FFTs. Explicitly we would evaluate the term in equation \ref{eq:6pnt_6connected} in the follow manner. First, for notational clarify, we define
\begin{align}
f^{(1),L}(M,z)=\sum\limits_i\sum\limits_{\bm{\ell}_1}X^{i}_{\ell_1}{C_{\bm{\ell}_1}^{-1}}{a^{L}}^{(1)},
f^{(1),NL}(M,z)=\sum\limits_i\sum\limits_{\bm{\ell}_1}X^{i}_{\ell_1}{C_{\bm{\ell}_1}^{-1}}{a^{NL}}^{(1)}.
\end{align}
 $f^{(2),L}(M,z)$ and $f^{(2),L}(M,z)$ are defined in an analogous manner. Then equation \ref{eq:6pnt_6connected} can be written as
\begin{align}
\sigma^2_{6pnt}\propto \int_0^\infty \mathrm{d} z \frac{\mathrm{d}^2V}{\mathrm{d}z\mathrm{d}\Omega} \int_0^\infty \mathrm{d} \ln M \frac{\mathrm{d} n}{\mathrm{d} \ln M}f^{(1),L}(M,z)f^{(1),L}(M,z)f^{(1),NL}(M,z)f^{(2),L}(M,z)f^{(2),L}(M,z)f^{(2),NL}(M,z). 
\end{align}
 It is easily seen that the ensemble average of equation \ref{eq:ensmbleSixPointT1} generates the corresponding term in equation \ref{eq:sixPointTerms}. The calculation of the n$^{\rm{th}}$ order HOD moments required for the DSFG terms is described in Appendix \ref{app:NorderHODmoments}. By using two different Gaussian simulations we find that the majority of the contributions converge very rapidly, to within 10 percent for a single pair of simulations. We found that a handful of contributions require large numbers of simulations to converge; these terms arise from Poisson contributions to the trispectrum term such as
\begin{align} 
C_{\ell_1,\ell_4}T_{\ell_2,\ell_3,\ell_5,\ell_6}=\sum\limits_{l} \int_0^\infty \mathrm{d} z \frac{\mathrm{d}^2V}{\mathrm{d}z\mathrm{d}\Omega}\frac{\mathrm{d}I^{(3)}(\nu_1,\nu_2 ,\nu_5)}{\mathrm{d}z} \int_0^\infty \mathrm{d} \ln M \frac{\mathrm{d} n}{\mathrm{d} \ln M} u(\bm{\ell}_2+\bm{\ell}_3+\bm{\ell}_5)X^l_{\bm{\ell}_6}. 
\end{align}
As these terms are sub dominant, we neglect their contribution so that we can use orders of magnitude fewer simulations to obtained converged errors.

\section{N$^{\rm{th}}$ order HOD moments}\label{app:NorderHODmoments}
When calculating $n$ point functions of DSFG fluctuations we need expectations of drawing $n$ different galaxies from a cluster, {\it i.e.}:
\begin{equation}
\left \langle N(N-1)(N-2)..\left(N-(n-1)\right) \right \rangle
\end{equation}
Simulations show that the number of galaxies in a cluster has first a step, then a plateau and then a power law \citep{Kravtsov2004,Berlind2003}. This is understood as the galaxies being split into two: a central and satellite galaxies (with the central galaxy not necessarily being in the center). As the mass of the cluster increases the probability of having a central galaxy increases, this is the step and plateau, above a certain mass the cluster can also host satellites, whose number follows a power law. The HOD model parameterizes the probability of central and satellite galaxies. Only clusters with central galaxies can host satellite galaxies, so the probability distribution of satellite is conditional on the central galaxies, ie:
\begin{equation}
P(N_{s})=\sum\limits_{N_{c}}P(N_{s} | N_{c})P(N_{c})=P(N_{s} | N_{c}=1)P(N_{c}=1),
\end{equation}
where $N_{c}$ is the number of central galaxies and $N_{s}$ is the number of satellites. We model the distribution of central galaxies as a Bernoulli distribution, as given in eq. \ref{eq:nCentralDef}, and the satellite galaxies with a Poisson distribution, as described in eq. \ref{eq:nSatDef}. With these pieces we can compute the desired expectations values. It can be shown that for $n\geq2$ the expectation values can be written in the following form:
\begin{align}
\langle N(N-1)(N-2)..(N-(n-1))\rangle&= \left\langle (N_c+N_s)(N_c+N_s-1)(N_c+N_s-2)..\left(N_c+N_s-(n-1)\right) \right\rangle\nonumber \\&= \left\langle f(N_c,N_s)N_c(N_c-1)+ n N_c\prod\limits_{i=0}^{i=n-2}\left(N_s-i)\right)+\prod\limits_{i=0}^{i=n-1}\left(N_s-i)\right) \right\rangle.
\end{align}
The first term in the above always vanishes as $N_c \in {0,1}$ and using the Poisson statistics of the satellites we find:
\begin{equation}
\langle N(N-1)(N-2)...(N-(n-1))\rangle= n \langle N_c\rangle \langle N_s\rangle^{n-1}+\langle N_s\rangle^n
\end{equation}

\section{Cosmological constraints from the thermal Sunyaev Zel'dovich bispectrum}\label{app:sig8}

\begin{figure} 
  \centering
    \includegraphics[width=.49\textwidth]{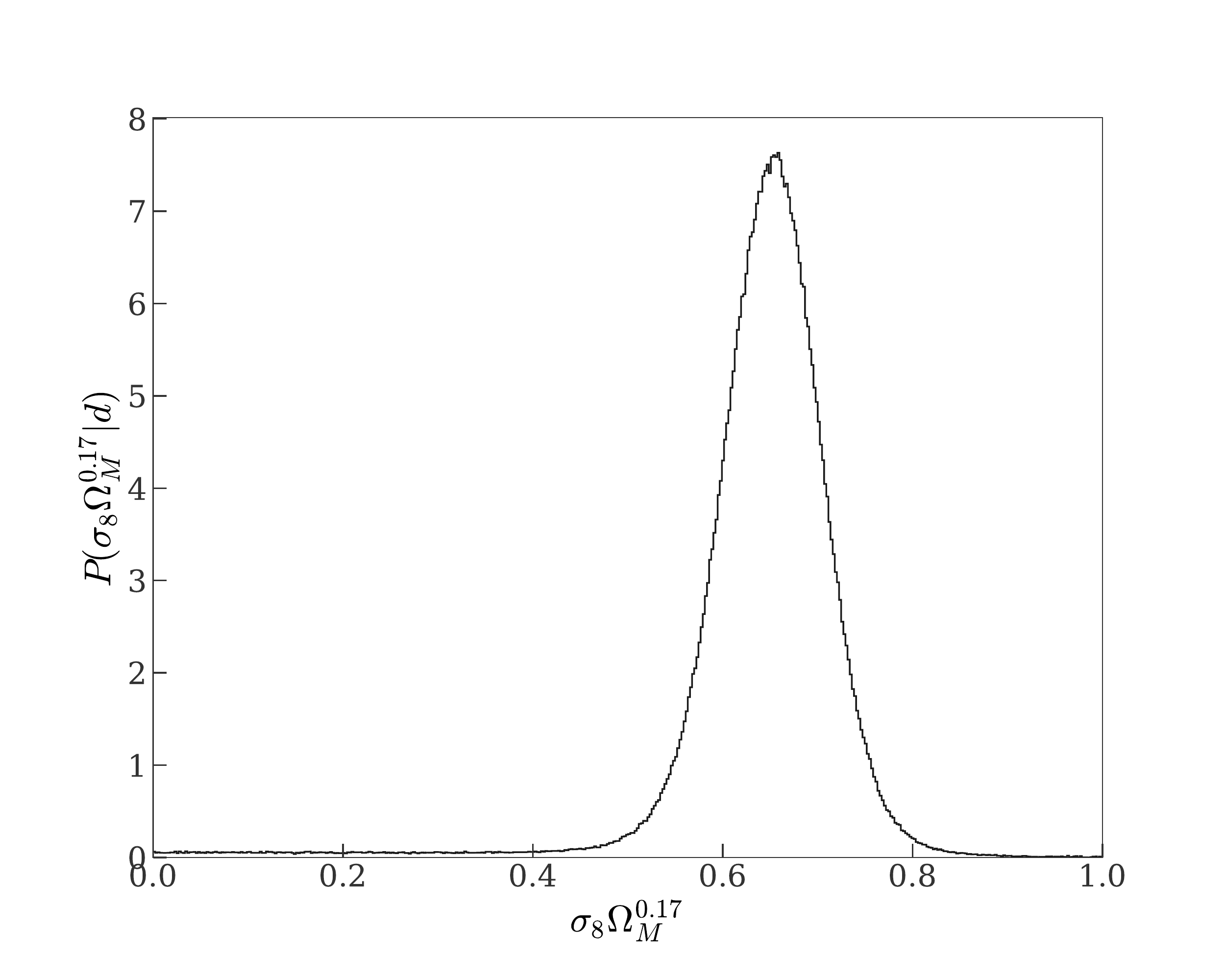}

\caption{ The posterior distribution of  $\sigma_8 \Omega_M^{0.17}$.}
    \label{fig:posterior}
\end{figure}

Previous work by \citet{bhattacharya2012,Crawford2014} has shown that measurements of the thermal Sunyaev Zel'dovich bispectrum are sensitive probes of cosmological parameters. Ideally one would perform a full Markov Chain Monte Carlo analysis to constrain cosmological and astrophysical parameters from our bispectrum amplitudes.  Unfortunately this is computationally very challenging and will be the subject of on going work. Instead we used scaling relations to model the cosmological and astrophysical dependence of bispectrum measurement. 

In this work we assumed that the cosmological dependence of measured tSZ amplitude is the same as the amplitude of our tSZ bispectrum template, $b^{\mathrm{tSZ-tSZ-tSZ}}_{\ell_1,\ell_2,\ell_3}$  evaluated at $\ell_1=\ell_2=\ell_3=3000$. We found that the cosmology dependence of the bispectrum amplitude was
\begin{equation}
 \bar{A}_{\mathrm{tSZ-tSZ-tSZ}} \propto b^{\mathrm{tSZ-tSZ-tSZ}}_{3000,3000,3000}\propto \sigma_8^{10.8}\Omega_M^{1.85}\Omega_b^{0.265}n_s^{-1.59}H_0^{-0.88},
\end{equation}
 where $\Omega_b$ is the baryon density,  $H_0$  the Hubble constant,  $n_s$ is  the scalar spectral index and $\Omega_M$ is the matter density.
 
For the tSZ power spectrum, scaling relations are sensitive to scale \citep[e.g.][]{Komatsu2002} and this will also be true for the bispectrum, with the added complication that it is likely that different bispectrum configurations will exhibit different dependencies. We chose the bispectrum amplitude at $\ell=3000$ as, when the amplitude of the tSZ effect is determined without fitting the other templates, the constraint is driven by equilateral-like configurations at a scale of around $\ell \sim 3000$. We found that the scaling relation coefficients only changed weakly with small changes in the reference scale. It should be noted that it is harder to identify which scales drive the amplitude constraint when jointly fitting all the templates and that if it is driven by the large scales $\ell<1000$ then the scaling coefficients could be inaccurate.

We assume a  Gaussian distribution for the measured amplitude
\begin{equation}
\ln P(A_{\mathrm{tSZ-tSZ-tSZ}} | \Theta) \propto  -\frac{1}{2}  \frac{(A _{\mathrm{tSZ-tSZ-tSZ}}-   \bar{A}_{\mathrm{tSZ-tSZ-tSZ}}(\Theta) )^2}{\sigma(\Omega_m,\sigma_8,H_0,\Omega_b,n_s)^2}
\end{equation}
where $\sigma^2$ is the template variance including the non-Gaussian contributions. We assume that the Gaussian contribution to the variance is independent of cosmological parameters, which is justified as the Gaussian variance is primarily determined by the number of modes combined with the CMB and detector noise, and we assume the cosmological dependence of the non-Gaussian contributions is the same as the cosmological dependence of $\bar{A}_{\mathrm{tSZ-tSZ-tSZ}}$. The cosmological dependence of the non-Gaussian contribution to the variance is very complicated as it has contributions from the three, four and six point terms and the dependence assumed here is a simple approximation. Further the true likelihood is positively skewed as the dominant source of variance is the non-Gaussian contribution from fluctuations in the number of massive halos, which is a Poisson-like process. In future work we will constrain the exact form of the likelihood and the cosmological dependence of the variance with simulations. Through the bispectrum amplitude measurement we constrain a combination of $\sigma_8$ and $\Omega_M$, specifically $\sigma_8 \Omega_M^{0.17}$. Using priors from \cite{planck2014-a15} and priors that the cosmological parameters must be positive, we marginalize over the following cosmological parameters: the baryon density, $\Omega_b$; the Hubble constant, $H_0$; and the scalar spectral index, $n_s$. Finally we assume an astrophysical modeling uncertainty of $35\%$.  This level of astrophysical uncertainty is consistent with the uncertainty used in previous work on the tSZ skew and tSZ bispectrum \citep{bhattacharya2012,Hill2013} and corresponds to an uncertainty on the hydrostatic bias parameter (1-b) of $1-b=1.0_{-0.23}^{+0.19}$ which is comparable to current constraints  \citep[see Fig 8 of ][ for a summary of current constraints on hydrostatic mass bias]{Miyatake2018}.  We jointly fit the tSZ amplitude with the other secondary templates which provides some robustness to possible contamination. However, the large degeneracies seen in section \ref{section:Results} means that our results are sensitive to the theoretical modeling; this is discussed further in section \ref{section:conclusions}. We do not include any uncertainty in our $\sigma_8 \Omega_M^{0.17}$ constraint from the joint modeling. The resulting constraint on $\sigma_8 \Omega_M^{0.17}$ is
\begin{equation}
\sigma_8 \Omega_M^{0.17}=0.65^{+0.05}_{-0.06}.
\end{equation}
In figure \ref{fig:posterior} we plot the posterior; we see a long tail extending to zero, which arises as our measured amplitude is only $3.2 \sigma$ from zero.

\bibliographystyle{act}
\bibliography{projectBib,Planck_bib}
\end{document}